\documentclass[longauth]{aa} 

\usepackage{graphicx}
\usepackage{txfonts}
\usepackage{amsmath}
\usepackage{graphicx} 
\usepackage{lscape}
\usepackage{longtable}
\usepackage{amssymb}	
\usepackage{rotating}
\usepackage{lineno}
\usepackage{natbib}
\usepackage[colorlinks,linkcolor=blue,anchorcolor=blue,citecolor=blue]{hyperref}

\newcommand{\fcont}{{$f_\mathrm{cont}$}}

\newcommand*{\LamMCMF}{$\lambda_{\mathrm{MCMF}}$}

\begin{document}

   \title{The eROSITA Final Equatorial-Depth Survey (eFEDS): Optical confirmation, redshifts, and properties \\ of the cluster and group catalog}


   \author{M. Klein
          \inst{1},
          M. Oguri\inst{2,3,4},
          J.J. Mohr\inst{1,}\inst{5},
S. Grandis\inst{1},
V. Ghirardini\inst{5},
T. Liu\inst{5},
A. Liu\inst{5},
E. Bulbul\inst{5},
J. Wolf\inst{5,19},
J. Comparat\inst{5},
M. E. Ramos-Ceja\inst{5}, 
J. Buchner\inst{5},
I. Chiu\inst{6,7,8},
N. Clerc\inst{9}, 
A. Merloni\inst{5},
H. Miyatake\inst{4,10},
S. Miyazaki\inst{11,12},
N. Okabe\inst{13,14,15},
N. Ota\inst{16,17},
F. Pacaud\inst{16},
M. Salvato\inst{5}
\and
S. P. Driver\inst{18}
          }

   \institute{Faculty of Physics, Ludwig-Maximilians-Universit{\"a}t, Scheinerstr. 1, 81679, Munich, Germany\\
              \email{matthias.klein@physik.lmu.de}
         \and
              Research Center for the Early Universe, University of Tokyo, Tokyo, 113-0033, Japan
         \and
             Department of Physics, University of Tokyo, Tokyo 113-0033, Japan 
         \and 
             Kavli Institute for the Physics and Mathematics of the Universe (Kavli IPMU, WPI), University of Tokyo, Chiba 277-8582, Japan
             \email{masamune.oguri@ipmu.jp}
         \and         
            Max Planck Institute for Extraterrestrial Physics, Giessenbachstrasse 1, 85748 Garching, Germany 
            \and
Tsung-Dao Lee Institute, and Key Laboratory for Particle
Physics, Astrophysics and Cosmology, Ministry of Education,
Shanghai Jiao Tong University, Shanghai 200240, China
\and
Department of Astronomy, School of Physics and Astronomy,
and Shanghai Key Laboratory for Particle Physics and Cosmology,
Shanghai Jiao Tong University, Shanghai 200240, China
\and
Academia Sinica Institute of Astronomy and Astrophysics (ASIAA), 11F of AS/NTU Astronomy-Mathematics Building, No.1, Sec. 4, Roosevelt Rd, Taipei10617, Taiwan
        \and
            IRAP, Université de Toulouse, CNRS, UPS, CNES, Toulouse, France
        \and
        Kobayashi-Maskawa Institute for the Origin of Particles and the Universe (KMI), Nagoya University, Nagoya, 464-8602, Japan
        \and
National Astronomical Observatory of Japan, 2-21-1 Osawa, Mitaka, Tokyo 181-8588, Japan
\and
SOKENDAI (The Graduate University for Advanced Studies), Mitaka, Tokyo, 181-8588, Japan
       \and
Physics Program, Graduate School of Advanced Science and Engineering, Hiroshima University, 1-3-1 Kagamiyama, Higashi-Hiroshima, Hiroshima 739-8526, Japan
\and
Hiroshima Astrophysical Science Center, Hiroshima University, 1-3-1 Kagamiyama, Higashi-Hiroshima, Hiroshima 739-8526, Japan
\and
Core Research for Energetic Universe, Hiroshima University, 1-3-1, Kagamiyama, Higashi-Hiroshima, Hiroshima 739-8526, Japan 
               \and
            Argelander-Institut f{\"{u}}r Astronomie (AIfA), Universit{\"{a}}t Bonn, Auf dem H{\"{u}}gel 71, 53121 Bonn, Germany 
\and            
    Department of Physics, Nara Women's University, Kitauoyanishi-machi, Nara, 630-8506, Japan
\and
           CRAR, The University of Western Australia, 7 Fairway, Crawley WA 6009, Australia      
\and           
   Exzellenzcluster ORIGINS, Boltzmannstr. 2, D-85748 Garching, Germany   }

   \date{xxx; accepted xx}

\titlerunning{eFEDS: optical confirmation and redshifts of the clusters and groups Catalog}
\authorrunning{Klein et al.}

 
  \abstract
   {In 2019, the eROSITA telescope on board of the Russian-German satellite Spectrum-Roentgen-Gamma (SRG), has started to perform a deep all-sky X-ray survey with the aim of identifying $\sim100,000$ clusters and groups over the course of four years. As part of its performance verification phase a $\sim140$ deg$^2$ survey called eROSITA Final Equatorial-Depth Survey (eFEDS) was performed. With a depth typical of the all-sky survey after four years, it allows tests of tools and methods as well as improved predictions for the all-sky survey.}
   {
   As part of this effort, a catalog of 542 X-ray extent selected galaxy group and cluster candidates was constructed. In this paper we present the optical follow-up with the aim of providing redshifts and cluster confirmation for the full sample. Further we aim to provide additional information on cluster dynamical state, cluster richness and cluster optical centre. Finally we aim to evaluate the impact of optical cluster confirmation on the purity and completeness of the X-ray selected sample.
   }
   {We use optical imaging data from the Hyper Suprime-Cam Subaru Strategic Program and from the Legacy Survey to identify optical counterparts to the X-ray detected cluster candidates. We make use of the red sequence based cluster redshift and confirmation tool MCMF as well as the optical cluster finder CAMIRA to derive cluster redshifts and richnesses. MCMF provided probabilities of an optical structure being a chance super position with the X-ray candidate is used to identify the best optical counter part as well as to confirm an X-ray candidate as a cluster. The impact of this confirmation process on catalog purity and completeness is estimated using optical to X-ray scaling relations as well as simulations. The resulting catalog is further matched with public group and cluster catalogs.
   Optical estimators of cluster dynamical state are constructed based on density maps of the red sequence galaxies at the cluster redshift.}
   {While providing redshift estimates for all 542 candidates, we construct an optically confirmed sample of 477 clusters and groups with a residual contamination of 6\%.
   Of these, 470 (98.5\%) are confirmed using MCMF, and 7 systems are added through cross matching with spectroscopic group catalogs. Using observable to observable scaling and the applied confirmation threshold, we predict $8\pm2$ real systems have been excluded with the MCMF cut required to build this low contamination sample. 
   This number is in good agreement with the 7 systems found through cross matching that were not confirmed with MCMF. The 
   predicted redshift and mass distribution of this catalog is in good agreement with simulations.   Thus, we expect those 477 systems to include $>99\%$ of all true clusters in the candidate list. Using an MCMF independent method, we confirm the catalog contamination of the confirmed subsample to be $6\pm3\%$. Applying the same method to the full candidate list yields $17\pm3\%$, consistent with estimates coming from the fraction of confirmed systems of $\sim17\%$ and expectations from simulations of $\sim 20\%$. We further present a sample of merging cluster candidates based on the derived estimators of cluster dynamical state.}
   {}


   \maketitle
%
\section{Introduction}
Galaxy clusters are 
the most massive 
collapsed halos in the Universe. 
Their abundance is sensitive to cosmological parameters, making them valuable
cosmological probes 
\citep[e.g.,][]{vikhlinin09,mantz10a,rozo10,PlanckSZcosmology,bocquet18,IderChithamCODEX2020,DESY1clucosmo}.

Furthermore, galaxy clusters are exceptional astrophysics laboratories to study galaxy evolution, the dark matter self interaction cross section, cosmic ray acceleration and many other physical quantities \citep[e.g.,][]{dressler80,moore96,clowe06,vanweeren10,harvey15}. 

An important topic for many cluster-related studies is understanding the cluster selection function and the purity of the cluster catalog. 
Cluster catalogs derived from X-ray observations have the advantage that the X-ray emission from galaxy clusters depends on the square of the electron density of the intra-cluster medium (ICM), which reduces the impact of projection effects of non-collapsed systems into the cluster catalog.
X-ray surveys with a controlled selection function are therefore an excellent source for cluster based studies \citep[e.g.,][for reviews]{rosati02,allen11}. Previous works using either small area deep data \citep{COSMOS1,COSMOS2,XXL1,XXL2} or large but shallow surveys \citep{MCXC,MARDY3,codex19} have produced useful cluster catalogs, some including as many as a few thousand X-ray selected clusters. The largest number of X-ray clusters were found using the shallow but large area ROSAT all-sky survey \citep[RASS;][]{Truemper82,RASS}.

With eROSITA \citep{Predehl2020,Merloni12}, the next generation X-ray survey telescope recently started its operation and its journey to produce a new high quality all-sky X-ray survey. As part of its performance verification program, prior to the start of the all-sky survey, a medium area ($\sim 140$ square degrees) survey was performed. The average exposure time of $\sim 1.3$ ks after vignetting corrections is comparable to that reached in the final all-sky survey in the equatorial regions.
The field location was chosen to lie within surveys with deep, multi-band imaging data from the Hyper Suprime-Cam Subaru Strategic Program \citep[HSC-SSP;][]{HSC-SSP} and the Legacy Survey \citep[LS;][]{legacysurveys}. It further overlaps the Sloan Digital Sky Survey \citep[SDSS;][]{SDSS} and one of the Galaxy And Mass Assembly \citep[GAMA;][]{GAMA} fields, both of which provide significant amounts of spectroscopic data.
This makes the eROSITA final equatorial depth survey (eFEDS) a good and early test case to validate and improve the tools and methods needed for the all-sky survey scientific exploration.

This paper is part of the eROSITA early data release and focuses on the optical confirmation and redshift estimates of X-ray selected cluster and group candidates described in Liu et al. (submitted). The goal is to provide information for a clean and reliable cluster catalog with well understood contamination and incompleteness due to optical cleaning.   For each cluster we also provide optical centres, estimators of cluster dynamical state and richness measurements as an additional mass estimator. 


\begin{figure*}
\begin{center}
\includegraphics[width=0.99\linewidth]{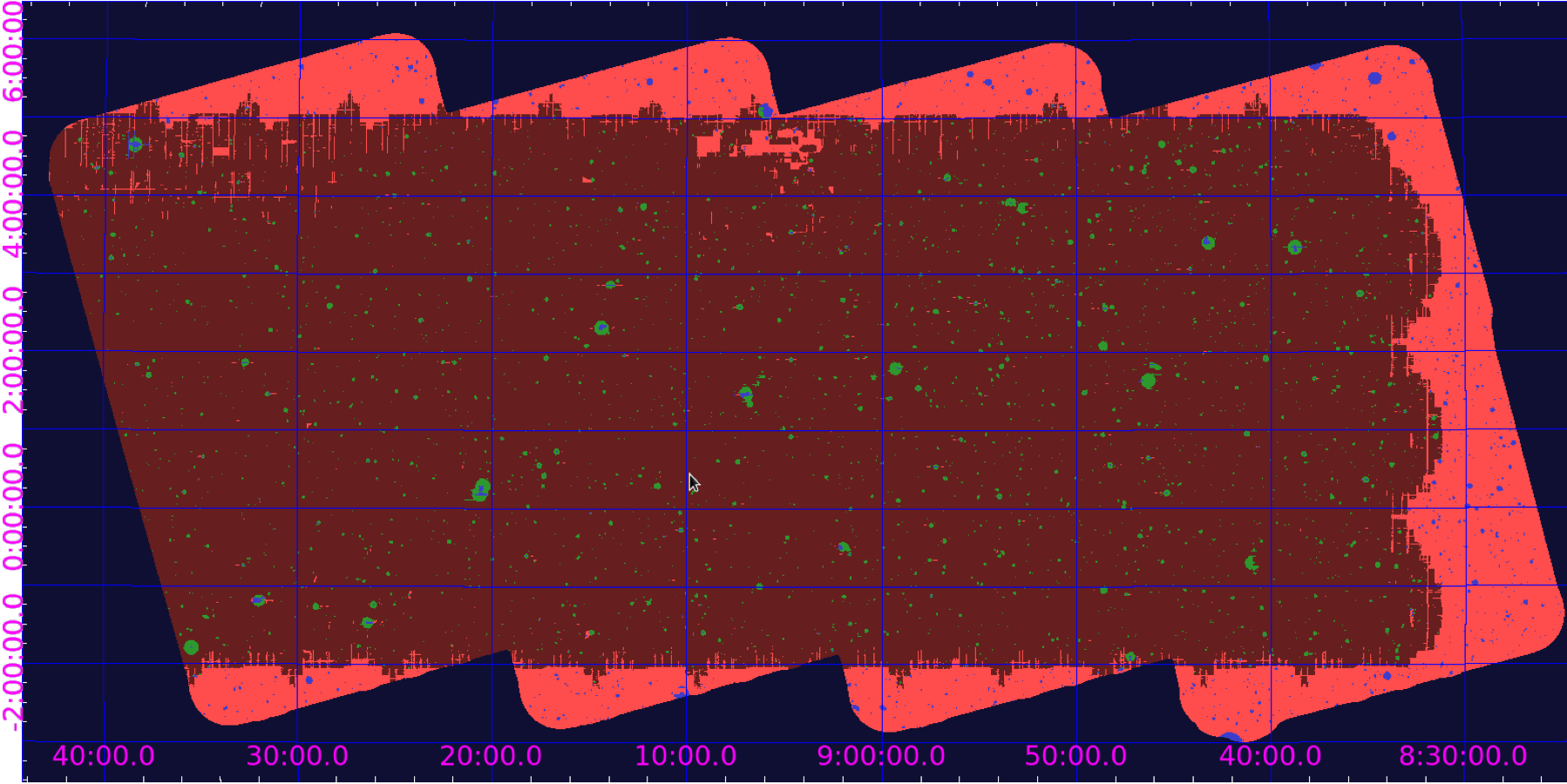}
\caption{eFEDS footprint and its coverage with HSC and LS data. Dark red shows HSC and LS coverage, light red shows area only covered by LS. Green and blue indicate masked regions, mostly caused by bright star masks. Besides masked regions the full eFEDS footprint has coverage by either HSC or LS.}
\label{fig:footprint}
\end{center}
\end{figure*}

\section{Datasets}
We restrict the description of datasets to the three main resources used in this work: the X-ray data from eROSITA and the optical imaging surveys HSC-SSP and LS. Each of them are described in dedicated papers on the data processing and data products. We therefore touch only on the most crucial aspects 
needed for this paper.
The eFEDS footprint and its coverage with optical data is shown in Fig.~\ref{fig:footprint}. 

\subsection{X-ray dataset}
The extended ROentgen Survey with an Imaging Telescope
Array \citep{Predehl2020} onboard the Spectrum-Roentgen-Gamma (SRG) consists of seven telescope modules (TMs), acting as seven separate telescopes observing the same $\sim 1$ degree diameter circular patch of the sky. The field-of-view average angular resolution is $\sim 26$\arcsec (HEW at 1.49~keV) and $\sim 18$\arcsec on-axis \citep{Predehl2020} and its effective area in the 0.5--2~keV band of the seven telescope modules is similar to that of the combined PN+MOS instruments on \emph{XMM-Newton}. 

The eFEDS field was observed by eROSITA in the first quarter of November 2019 as part of the Performance Verification phase (Brunner et al., submitted). The total time spent on this field is $\sim350\,$ks, corresponding to and average exposure time of $\sim2.3\,$ks and $\sim$1.3\,ks after accounting for vignetting.
The data were processed with the eROSITA Standard Analysis Software System (eSASS, Brunner et al., submitted). 

The X-ray cluster candidate list contains 542 sources, selected by requiring extent likelihoods greater 6, detection likelihood greater than 5 and a minimum source extent of 2 pixels (8\arcsec). From extensive X-ray simulations of the eFEDS field \citep{comparat20} we expect that $\sim80.3$\% of these sources correspond to genuine galaxy clusters.
For more details we refer the reader to the dedicated paper on the X-ray catalogs (Liu et al., submitted and Brunner et al., submitted).

\subsection{Hyper Suprime-Cam Subaru Strategic Program (HSC-SSP)}

The Hyper Suprime-Cam Subaru Strategic Program \citep{HSC-SSP} is an imaging survey conducted using the Hyper Suprime-Cam (HSC) camera on the 8.2 m Subaru telescope at Mauna Kea, Hawaii.
HSC is a wide-field (1.7 degree diameter) optical imager \citep{HSC-camera} installed at the prime focus of the Subaru telescope. The HSC-SSP survey consists of three different layers named Wide, Deep, and Ultradeep. Of interest here is the Wide layer conducted in five broad bands ($grizy$) over a total area of 1,400 square degrees to $5 \sigma$ depths of about 26.6, 26.2, 26.2, 25.3, and 24.5. 
Details of the data processing and source detection can be found in \citet{HSC-software}, \citet{HSC-DR1}, and \citet{HSC-DR2}.

Using the most recent S20A data from HSC-SSP results into a coverage of $\sim95$\% of the eFEDS footprint in $griz$-band. 
In contrast, the public data release 2 \citep[PDR2;][]{HSC-DR2} only contains data out to S18A (2018).
So, in this work we use the non-public data stemming from observations until 2019 and 2020, which are referred to as S19A and S20A, respectively.  
Throughout the paper we utilize {\tt cmodel} magnitudes derived from light profile fitting as total magnitudes of galaxies, while colors of each galaxy are derived from the point spread function (PSF)-matched aperture photometry without deblending with the target PSF FWHM of $1\farcs3$ and the aperture diameter of $1\farcs5$ in order to mitigate an issue of deblending failure in crowded areas \citep{CAMIRA-HSC,HSC-DR1}. We only use galaxies with $z$-band {\tt cmodel} magnitudes after the Galactic extinction correction \citep{SDF} brighter than 24, which correspond to 10$\sigma$ detection significance for extended sources, and their errors smaller than 0.1 mag. We select extended objects based on the star-galaxy separation in $i$-band images ({\tt i\_extendedness\_value}) because $i$-band images tend to be taken in better seeing conditions for weak lensing shape measurements.

\subsection{Legacy Survey (LS)}
The DESI Legacy Imaging Surveys \citep[LS;][]{legacysurveys}, is a combination of four imaging surveys, the 9,000 $\text{deg}^2$ $grz$-band DECam based DECaLS survey, the 5,000 $\text{deg}^2$ BASS and MzLS surveys providing photometry in gr and $z$-band, respectively, and the WISE and NEOWISE surveys in the mid-IR at 3.4$\mu$m and 4.6$\mu$m. Data release 8 of LS further includes archival observations and covers an area of 19,000 $\text{deg}^2$ in total. Source detection and photometry is performed using \texttt{the tractor}\footnote{http://ascl.net/1604.008} software \citep{tractor} in optical bands. On the mid-IR data forced photometry with deblending is performed to take the PSF into account. The eFEDS footprint lies within the DECaLS footprint which shows median 5$\sigma$ depth's of 23.72, 23.27, and 22.22 in $grz$-band.

\section{Methods}

For the definition of the cluster and group candidate catalog a compromise has to be made between catalog purity and completeness and-- related to that-- catalog size.
By examining simulated X-ray surveys of eFEDS--like fields (Liu et al., submitted and Brunner et al, submitted) that use the same definition of the candidate list, we expect about $\sim19.3$\% of the cluster candidates 
to be non-cluster contaminants.  Furthermore, the X-ray selection function strongly depends on redshift, reaching mass ranges well within the group regime at low redshifts while probing relatively high mass systems at the high redshift end.

Therefore, the optical follow-up has to tackle two challenges: (1) identifying those systems that are likely not real clusters at all,  and (2) assigning the best optical counterpart in cases where there are multiple systems along the line of sight. To this end we have developed the multicomponent matched filter (MCMF) cluster confirmation tool \citep{MCMF1,MARDY3}.

In addition to applying MCMF, we execute a forced run of the CAMIRA optical cluster finder \citep{CAMIRA,CAMIRA-HSC} and then cross match to the standard CAMIRA optical cluster catalog using optical data from the HSC-SSP. 

\subsection{MCMF}\label{sec:mcmf}
The MCMF method is described in detail elsewhere \citep{MCMF1,MARDY3}. In this section we therefore restrict ourselves to the basic description and to modifications compared to previous work.
The core of MCMF makes use of the red sequence of cluster galaxies \citep{gladders00} and calculates the weighted number-- called richness ($\lambda$)-- of excess galaxies within a certain magnitude and radial range around the X-ray position.

MCMF calculates the richness in 230 redshift bins out to $z=1.3$. For each redshift bin an aperture corresponding approximately to $r_{500}$ is estimated, based on the X-ray count rate, the redshift bin and an observable-mass scaling relation. 

In this work we use the luminosity-mass scaling relation given in \cite{bulbul19},
\begin{eqnarray}
\label{eq:xray_type_II}
L_{500,0.5--2.0 \mathrm{~keV}} = A_\mathrm{X}\left( \frac{M_{500}}{M_\mathrm{piv}} \right)^{B_\mathrm{X}} 
\left(\frac{E(z)}{E(z_\mathrm{piv})}\right)^2
\left( \frac{1+z}{1+z_\mathrm{piv}} \right)^{\gamma_\mathrm{X}},
\end{eqnarray}
where $A_\mathrm{X}$, $B_\mathrm{X}$ and $\gamma_\mathrm{X}$ were found to have best fit values of $4.15\times10^{44}$~erg s$^{-1}$, 1.91 and 0.252 respectively. The pivot mass $M_\mathrm{piv}$ and redshift $z_\mathrm{piv}$ are $6.35\times10^{14} M_\odot$ and 0.45. 
For the richness estimate we solve this equation for $M_{500}$ to obtain $r_{500}$ using a simplified count rate to luminosity conversion. For this conversion we take the count rate from the source detection and assume a MEKAL model with fixed temperature of 3~keV and metallicity of 0.3 $Z_{\odot}$ and cosmology of $\Omega_M=0.3$, $\Omega_\Lambda=0.7$, $H_0=70~{\rm km\, s}^{-1}{\rm Mpc}^{-1}$ to obtain an approximate $L_{500}$. 
With that, the conversion from count rate to our approximate $M_{500}$ solely depends on redshift and hydrogen column
density, where the latter is unimportant for the column densities found in the eFEDS area. 

The distribution of richness as a function of redshift obtained in this way is then searched for peaks, 
and up to three potential candidates are registered. The same approach is then repeated with randomized positions that exclude regions around the real cluster candidates. With the set of richnesses and redshifts of real candidates and from random lines of sight we calculate for each candidate $i$ the estimator $f_{\mathrm{cont},i}$ which is defined as,
\begin{equation}
   f_{\mathrm{cont},i}=\frac{\int_{\lambda_i}^{\infty} f_\mathrm{rand}(\lambda,z_i) d\lambda}{\int_{\lambda_i}^{\infty}
f_\mathrm{obs}(\lambda,z_i) d\lambda},
\end{equation}
where $f_\mathrm{rand,z}$ is the richness distribution of randoms at the cluster candidate redshift $z_i$, $f_\mathrm{obs}(\lambda,z_i)$ is the richness distribution of true candidates and $\lambda_i$ is the richness of the cluster candidate. Because the number of candidates and independent randoms are limited, the calculations are done in multiple redshift bins and estimates for specific redshifts are then derived using interpolation.

The estimator $f_{\mathrm{cont},i}$ is correlated with the probability of a source being a chance superposition, and it can therefore be used to control the overall sample contamination. Selecting all sources $f_{\mathrm{cont},i}<0.3$ means that we statistically allow for a 30\% contamination, caused by sources showing similar richnesses and redshifts as the candidates that nonetheless have no physically associated extent selected X-ray source.  These are random superpositions of X-ray candidates with optical systems that just happen to lie along the line of sight.
As long as the contaminants allowed by the optical selection and the contaminants allowed by the X-ray selection are uncorrelated, the applied cut in \fcont\
can be seen as a factor describing the reduction of the initial contamination of the cluster candidate catalog.
Further, the estimate of \fcont\ for each of the peaks found for a single cluster candidate allows one to select the most likely counterpart as the one showing the lowest \fcont. Ambiguity only arises if multiple sources show very low \fcont\; this is typically the case for $\sim$2\% of the sources and can be traced by the difference between the lowest and the second lowest \fcont\ estimate of a given source.

We note here that all X-ray masses called approximate $M_{500}$ in this paper are referring to the estimates used for the richness measurement. A bias as well as redshift and mass trends to true halo mass can be expected. This can impact the estimated richness but has no significant impact on the redshift estimation. As \fcont\ is derived as function of redshift and using randoms which share the same characteristics as the candidates, the estimate is robust against systematics in the adopted $M_{500}$ versus halo mass relation.  We also note here that in parallel to this work, X-ray count rates and luminosities are re-extracted and measured within $r_{500}$ (Bahar et al., in prep).  Additionally, a new luminosity-mass relation is obtained using HSC weak lensing (Chiu et al., in prep.).

\subsubsection{MCMF on HSC}
Similar to our previous work using data from the Dark Energy Survey (DES), we make use of the $griz$-bands. We omit the $y$-band because it does not significantly contribute to the MCMF performance.  This choice also allows us to maximise the footprint with full color coverage. We make use of clusters with spectroscopic redshifts over the full HSC-SSP area to calibrate $\lambda$ versus redshift peak profiles and the red sequence models.
The HSC flagging to mask regions around bright stars are chosen quite conservatively due to drivers from weak gravitation lensing studies \citep{coupon18}, and therefore the masked regions cove a significant fraction of the footprint. Originally, we performed two MCMF runs, one with, one without using the flags. We did not find significant issues with the run ignoring the near bright source flag and
therefore decided to use this setting as default. However, we provide a flag if a source center is lying within a flagged region. We also report the fraction of area within $r_{500}$ that is flagged.

\subsubsection{MCMF on Legacy Survey}
For the Legacy Survey we make use of the $grz$-bands and the W1 band from unWISE \citep{unWISE1,unWISE2}. MCMF was run twice: once only with $grz$-band, and once adding the W1 band. The $z$-W1 color provides improved redshifts at the high redshift end and the tractor photometry addresses the significant blending that occurs in the center of clusters in the W1-bands. The star-galaxy separation is performed out to $z=21.7$ mag, and the residual stellar contamination is removed statistically by using a local background for the richness measurement.

\begin{figure}
\begin{center}
\includegraphics[width=0.99\linewidth]{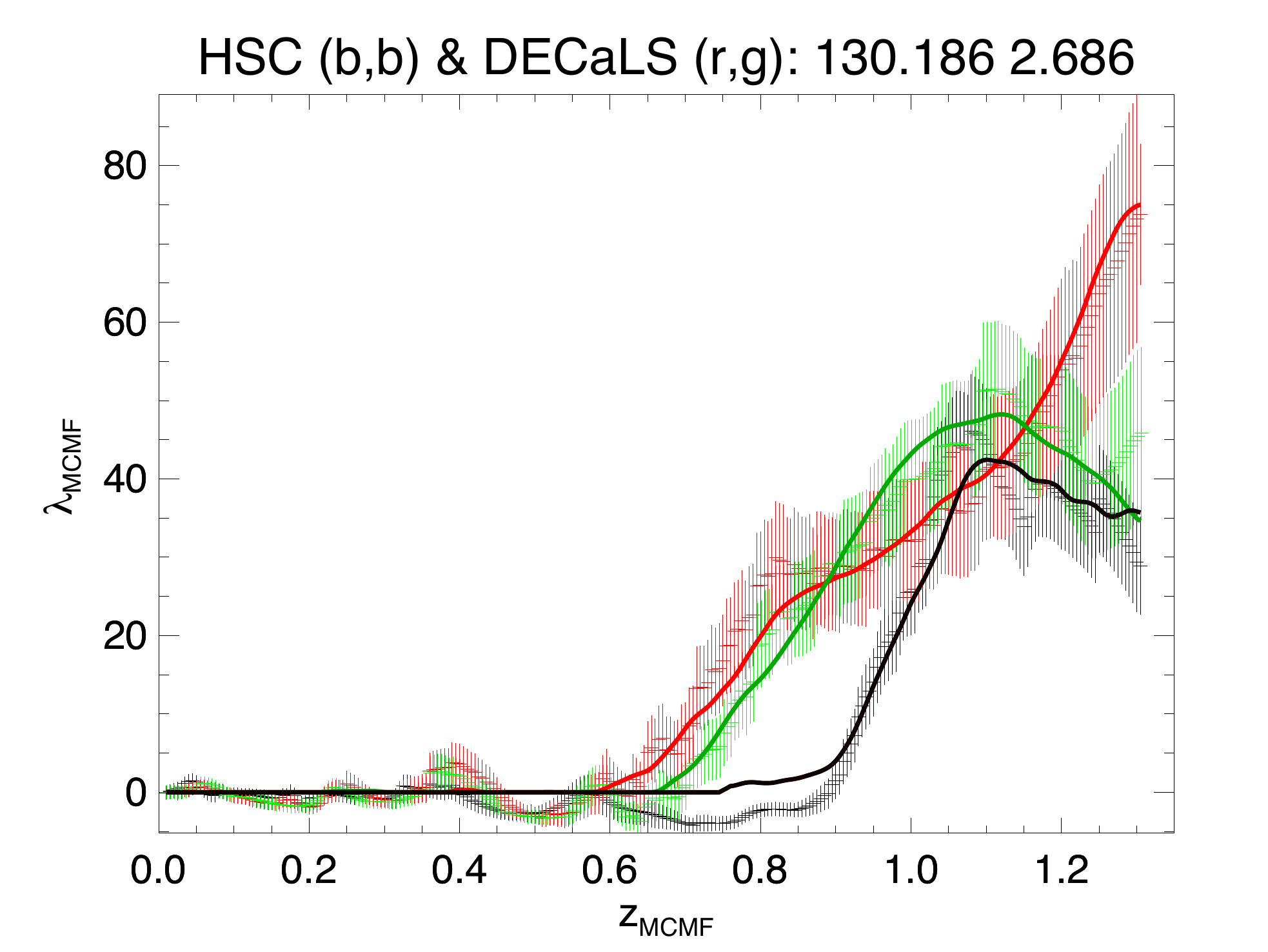}
\caption{MCMF richness versus redshift plot for the $z=1.1$ cluster eFEDS~J084044.7+024109. In red and green the output of MCMF on LS data, using $grz$ and $grz$W1, respectively. The run on HSC is shown in black. The continuous lines show the best fit peak profile at the cluster redshift.}
\label{fig:id8131}
\end{center}
\end{figure}

An example of the redshift estimate of the $z=1.1$ cluster eFEDS~J084044.7+024109 is shown in Fig.\ref{fig:id8131}. MCMF fits so-called peak profiles to the $\lambda$ versus redshift distribution to obtain best fit redshift and richness estimates. 
MCMF finds consistent redshifts in all three MCMF runs. The MCMF runs on the different optical data result in vastly different peak shapes.
Nonetheless, the peak shapes predicted and observed peaks for each optical dataset are in good agreement. At the high redshift end, adding the w1 band to the LS MCMF run significantly changes the peak profile from a steady rising function to a peaked profile. 
This demonstrates how the $w1$ band data helps improving the MCMF performance at high redshift.

\subsubsection{Combining HSC and LS results}

\begin{figure}
\begin{center}
\includegraphics[width=0.99\linewidth]{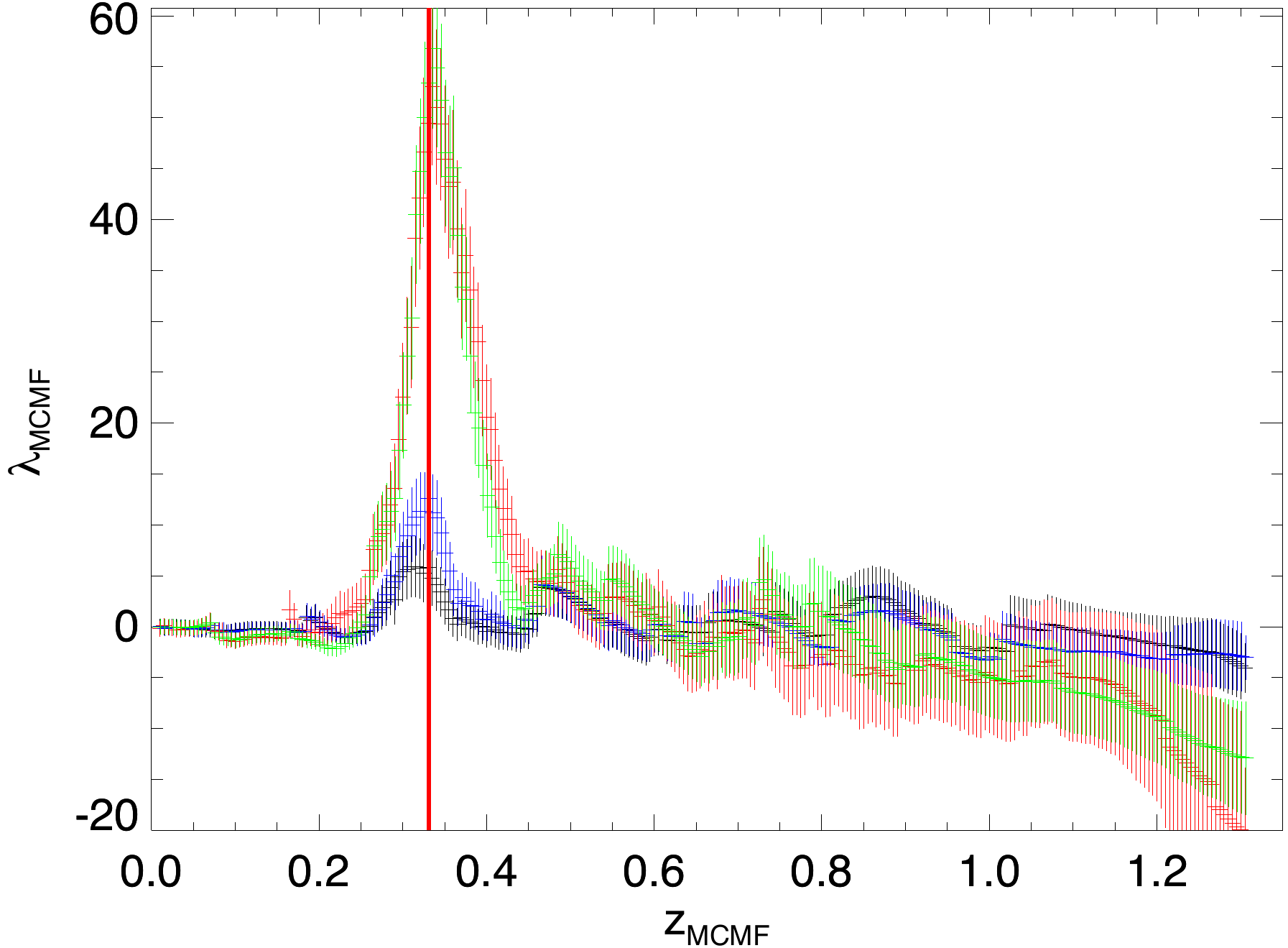}
\caption{Similar to Fig.\ref{fig:id8131}, richness versus redshift plot for eFEDS~J092739.8-010427 for four MCMF runs. The runs using LS (red and green) both show a clear cluster peak and agree well with one other. The MCMF runs using HSC S19A data (black) and S20A data (blue) show a significantly smaller peak due to local offsets in the photometric zero point.}
\label{fig:2636}
\end{center}
\end{figure}

Similar to what is shown in Fig.\ref{fig:id8131}, the structures found in the MCMF runs on the different datasets are usually in excellent agree with one other. Differences mostly occur either at the high redshift end where the shallower LS data and the missing $i$-band information causes redshift and richness estimates to be noisier. A second reason for mis-matches is the local bright star mask or patchy data at the field corners. A third obstacle is that the used S19A and even more frequent S20A HSC data suffer from photometric zero point issues at a few locations \citep[see][]{HSC-DR2}.
MCMF combines the red-sequence based weights of all three colors $g-r$, $r-i$, and $i-z$. If at least one color is significantly off in terms of expected width of the red sequence, the overall richness is significantly biased low. Because the observed red sequence width depends on redshift, the impact of the zero point issues is most prominent at low redshifts.
An example for this effect is shown in Fig.\ref{fig:2636}.

As the LS suffers less from this issue and provides homogeneous coverage over the full eFEDS footprint, we choose the MCMF run on LS in the $grz$W1 mode as default for cluster confirmation for $z<0.5$. For counterparts at higher redshifts, the HSC measurements were used. For sources that suffer from a local lack of data in either of the surveys  the entries from the least affected MCMF run are used.

The combined redshift, richness and confirmation estimates are provided alongside the key entries of the MCMF runs on both surveys in the results table.
The best counterpart is chosen as the one showing the lowest \fcont\ estimate.

\subsubsection{Spectroscopic redshifts}
The location of the eFEDS field was partially chosen because of its overlap with spectroscopic surveys such as the 2MRS \citep{2MRS}, SDSS \citep{SDSS} and GAMA \citep{GAMA} surveys. We make use of the public data and list spectroscopic redshifts for clusters that have either a measurement for the brightest cluster galaxy (BCG) or for at least three cluster members within $r_{500}$.

\subsubsection{Optical estimators of cluster dynamical state}\label{sec:optdynest}

As in \citet{MARDY3}, we calculate estimators of cluster dynamical state based on the optical data. The estimators in this work
are mostly identical to those computed in our previous work. We therefore
only briefly summarise the provided estimators.

A first set of three estimators are closely related to those described in \cite{Wen13}. The main difference between \cite{Wen13} and those provided here is the different kind of galaxy density map.
In \citet{Wen13} the maps used for the dynamical state estimators are based on galaxy positions and $r$-band luminosities of sources with redshifts within 4\% of the cluster redshift. In this work we use the galaxy density maps from the 
the MCMF pipeline, which includes galaxies consistent with the red sequence at the cluster redshift smoothed with a fixed 125~kpc Gaussian kernel.

The asymmetry factor $\alpha$ is defined as the ratio of the 'difference power' over the 'total fluctuation power' 
within $r_{500}$
\begin{equation}
  \alpha=\frac{\sum_{i,j}[I(x_i,y_j)-I(-x_i,-y_j)]^2/2 }{\sum_{i,j}I^2(x_i,y_j)},
\end{equation}
where $I(x_i,y_j)$ is the value of the density map at cluster centric position $x_i$, $y_j$.
The normalized deviation $\delta$ is based on a fit of a 2D King model \citep{king62}
\begin{equation}
  I_\mathrm{2Dmodel}(x,y)= \frac{I_0}{1 + (r_\mathrm{iso}/r_0)^2},
\end{equation}
where $I_0$ is the peak intensity at the cluster center, $r_0$ is the characteristic radius and $r_\mathrm{iso}$ describes the profile isophote with
$r_\mathrm{iso}^2= (x \cos \theta + y \sin \theta)^2 + \epsilon(-x\sin \theta + y \cos \theta)^2$.
The estimator $\delta$ is then the normalized deviation of the residual map within $r_{500}$ after 
subtraction of the 2D model
\begin{equation}
\delta=\frac{\sum_{i,j}[I(x_i,y_j)-I_\mathrm{2Dmodel}(x_i,y_j)]^2}{\sum_{i,j}I^2(x_i,y_j)}.
\end{equation}
The last estimator is the ridge flatness $\beta$ and is derived by fitting a 1D king profile $I_\mathrm{1D}=I_0/(1+(r/
r_0)^2)$ to different sectors of the galaxy density map. Using the concentration estimator $c_\mathrm{King}$ as 
$c_\mathrm{King}=r_{500}/r_0$, we search for the lowest concentration out of thirty-six $10^\circ$ wide angular 
wedges centered on the cluster and call this the 
concentration of the ridge $c_\mathrm{King,R}$. The ridge flatness is then defined with respect to the median of 
the derived concentrations as
\begin{equation}
  \hat{\beta}=\frac{c_\mathrm{King,R}}{\tilde{c}_\mathrm{King}}.
\end{equation}
To ensure positive scaling with the other estimators we simply redefine the original estimator in \cite{Wen13} to $\beta=1-\hat{\beta}$, while $\hat{\beta}$ corresponds to the original definition.

While the work by \cite{Wen13} is based on optically selected clusters and thus always starts at the optical center, our work starts with X-ray selected clusters and therefore X-ray centers. To obtain the aforementioned estimator we therefore leave the 2D profile free to re-center. Assuming that the X-ray center provides a good estimate of the position of the ICM and the optical center that of the dark matter, then the offset between both centers yields valuable information about the cluster dynamical state. In practice, not only gas to halo center offsets are driving offsets but complex shapes of the galaxy distributions, which in turn are again correlated with cluster dynamical state. We therefore list the center offset in terms of approximate $r_{500}$ as an additional proxy of dynamical state.

The estimators so far are either probing disagreement between the model and data ($\delta$) or asymmetry ($\alpha$, $\beta$). 
Systems that are symmetric and well described by the model are therefore down weighted by the these estimators. Given the freedom of the 2D model and the resolution of the smoothed map, merging systems may be decently fit by the model, but with unusual fit parameters. Besides the offset between 2D model centre and X-ray position, the estimator most sensitive to merging is the 2D profile ellipticity. High ellipticities found by the model could be either related to very elliptical halos, that could arise from a cluster merger event or when the code aims to fit two merging clusters with one model centered between both systems.

All the aforementioned estimators are based on the fit of profiles to the galaxy density map. Therefore they strongly share systematics related to the fit of the data. Further one might be preferentially interested in mergers with at least two pronounced over-densities.
To address this we use the galaxy density map and use the source detection tool {\tt SExtractor} \citep{bertin96} to identify over densities in the map. We select the nearest
over-density that has a $FLUX\_ISO$ measurement of at least $25\%$ of that over-density that is assumed to be the counterpart of the X-ray source. The $FLUX\_ISO$ measurement of {\tt SExtractor} can be interpreted in this context as a noisy richness estimate 
that should scale with the mass of the structure. For all cluster candidates reaching this threshold, we list the $FLUX\_ISO$ ratio and the offset distance in units of $r_{500}$ of the main cluster to the nearest such over density.
As a last bit of information for selecting merging systems we also provide the approximate X-ray mass and distance of the nearest neighbor in our MCMF confirmed cluster catalog.
This can be seen as the analogue to the optical based estimator.
While the optical estimator may fail due to noise and projections of nearby non-collapsed systems, the X-ray estimator alone might be triggered by source splitting or projection of AGN flux into a relaxed system. But the combination of both estimators should produce a rather clean sample of multiple systems within a certain radius. This is especially true if both estimators point to the same neighboring system, indicated through similar values for the nearest neighbor.

\subsection{CAMIRA}\label{sec:camira}

CAMIRA is a cluster finding algorithm based on over-densities of red-sequence galaxies \citep{CAMIRA}. Multi-band galaxy colors are fitted with a stellar population synthesis model of \citet{BC03} with a red-sequence galaxy template as a function of redshift. Specifically, a formation redshift of galaxies is fixed to $z=3$ and the stellar mass dependence of the metallicity is included to reproduce the tilt of the red-sequence. To improve the accuracy of the model, corrections of colors that are derived by fitting the model to galaxies with spectroscopic redshifts are included. The likelihood of fitting as a function of redshift is converted to a number parameter, from which the three-dimensional richness map is constructed using a spatial filter with a transverse distance of $\sim 1$~Mpc. Cluster candidates are selected from peaks in the richness map. A massive red-sequence galaxy near each peak is selected as the BCG of the cluster, and richness and photometric redshift (photo-$z$) estimates are repeated around the BCG. For more details of the CAMIRA algorithm, please refer to \citet{CAMIRA}.

\citet{CAMIRA-HSC} apply the CAMIRA algorithm to the HSC-SSP S16A dataset covering $\sim 230$~deg$^2$ to construct a catalog of $\sim 1900$ clusters at redshift $0.1<z<1.1$ and richness $N>15$. The comparison with spec-$z$'s of the BCGs indicates that the cluster photo-$z$'s are accurate with the bias and scatter in $\Delta z/(1+z)$ being better than 0.005 and 0.01, respectively, over a wide redshift range. A careful weak lensing analysis of these CAMIRA HSC-SSP clusters by \citet{CAMIRA-WL} suggests that the richness correlates with the halo mass with a relatively low scatter and that the richness threshold of 15 corresponds to the halo mass threshold of $\sim 10^{14}h^{-1}M_\odot$ \citep[see also][]{chiu20a,chiu20b}. The richness and mass relation shows a tight correlation irrespective of dynamical states, in contrast to the $Y_{\rm SZ}$-mass and X-ray luminosity-mass relations \citep{Okabe19}.

While the default set-up of CAMIRA finds clusters of galaxies around peaks of the three-dimensional richness map, CAMIRA can also find clusters of galaxies around positions provided by external catalogs. In this mode, peaks of the three-dimensional richness map are simply replaced by the positions of the external catalogs. CAMIRA then scans the richness at these positions as a function of redshift to identify cluster candidates, and searches the BCGs around the cluster candidates. This {\it forced} CAMIRA algorithm has successfully been applied to the ACT-DR5 Sunyaev-Zel'dovich (SZ) cluster catalog \citep{ACTDR5} and the XXL X-ray cluster catalog \citep{2021arXiv210313434W} to construct CAMIRA cluster catalogs around the SZ and X-ray peaks, respectively.


\begin{figure*}
\begin{center}
\includegraphics[width=0.64\linewidth]{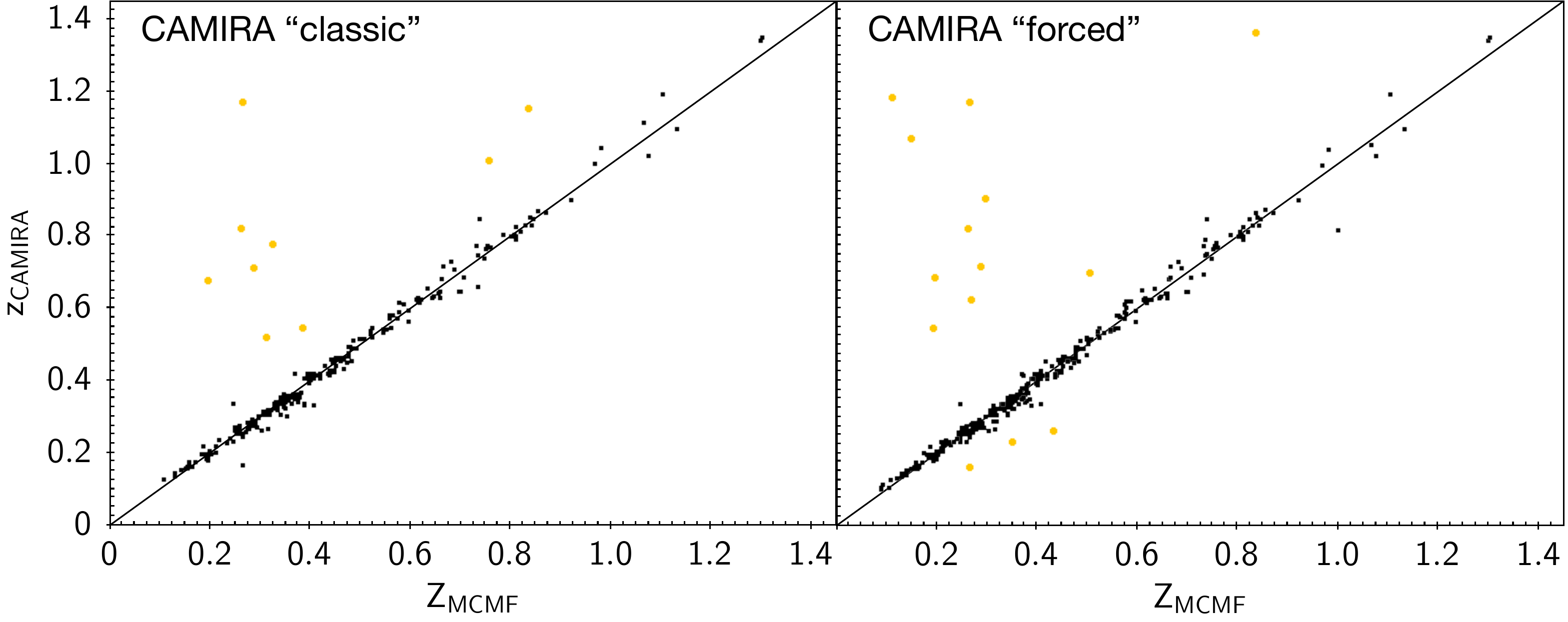}
\includegraphics[width=0.35\linewidth]{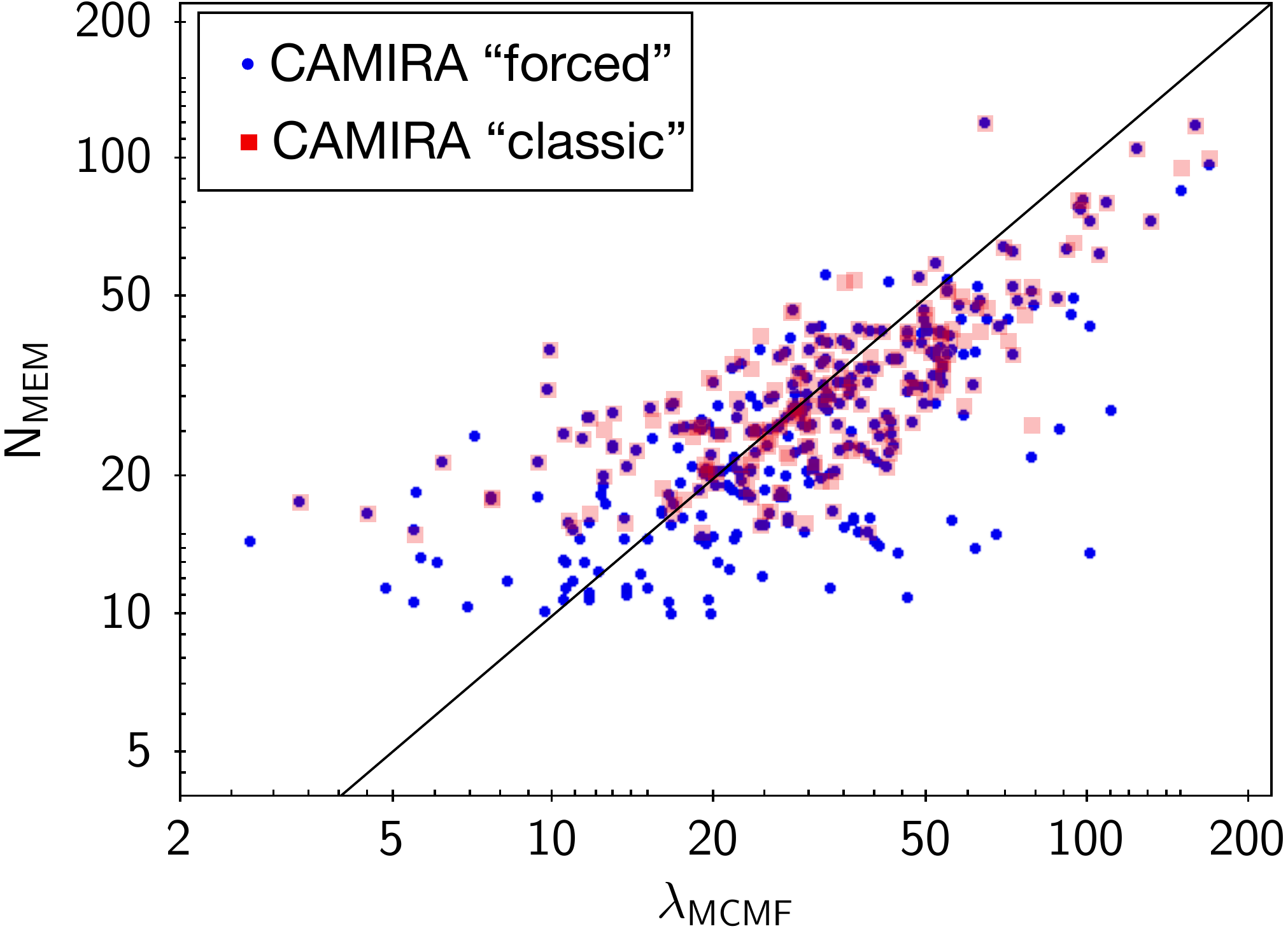}
\caption{Redshift and richness comparison between CAMIRA and MCMF. Left: photo-$z$ comparison between default CAMIRA and MCMF and the same comparison using the forced CAMIRA run. Outliers are highlighted as larger colored points. Right: Comparison of the richness estimate of CAMIRA ($N_\text{MEM}$) and MCMF $\lambda$ for sources with consistent redshifts. Continuous lines show the one to one relation.}
\label{fig:photozcomp}
\end{center}
\end{figure*}

\section{Application and results}

\subsection{Comparing CAMIRA and MCMF results}
The CAMIRA and MCMF codes both make use of the red sequence of galaxies in galaxy clusters and its dependency on redshift. Therefore we do expect similar performance in key properties such as redshift and richness estimates. As discussed in Sect.\ref{sec:camira}, CAMIRA can run in two modes, one as a classical cluster finder and the other using the X-ray position as prior. In this paper we adopt two CAMIRA cluster catalogs constructed using these two modes for our analysis.
We compare the MCMF results with both CAMIRA runs, as one can provide insights on the optical cluster finding algorithm while the other can maximise the information on the X-ray cluster candidates. 

For the match with the normal catalog we use a maximum offset of 2 arcminutes and keep the nearest match to the X-ray position, finding 239 matches. Out of those all but three are \fcont$ < 0.3$ and therefore assumed to be confirmed. The three remaining sources do have similar redshifts in both MCMF and CAMIRA but are low richness and therefore likely not a proper counterpart of the X-ray source.
The redshifts of the 236 \fcont$<0.3$ found by both codes are shown in Fig.\ref{fig:photozcomp}. We investigate the 7 most extreme photo-$z$ outliers, shown in yellow in Fig.\ref{fig:photozcomp}. We find for all these systems, that the second ranked MCMF system is consistent with the system matched using CAMIRA and that the first ranked MCMF system is the better match to the X-ray source. Those systems show low richnesses and therefore likely fall below the cut of 15 member galaxies in the normal CAMIRA catalog.

In case of the forced CAMIRA run we find 324 sources with richness $\lambda$>10 and 316 that show \fcont$<0.3$. Again, the \fcont$>0.3$ systems with CAMIRA matches are also found by MCMF either as best or second best counterpart.
The situation for \fcont$<0.3$ systems is also similar to the normal CAMIRA run; MCMF usually also finds the CAMIRA systems, but ranks another-- typically lower redshift-- cluster as the better counterpart. This is because with \fcont\ MCMF takes the redshift dependent X-ray selection into account.

In the third panel of Fig.\ref{fig:photozcomp} we show the richness comparison between CAMIRA and MCMF. Besides scatter between both estimates a divergence from the one to one scaling is visible. This can be explained by the fixed aperture used in CAMIRA for its richness estimate while MCMF uses a scaling proportional to the expected size of $r_{500}$.

We summarise the comparison between CAMIRA and MCMF as follows. The photometric redshifts are highly consistent with each other if the same systems are matched. Outliers in redshift usually arise when the system found by MCMF is at low richness where CAMIRA becomes incomplete. Judging from the last panel of Fig.\ref{fig:photozcomp} this is in the range of $20<\lambda<40$, depending on the cut in CAMIRA richness. This, the smaller footprint covered by HSC alone and the patchiness of the HSC data at the footprint visible in Fig.\ref{fig:footprint} can explain why only 324 out of 542 candidates do have a potential CAMIRA based counterpart. We therefore use MCMF results for the baseline confirmation and redshifts and provide CAMIRA estimates only as additional information in the catalog.

\subsection{General properties of the MCMF confirmed cluster catalog}

\begin{figure}
\begin{center}
\includegraphics[width=0.99\linewidth]{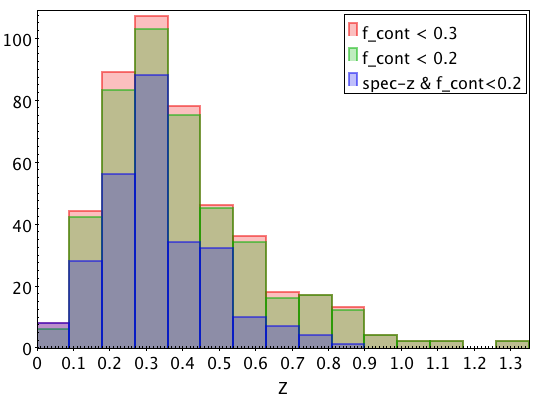}
\caption{Redshift distribution of the MCMF confirmed eFEDS clusters for two cuts in \fcont\ and for the subset of clusters with \fcont$ < 0.2$ and spectroscopic redshifts.}
\label{fig:redshiftd}
\end{center}
\end{figure}

\begin{figure}
\begin{center}
\includegraphics[width=0.99\linewidth]{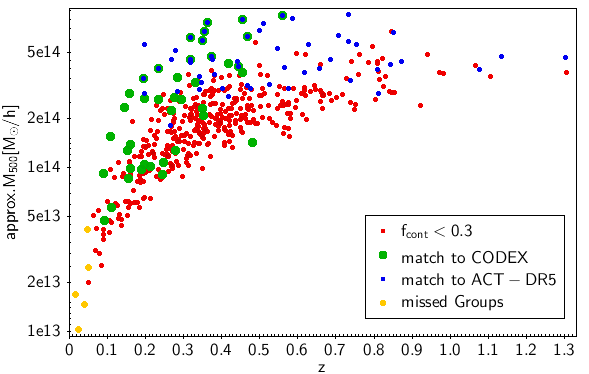}
\caption{Approximate $M_{500}$ versus redshift plot for \fcont$ < 0.3$ clusters. Only 81 of the 470 systems show a match in the ICM based CODEX and ACT-DR5 surveys, nicely illustrating the increase in number of ICM selected clusters over the eFEDS field.} 
\label{fig:massvsredshift}
\end{center}
\end{figure}

In total we find 470 (446) clusters with \fcont$<0.3$ (0.2), and $\sim60$\% of these systems have a spectroscopic redshift. The redshift distribution
shown in Fig.\ref{fig:redshiftd} peaks at $z=0.3$ and reaches out to $z=1.3$. To provide the reader with some indication of the mass and redshift reach, we show in Fig.\ref{fig:massvsredshift} the distribution of the sources in redshift versus approximate $M_{500}$, where approximate $M_{500}$ is that used in Sect.~\ref{sec:mcmf} to estimate cluster richness. Color coded in blue are matches to the SZ based ACT-DR5 cluster catalog \citep{ACTDR5} and in large green points matches with the ROSAT telescope X-ray based CODEX catalog \citep{codex19}. Highlighted in yellow are eFEDS clusters found by matching with spectroscopy based group catalogs (Sect.~\ref{sec:crossmatches} and Table~\ref{tab:missing}). Details to the matches between eFEDS and other catalogs are provided in Section~\ref{sec:crossmatches}.
The distribution nicely illustrates the shape of the X-ray selection for a sample following a flux limit and the high sensitivity to low mass systems at low redshift.  Note also the redshift range is competitive with SZ selected cluster catalogs.

\begin{figure}
\begin{center}
\includegraphics[width=0.99\linewidth]{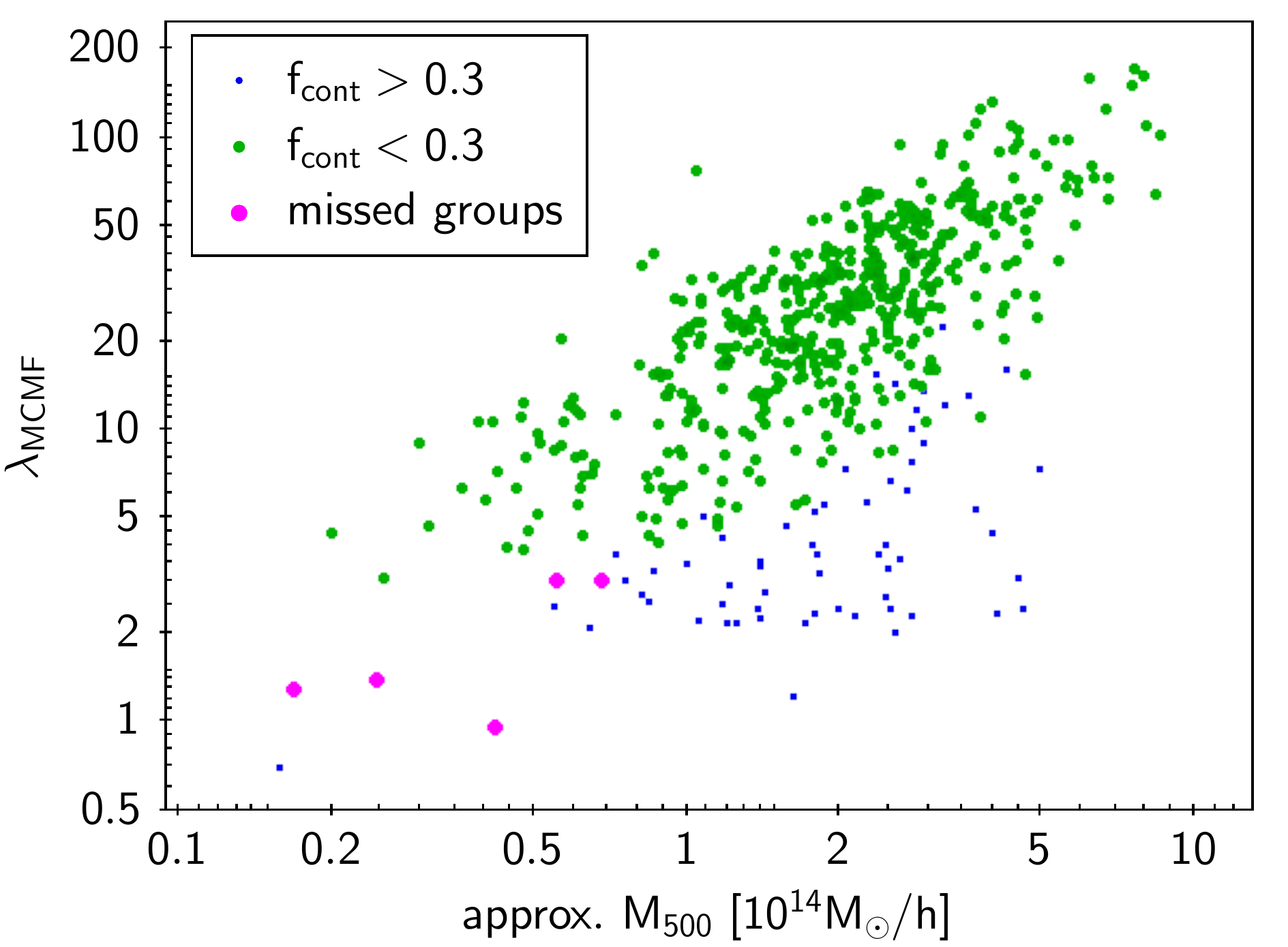}
\caption{Richness versus approximate $M_{500}$ for eFEDS sources. MCMF confirmed (\fcont$< 0.3$) are shown in green, \fcont$>0.3$ systems are shown in blue.  MCMF unconfirmed (missed) groups found by catalog cross-matching are shown in magenta using a forced richness estimate at the group redshift.} \label{fig:lambdavsmass}
\end{center}
\end{figure}

Finally we show in Fig.~\ref{fig:lambdavsmass} the distribution of candidates in the $\lambda-\mathrm{approx.} M_{500}$ plane. Highlighted in green are confirmed systems, while unconfirmed systems are shown in blue. Although \fcont\ is not using this scaling relation to clean the candidate list, the cleaning obviously tends to exclude systems below the main relation. We further highlight in magenta sources not confirmed by MCMF but found by cross matching with group catalogs. The richness measurement for those systems are performed by fixing the redshift to the redshift of the corresponding group. A decent scaling between both observables seems to exist down to the group mass regime. 

Throughout the paper we define the cluster catalog by cutting the cluster candidate list at a certain \fcont\ (typically 0.2 or 0.3). We provide \fcont\ estimates for all candidates to allow users to adapt the level of catalog contamination and completeness to their needs. We do not recommend to apply a higher threshold than 0.3, because the vast majority of \fcont$>0.3$ systems are not clusters and consequently the redshifts provided are likely not associated with a true, X-ray selected cluster. As outlined in Sect.\ref{sec:xmatchsum} we manually add known missed systems to the sample by flagging them with \fcont\ values of $-2$ and $-1$. With that we expect the \fcont$<0.3$ cluster catalog to include $>99$\% of the true clusters in the candidate catalog with an estimated contamination level of 6\% by non-cluster sources. 

\subsection{Optical dynamical state estimates} \label{sec:mergers}

\begin{figure*}
\begin{center}
\includegraphics[width=0.99\linewidth]{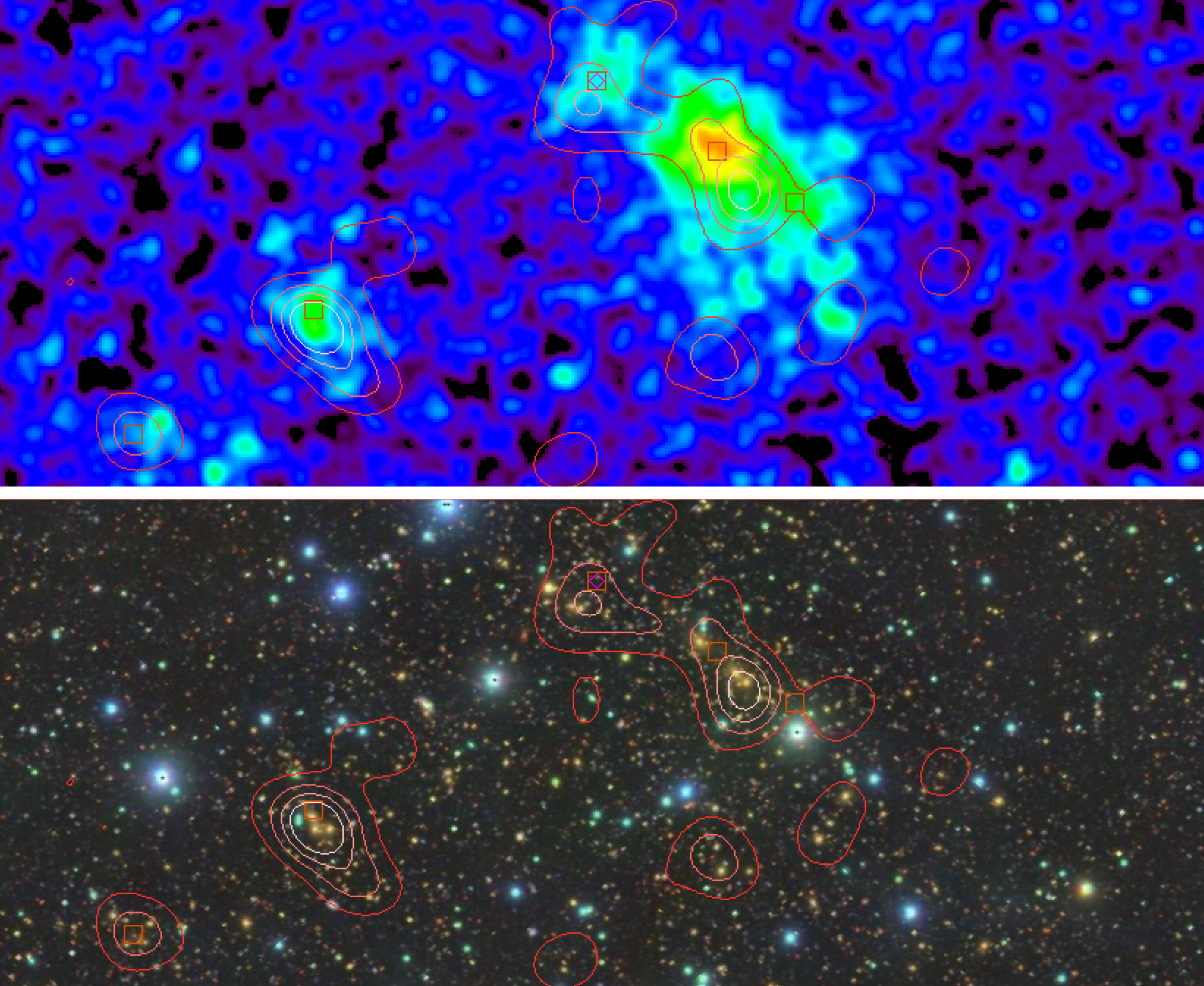}
\caption{Region around eFEDS~J092220.4+034806, one of the three most unrelaxed systems with $\lambda>50$ according to optical dynamical state estimators. Top: Smoothed 0.5--2.0~keV X-ray surface brightness map of a 7$\times$2.8~Mpc region around the merging system. Contours show red sequence based galaxy density at the cluster redshift. Boxes show extent selected sources at the system redshift. Bottom: HSC $grz$-color composite image of the same region.}
\label{fig:z0269merger}
\end{center}
\end{figure*}

\begin{figure}
\begin{center}
\includegraphics[width=1.\linewidth]{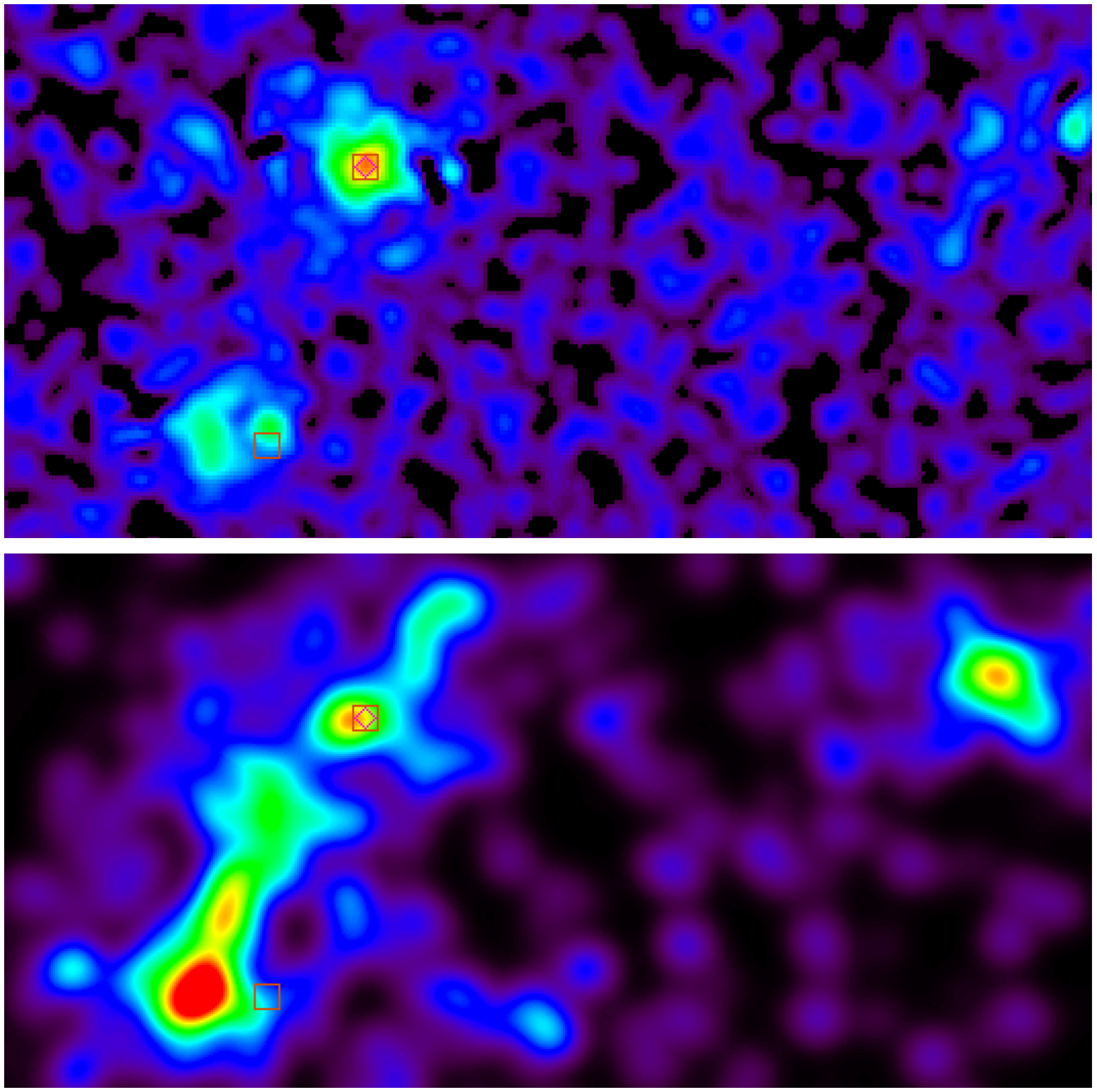}
\caption{Region around eFEDS~J085620.8+014649, another one of the three most unrelaxed system with $\lambda>50$ according to optical dynamical state estimators. Top: Smoothed 0.5--2.0~keV X-ray surface brightness map indicates two main clumps separated by 5~arcmin (2~Mpc). Boxes show extent selected sources at the system redshift. Bottom: Red sequence galaxy density map at $z=0.73$ over the same region. The galaxy density map suggests a connection between the two clusters. Additionally, a third cluster is clearly visible about 3 Mpc away from the main pair.}
\label{fig:z073merger}
\end{center}
\end{figure}

The different estimators that depend to the shape of the galaxy distribution and its location relative to the X-ray center are well correlated. Two examples are shown in Fig.\ref{fig:optest}, for a complete set for all combinations placed in Fig.~\ref{fig:dynestimatorcorr} in Appendix~\ref{app:dyns}.
With the lack of a simulation-based calibration, a simple way to select merging systems is selecting the most unrelaxed systems from a simple linear combination of these estimators. In Fig.~\ref{fig:z0269merger} and Fig.~\ref{fig:z073merger} we show two of the most unrelaxed clusters with richness $\lambda$>50.

In Fig.~\ref{fig:z0269merger} we show the region around eFEDS~J085620.8+014649 at $z=0.269$, including three other detected extended sources. The main system shows clear indications of merging with a complex morphology in X-ray surface brightness. The main cluster harbours at least two massive central galaxies and shows two peaks associated with two peaks in the X-ray surface brightness map.
The second cluster merger shown in Fig.~\ref{fig:z073merger} is at $z=0.73$. In X-rays one can see two distinct main clusters both detected as extended. In the optical both clusters seem to be connected by a bridge of galaxies, which is likely responsible for this system being classified as an extreme merger from the estimators.
The X-ray data are likely not deep enough to detect a possible connecting structure.

Another way to select interesting merger candidates is by looking at the distance to nearest neighbour. This can be done using both X-rays and optical data. In the right panel of Fig.~\ref{fig:optest} we show the distance to the nearest neighbour in optical versus nearest extended neighbour in X-rays in units of $r_{500}$. Sources aligned on the one-to-one line likely correspond to the same structures. A rather clean sample of cluster pairs can be obtained when requiring low offsets in both X-ray and optical. We note here that cluster mergers can significantly impact the distribution of the ICM and its emission in X-rays. Having a nearby X-ray and optical neighbour may therefore lead to such a selection preferentially selecting cluster pairs in an early merging or a pre-merging phase. On the other hand the X-ray source detection tool assumes circular models, and significant elongation or complex morphology causes the code to identify multiple detections. In the right panel of Fig.~\ref{fig:optest}, we highlight merger candidates in red if their nearest optical and X-ray neighbour is closer than $2.5\times r_{500}$. This selection should include the majority of systems with overlapping virial radii. For convenience, those systems are listed in table~\ref{tab:Mergpairs}.

Besides providing group and cluster identification, it is well known that X-ray observations can be used to identify cluster mergers and estimate their dynamical state. A study using eFEDS X-ray data is in preparation (Ghirardini et. al in prep.), but was not yet ready for comparison with the optical estimates provided here. X-ray and optical estimators tend to probe different merger characteristics and are prone to different systematics. 
We therefore expect substantial scatter between different estimators of dynamical state. On the other hand, the combination of information from both X-ray and optical estimators can enable one to select specific merger types. One example of that could be cluster pairs in pre- and post-collision state. Pairs can be selected with joint X-ray and optical information, where the X-ray data adds the critical information on the dynamical state of the ICM, indicating a recent collision.

\begin{figure*}
\begin{center}
\includegraphics[width=0.33\linewidth]{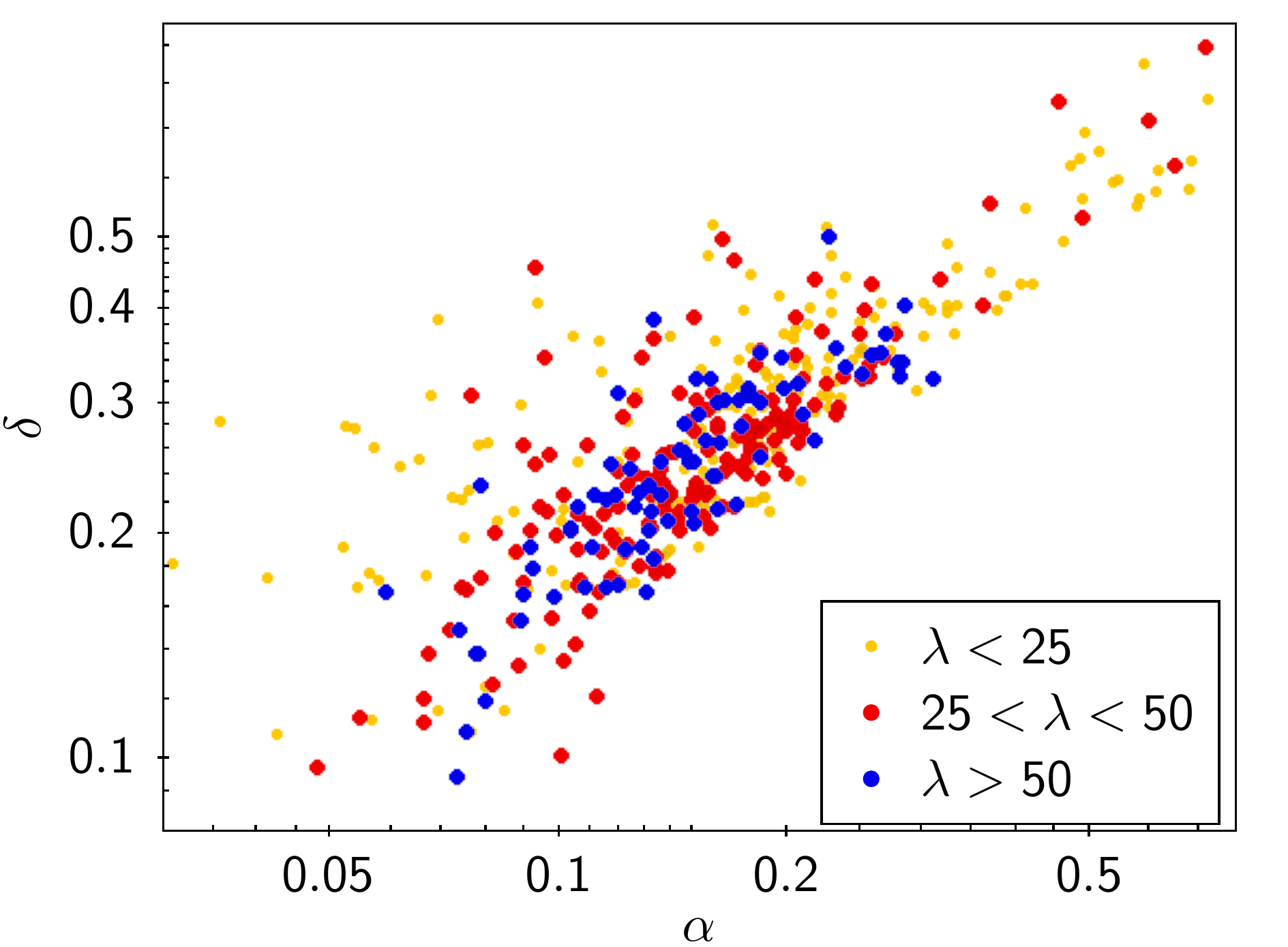}
\includegraphics[width=0.33\linewidth]{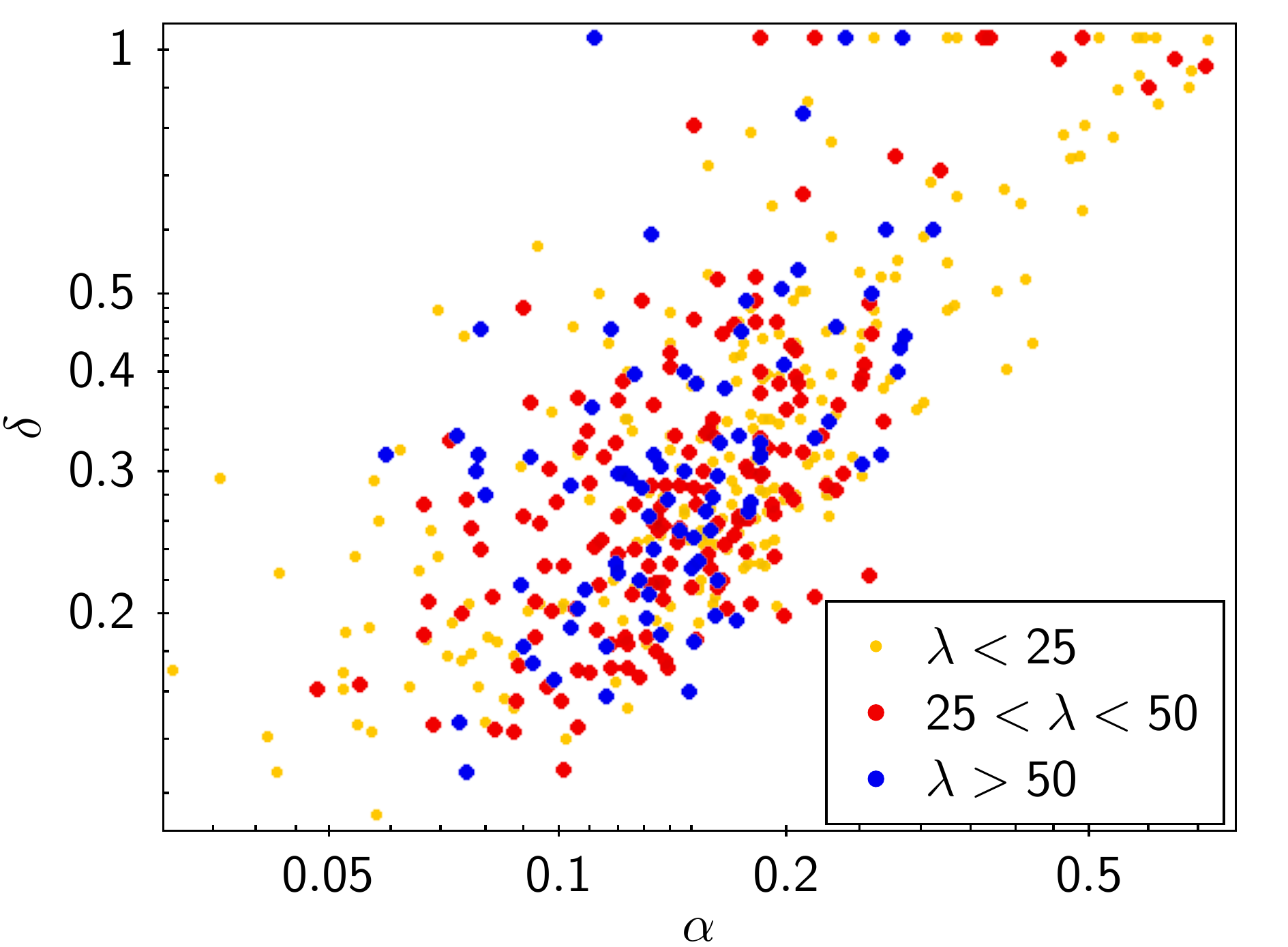}
\includegraphics[width=0.33\linewidth]{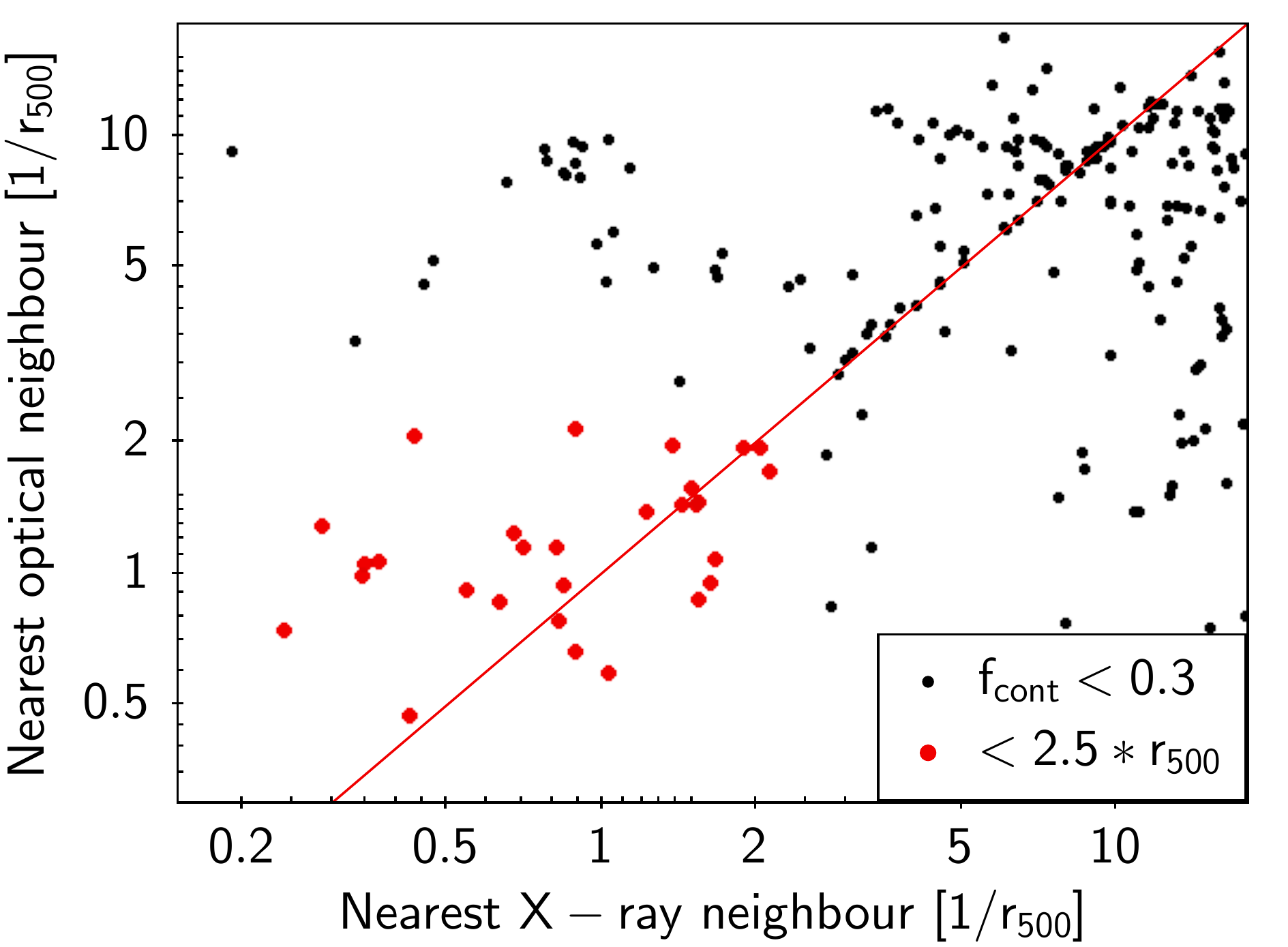}
\caption{Dynamical state estimators: Left and middle show a comparison between the three estimators defined in \cite{Wen13}, color coded are three different richness bins. Right: Distance to the next neighbour found in the galaxy density map versus distance to the next neighbour in the X-ray cluster catalog. The one-to-one relation is indicated as red line. Clusters showing a near neighbour within $\sim2.5*r_{500}$ in both X-ray and optical are likely cluster pairs, potentially merging.} 
\label{fig:optest}
\end{center}
\end{figure*}

\section{Catalog validation and performance of optical confirmation}

In this section we investigate the performance of the optical confirmation, estimating the incompleteness due to optical cleaning and the final catalog purity.
To do that we perform a set of five steps or investigations, building up from the experience from the previous test. The first and most lengthy test is using cross matches to other catalogs. With that we check for clusters and groups that are not confirmed by MCMF or do show a discrepant redshift estimate.
The second test compares the fraction of confirmed systems with those expected from simulated observations of the eFEDS field. The third check is an estimate of the incompleteness caused by the optical cleaning and a comparison with the missed systems found through cross-matching in the first test. For that a scaling relation fit between the approximate $M_{500}$ coming from X-rays and the MCMF richness is performed. 
The fourth step is adding in information from the tools dedicated to identifying point like counterparts to X-ray candidates. This allows us to define cleaner or more contaminated subsamples of the cluster catalog.
Finally, as a last step we estimate the catalog contamination by making use of the scatter distribution around the derived scaling relation and a clean sample defined in the previous step. The results are then compared to those found in the previous steps.

\subsection{Match to public catalogs}\label{sec:crossmatches}
Cross matching with public cluster catalogs allows us to assess the performance of the cluster confirmation and redshift assignment. 
The eFEDS field lies inside multiple surveys for which cluster and group catalogs were constructed. In particular, it overlaps with the SDSS footprint from which optical cluster catalogs were constructed using photometric as well as spectroscopic data. Further, it overlaps the cluster catalogs of ACT-DR5 \citep{ACTDR5} and is naturally covered by all-sky cluster surveys such as Planck PSZ2 \citep{PSZ2} and ROSAT \citep{MCXC}. 

Given the large number of available cluster catalogs we restrict ourselves to the largest or most commonly used catalogs and to a subset of possible tests. The focus of the tests are on clusters that MCMF failed to confirm or where there are mis-assignments of redshifts. We do not aim to test the X-ray selection or the selection function of the other surveys we match with.

We perform two standardised tests of the data: redshift consistency and confirmation fraction of matched systems. Systems with inconsistent redshifts or high \fcont\ are individually checked. Readers not interested in the detailed discussion on the individual catalog can jump to the summary in Sect.\ref{sec:xmatchsum}.

\subsubsection{X-ray selected clusters}\label{xraysurvs}
The ROSAT all-sky survey \citep[RASS;][]{Truemper82,RASS} can be seen as a predecessor of the eROSITA all-sky survey (eRASS), which provides shallower X-ray observation over the whole sky with poorer angular resolution. While early work performed cuts to start with rather clean X-ray catalogs that need confirmation, more recent works \citep{MARDY3,codex19} make use of large optical photometric surveys to systematically identify clusters within all RASS detections. This approach results in cluster catalogs that are about ten times higher in source density than previous RASS based catalogs.

We match the eFEDS extended sources to two cluster catalogs the MCXC catalog \citep{MCXC} and the CODEX catalog \citep{codex19} using a maximum distance between X-ray positions of two arcminutes.

For MCXC we find that the only source in the eFEDS footprint, RXC J0920.0+0102 at a redshift of $z=0.017$, which is matched to the eFEDS source eFEDS~J092002.2+010220. This cluster is not confirmed by MCMF although a MCMF measurement on a substructure 130 arcsec from the cluster is close to being detected (\fcont=0.35, $z=0.033$). The very low redshift and estimated mass of $M_{500}=0.36 \times 10^{14}h^{-1}M_\odot$ makes it hard to confirm this cluster with MCMF. The deep imaging data used in MCMF are not optimised for very low redshift sources with their large angular extent. The size of the BCG shows a diameter of 5 arcmin in HSC, several hundred times larger than a typical galaxy in these surveys, and there are dozens of background galaxies shining through the BCG.

\begin{figure}
\begin{center}
\includegraphics[width=0.99\linewidth]{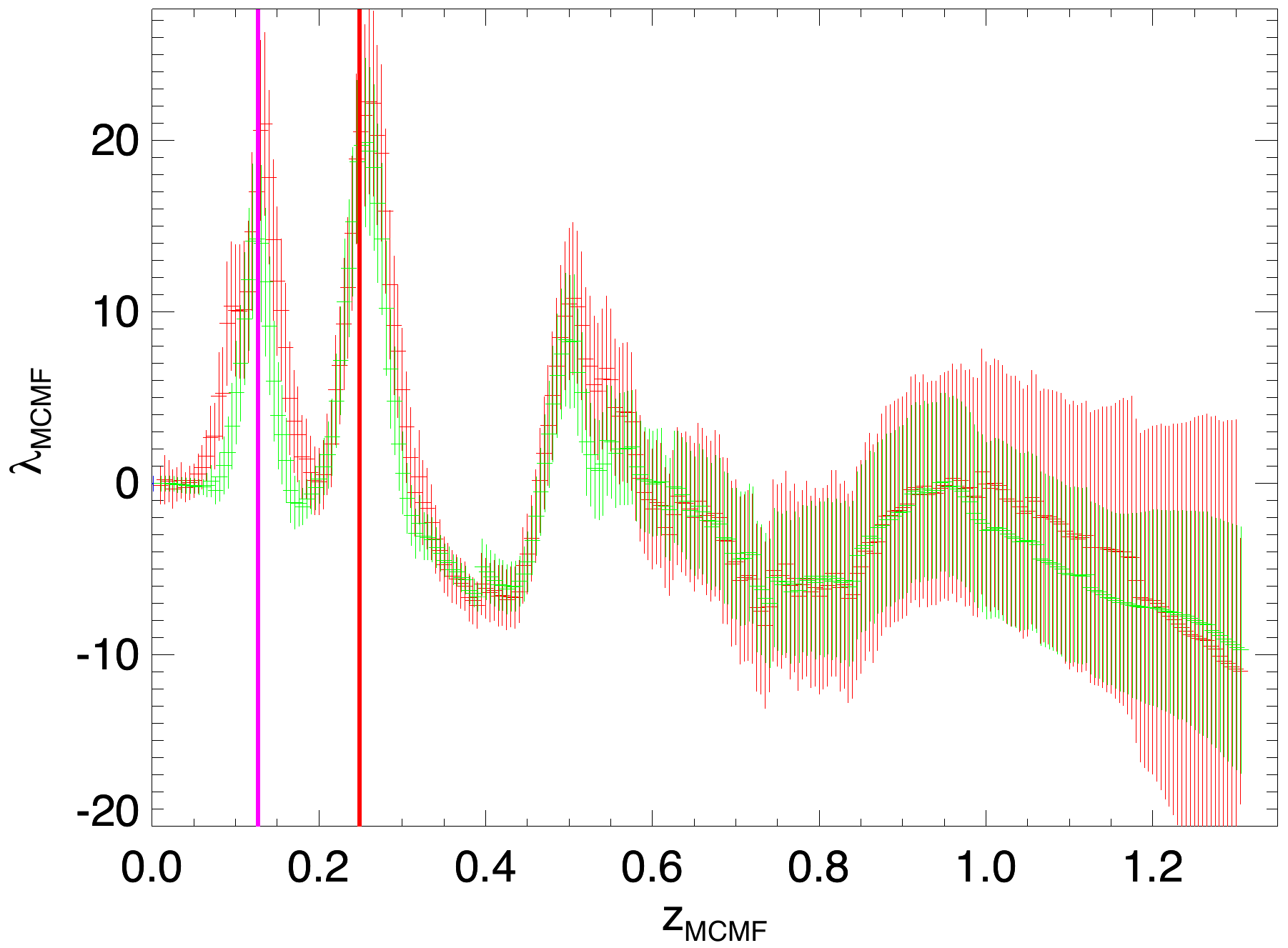}
\caption{The MCMF richness versus redshift plot for eFEDS~J091509.5+051521, one of the cases with photo-$z$ mis-match to the CODEX cluster catalog.
MCMF finds two counterparts highlighted as vertical lines with \fcont\ of 0.012 (red) and 0.017 (magenta). The redshift given in CODEX corresponds to the magenta line. Using richness from the LS $grz$W1-band run to predict X-ray count rates yield similar values for both peaks. Red points show case using LS $grz$ and green show LS $grz$W1 bands. Note, this cluster is not in the HSC-SSP footprint.}
\label{fig:091509}
\end{center}
\end{figure}

In the case of the CODEX catalog we find 43 positional matches, the aforementioned RXC J0920.0+0102 is not part of this catalog. All positional matches have \fcont$<0.02$ and are therefore clearly confirmed by MCMF. We find three matches
that show a redshift offset of more than 5\%. In two cases the optical system found in CODEX shows a large offset from the eFEDS X-ray source. The MCMF redshifts of both eFEDS sources are greater than 0.6 and therefore beyond the redshift reach of the SDSS data used to construct CODEX. The remaining cluster eFEDS~J091509.5+051521
belongs to an eFEDS source with two clusters along the line of sight with \fcont\ of 0.012 and 0.017 shown in Fig.\ref{fig:091509}. The second ranked cluster has a redshift consistent with CODEX. The difference in \fcont\ is too small for a clear single redshift assignment. Using the richness as a mass proxy to predict the number of X-ray photons coming from each cluster yields similar numbers for both clusters. It is therefore likely that both clusters significantly contribute to the X-ray detection.

Given the RASS based CODEX catalog and the great improvement regarding depth and angular resolution of eROSITA over ROSAT, one can also ask what the matched fraction of CODEX clusters over eFEDS is. For that we repeat the matching with more generous cut on the offset of 3 arcmin and limit the CODEX catalog to sources well within the eFEDS footprint to avoid border effects. We further allow multiple sources to match to one source so that the results are less affected by source splitting in either catalog. We find that only 45\% of the CODEX clusters do have a match with an extended eFEDS source.
Using the cleaning flag in CODEX that imposes a redshift dependent richness cut similar to \fcont\ the match fraction increases to 85\%, while decreasing the number of matched sources by 30\%. The unmatched CODEX sources in the clean subsample are usually caused by mismatches between the optical source and the X-ray source. As an example, the X-ray source given in CODEX is obviously a bright point source found in eFEDS. But at the position of the associated optical cluster there is indeed X-ray emission found, but that this source is not the same X-ray source as that listed in CODEX. Expanding the examination to the non-clean sample of non-matched sources reveals a further increased fraction of mis-matches mostly with bright point sources and an increased number of sources where no X-ray source is found either at the X-ray or the optical position given in CODEX.

If the eFEDS region is representative of the full CODEX catalog, then the CODEX catalog contains more than 50\% of sources where the X-ray flux is not associated to a real X-ray cluster. But, when applying the cleaning flag in CODEX the catalog becomes much cleaner with a purity of 85\% or above, including the fact that most of the non-matched clean sources indeed show X-ray emission at the optical position.

In summary, the match with CODEX raises awareness of multiple structures along the line of sight and the importance of the redshift limits of the optical survey.
The match with MCXC reveals a missed group at low redshift, highlighting the difficulty of confirming very nearby, low mass systems with photometric surveys. 

\subsubsection{SZ selected surveys}
The eFEDS survey overlaps the Planck and the ACT SZ surveys. The SZ
effect is redshift independent and, depending on the frequency, can be observed as a negative signal imprinted on the comic microwave background. The selection function of both surveys (Planck, ACT) are very different as source detection also depends on angular resolution and available frequency range. This results in a strongly redshift dependent selection in Planck similar to that seen in flux limited X-ray surveys, while ACT shows a flat selection with redshift, becoming even more sensitive at higher redshift as beam size and cluster size start to match-up.

We match the eFEDS catalog to the Planck PSZ2 catalog \citep{PSZ2} within 3 arcmin and find 10 matches. All matches show consistent redshifts and \fcont=0.

Matching the eFEDS catalog to the ACT-DR5 catalog we find 53 matches. All matches have \fcont$<0.07$ and therefore are MCMF confirmed systems. Noteworthy here is that this includes three eFEDS clusters with $z>1.0$, including the highest redshift eFEDs cluster with $z=1.3$.

We initially find three photo-$z$ mis-matches. eFEDS~J084441.4+021702 (ACT-CL J0844.6+0216) was assigned in ACT-DR5 a spectroscopic redshift assigned of a foreground galaxy at $z=0.56$. The MCMF identified cluster has a BCG with spectroscopic redshift $z=0.6515$. Here we expect the MCMF redshift to be the correct one.

\begin{figure}
\begin{center}
\includegraphics[width=0.99\linewidth]{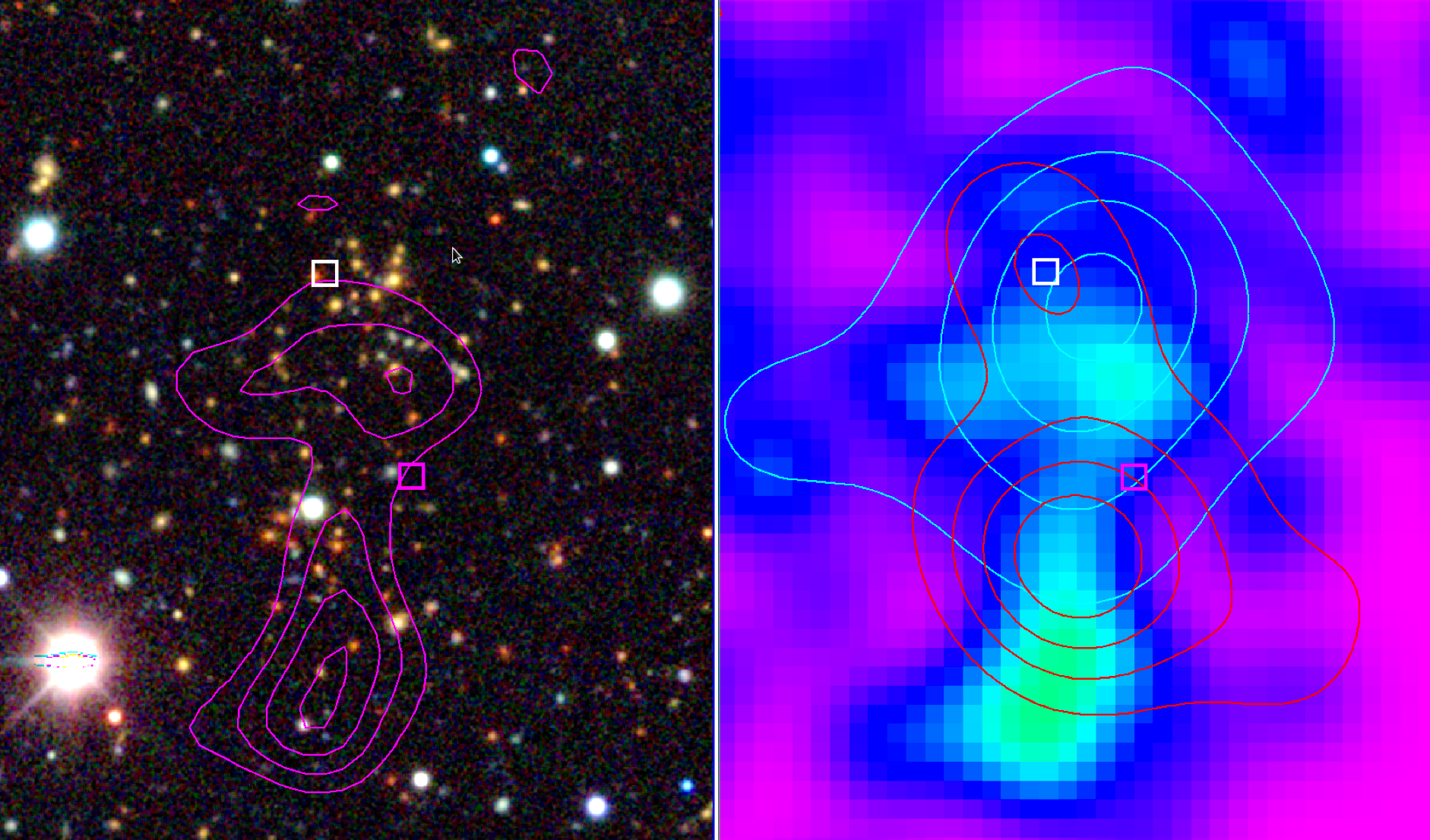}
\caption{The 2.5$\times$3~arcmin region around eFEDS~J083120.5+030950 (ACT-CL J0831.3+0310). Left: LS $grz$-band color image. Magenta contours are drawn from the X-ray surface brightness map, shown in the right panel. Magenta box shows the eFEDS position, the white box shows the  SZ position from ACT. Right: X-ray map, showing X-ray and SZ position with density contours for passive galaxies at $z=0.569$ (cyan) and $z=0.852$ (red).}
\label{fig:J083120}
\end{center}
\end{figure}

The second mis-match eFEDS~J083120.5+030950 (ACT-CL J0831.3+0310), appears in the X-ray map as two structures separated by 50 arcsec while detected as one cluster with its center in the middle and is shown in Fig.\ref{fig:J083120}. MCMF assignes \fcont=0 to both clusters and their richness are 50.1 for the $z=0.852$ cluster and 47.9 for the $z=0.569$ cluster. As both have the same \fcont\ the ranking of the clusters is somewhat arbitrary. The SZ center and redshift ($z=0.566$) fits well to the cluster at lower redshift.  The similar richness of both peaks further suggests that the majority of the X-ray photons are coming from the cluster with the lower redshift. We therefore change the order of the two peaks, assigning the lower redshift cluster to be the primary counterpart to the X-ray selected clusters.

\begin{figure}
\begin{center}
\includegraphics[width=0.97\linewidth]{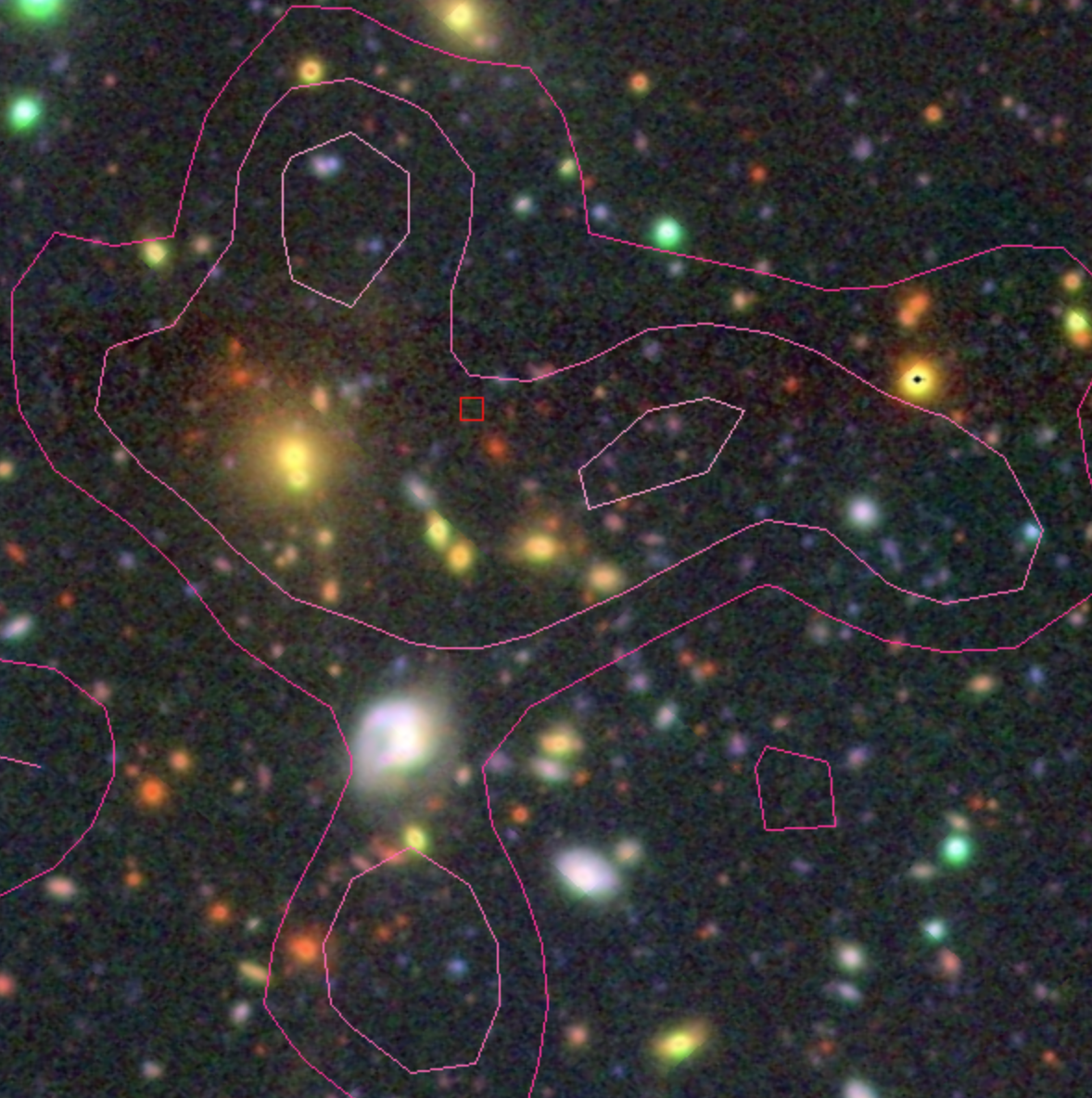}
\includegraphics[width=0.99\linewidth]{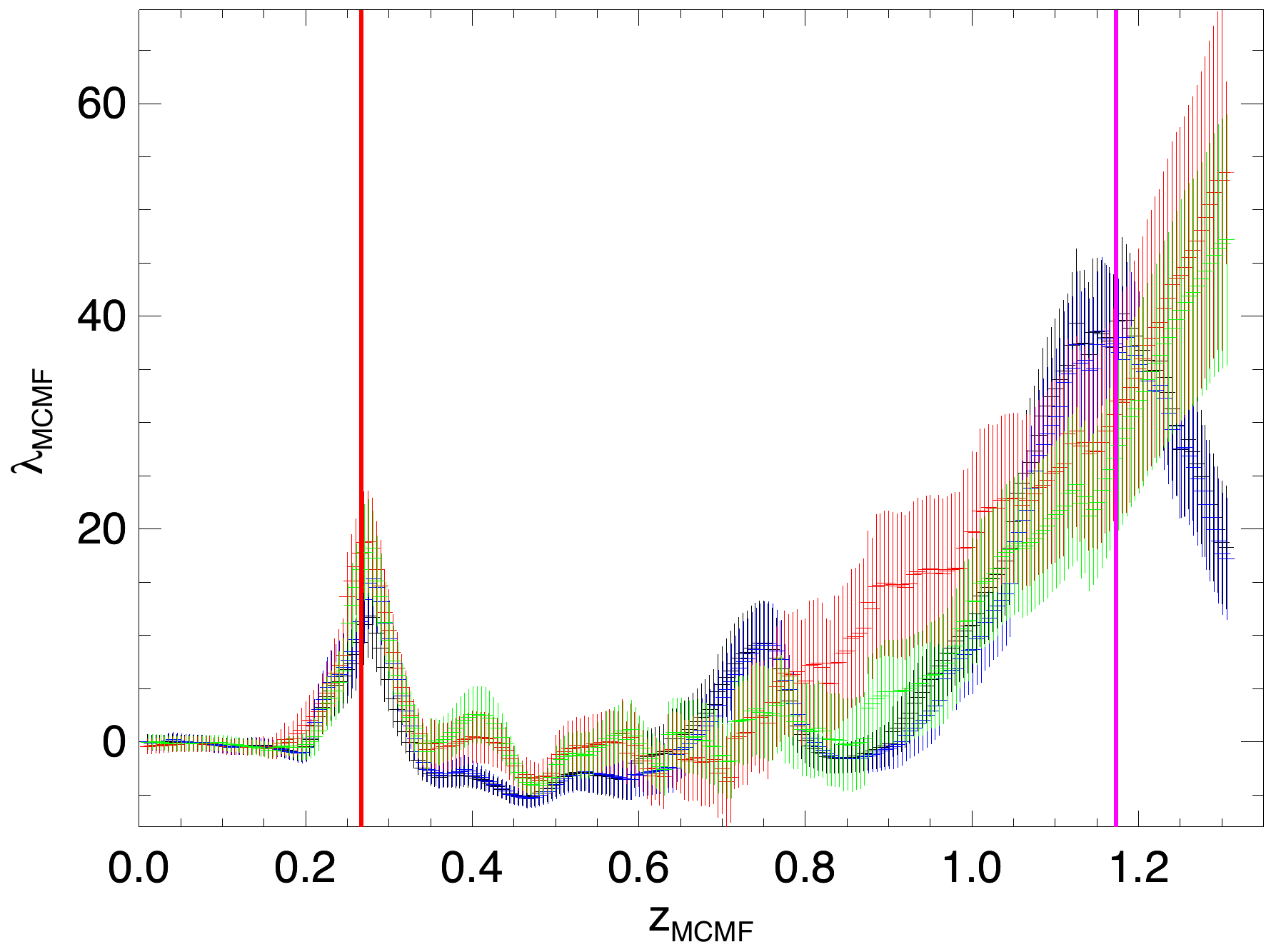}
\caption{The X-ray source eFEDS~J091647.0+015532. Top: HSC $grz$-color image of the central 2.5$\times$2.5~arcmin region with X-ray contours overplayed. Bottom:  MCMF richness versus redshift plot, one of three cases with photo-$z$ mismatch with positional matches from the ACT-DR5 cluster catalog.
MCMF finds two counterparts highlighted with vertical lines. The richer system at $z=1.17$ is consistent with the ACT cluster, different sensitivity on redshift for X-rays suggests the low redshift to be the better counterpart to the eFEDS source. Colors indicate different MCMF runs: HSC S19A (black), HSC S20A (blue), LS $grz$ (red) and LS $grz$w1 (green) bands.}
\label{fig:091647}
\end{center}
\end{figure}

The eFEDS cluster eFEDS~J091647.0+015532, with the positional match ACT-CL J0916.7+0155, has two optical counterparts with similarly low \fcont\ of 0.027 and 0.036. The richness versus redshift plot is shown in Fig.\ref{fig:091647}. The second ranked counterpart with $z=1.17$ fits well to the value listed in ACT of $z=1.15$. 
Given the different cluster sensitivity with redshift of X-rays and SZ it is possible that, although very close positional matches, the clusters that both surveys see are in fact different. This scenario seems to be the case here. The X-ray and richness fit well for the lower redshift cluster. But the X-ray based approx. $M_{500}$ putting all flux at $z=1.17$ suggest a three times higher mass than that found by ACT. The richness for the $z=1.17$ found by MCMF is consistent with the ACT based mass estimates. The richness of the low redshift cluster is about half of that at high redshift. So assuming the richness ratio is similar to the mass ratio of the clusters the ACT signal would be indeed dominated by the high redshift cluster. We conclude that although the redshifts of the ACT cluster and the eFEDS cluster don't agree, the redshift assignments seem to be correct for both cases.

\subsubsection{Match to optical cluster catalogs}
Besides having direct access to the most recent HSC based CAMIRA catalog, the eFEDS field overlaps SDSS and KiDS surveys used to construct cluster catalogs. In the case of SDSS the number of public cluster catalogs are indeed quite large. We therefore restrict our analysis to only three catalogs: redMaPPer \citep{redmapper} on SDSS DR8, AMF \citep{AMF} on SDSS DR9 and WHL \citep{WHL} on SDSS DR12.

The match to redMaPPer is done within a 2 arcminute radius and we find 163 matches. All systems have \fcont$<0.3$ and all except one show \fcont$<0.08$. The match with \fcont$=0.29$ is about 90 arcsec from the X-ray position at a location with no X-ray emission in the map. We therefore classify this match as a chance match given the adopted search region.
We find only one outlier in redshift, eFEDS~J091509.5+051521, which was already discussed in subsection \ref{xraysurvs} as two systems along the line of sight. Because redMaPPer on SDSS was used for the catalog creation of the CODEX catalog, this mis-match is not surprising.

The match to AMF is done using the same maximum offset as for redMaPPer. We find 129 sources, all show \fcont<0.3 and only one shows \fcont$>0.2$. The eFEDS cluster with the highest \fcont\ eFEDS~J084004.9+013751 shows a large X-ray to BCG offset of 102 arcsec. In fact, judging from the X-ray surface brightness map the X-ray source associated with the BCG and AMF counterpart seems to be separated from the eFEDS detected X-ray source and is not associated with a source in the extent selected sample.
We find seven matches with photo-$z$ offsets greater than 5\%. For four we find AMF to match the second ranked counterpart. Two show AMF redshifts between the second and first ranked MCMF counterpart. The small redshift differences between 1st and 2nd MCMF peaks of $\Delta z<0.2$ may have caused blending issues in AMF redshift estimation. The last cluster with photo-$z$ mis-match has an MCMF photo-$z$ and a spec-$z$ of $z=0.61$ and therefore beyond the reach of AMF, which assigned $z=0.51$.
In summary, all AMF matches are associated with MCMF confirmed systems and the redshift mis-matches typically come from different ranking of multiple possible counterparts or a photo-$z$ limitation of the AMF catalog.

\begin{figure}
\begin{center}
\includegraphics[width=0.99\linewidth]{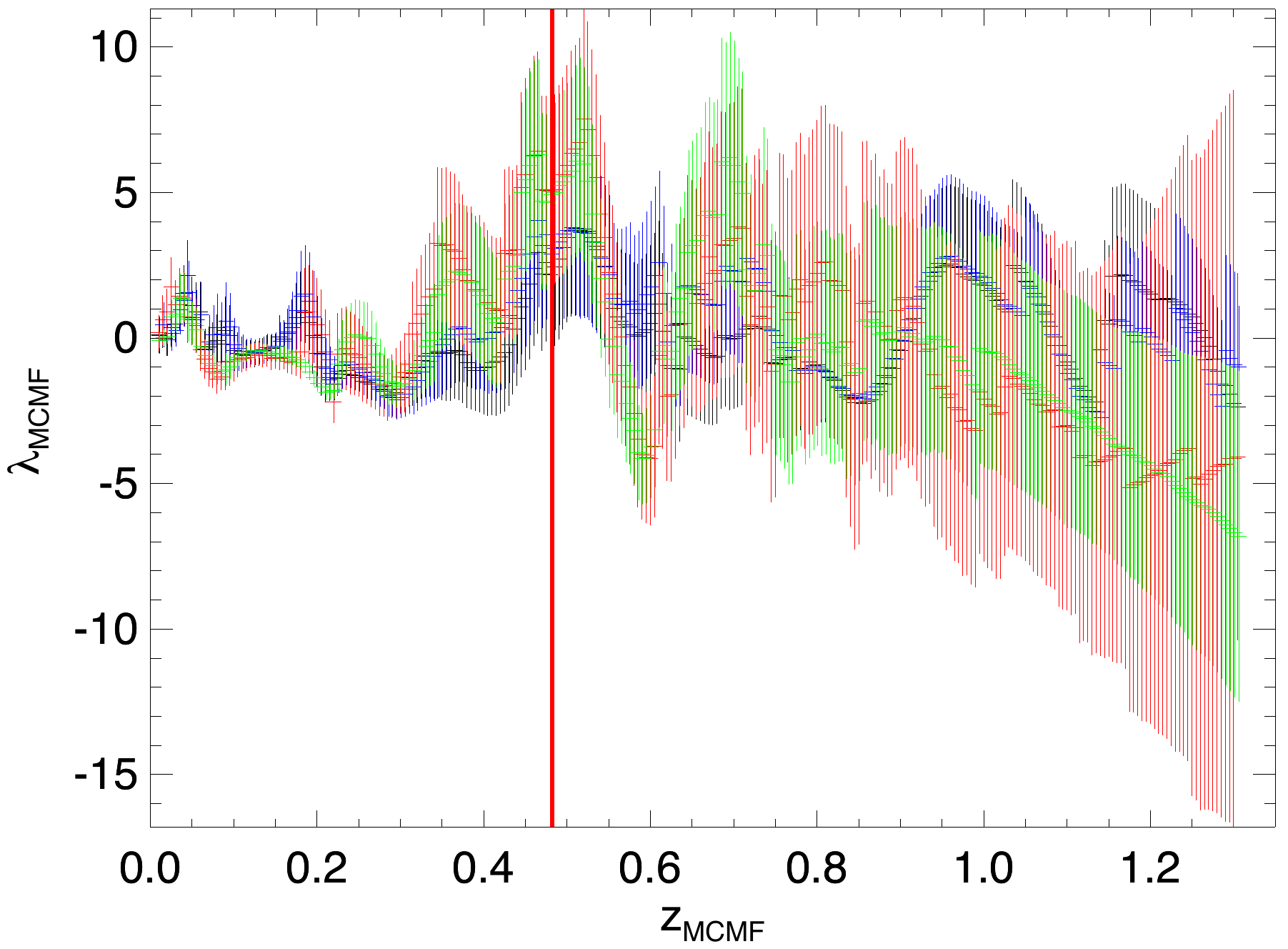}
\caption{The richness versus redshift plot of eFEDS~J084430.8+021737, a galaxy group at $z=0.05$ not confirmed by MCMF and found by cross-matching with the WHL catalog. The color coding is the same as in previous richness versus redshift plots.}
\label{fig:084430}
\end{center}
\end{figure}

Repeating the same exercise with the WHL catalog we find 244 matches, where 243 are matched to \fcont$<0.3$ sources and 239 show \fcont$<0.2$. From those five \fcont$>0.2$ systems three have consistent redshift and richness with the MCMF estimates. One outlier is eFEDS~J084004.9+013751, which was already discussed in the AMF cross match. The last one-- the only system with \fcont$>0.3$-- is eFEDS~J084430.8+021737, which is a $z=0.05$ group with an approximate mass off $M_{500}=2.45\times10^{13}h^{-1}M_\odot$. Looking at the $\lambda$ versus redshift plot in Fig.\ref{fig:084430} one can see a peak of richness 1 to 2 at redshift 0.05. Given the approximate X-ray based mass, the indicated richness is close to the expectation and illustrates the difficulty to identify eFEDS systems in the low redshift low mass regime with optical photometry. Additionally, there is the rich eFEDS cluster eFEDS~J084441.4+021702 just 2.5 arcmin away from this group and the assigned MCMF redshift of eFEDS~J084430.8+021737 corresponds to the residual signal of that cluster.
We find seven redshift outliers, including the aforementioned group at $z=0.05$ and the double system eFEDS~J091509.5+051521. In three of the remaining cases we find a second best counterpart with \fcont$<0.1$ consistent with WHL. The remaining two matches are obvious chance matches where the WHL system is more than 90 arcsec away with no associated X-ray emission.

\subsubsection{Group surveys}
As already indicated by the missed confirmation of RXCJ0920.0+0102 discussed above, it becomes hard to automatically confirm very nearby, X-ray selected systems using the photometric data from HSC or DECALS. The two main reasons for that are issues with the photometry of these nearby galaxies and the low mass of those systems. As visible in Fig.~\ref{fig:massvsredshift} and Fig.~\ref{fig:lambdavsmass}, the confirmed X-ray clusters reach down to low mass groups for which it is hard to distinguish between projected and bound galaxy over density using photometry alone.
Group and cluster catalogs based on spectroscopy allow one to probe this regime to evaluate and recover missed systems. We restrict this comparison to two catalogs coming from different surveys, one using SDSS data \citep{Tempel14} and one using redshifts from 2MASS selected galaxies.

With the available spectroscopic redshifts from SDSS, various group catalogs were constructed.
We picked here the SDSS DR10 based group catalog from \cite{Tempel14}. We find 35 matches with a 120 arcsec matching radius; of these, 33(32) show \fcont$<0.3$(0.2). From 33(32) sources with low \fcont, 25 have consistent redshifts, while all sources with \fcont$>0.3$ have inconsistent redshifts. From those 10 systems with inconsistent redshifts, we recover RXC J0920.0+0102, the $z=0.017$ group discussed in subsection \ref{xraysurvs} as well as the multiple system eFEDS~J091509.5+051521 discussed in the same section.
From the remaining clusters there is only one, eFEDS~J092821.1+041942, with a falsely assigned redshift. In this case MCMF assigns the redshift of a neighboring, more massive cluster (eFEDS~J092821.2+042149) to the X-ray source, while the actual best match is ranked second.
Besides RXC~J0920.0+0102, the only unconfirmed system with a match to the SDSS group catalog is eFEDS~J090806.5+032613. The optical investigation of that source does not provide a clear answer. The mass according to the SDSS catalog of the matched group is $1.8\times 10^{13}h^{-1}M_\odot$, which is at the lower limit to be detected in eFEDS. The three associated member galaxies are widely spread with 1.2 arcmin (140 kpc) being the distance to the nearest member to the X-ray center. Further, there are structures found in MCMF at $z=0.2$ and $z=0.84$, both containing at least one galaxy with a spectroscopic redshift. We therefore categorise the system as an unclear case.

\begin{figure}
\begin{center}
\includegraphics[width=0.99\linewidth]{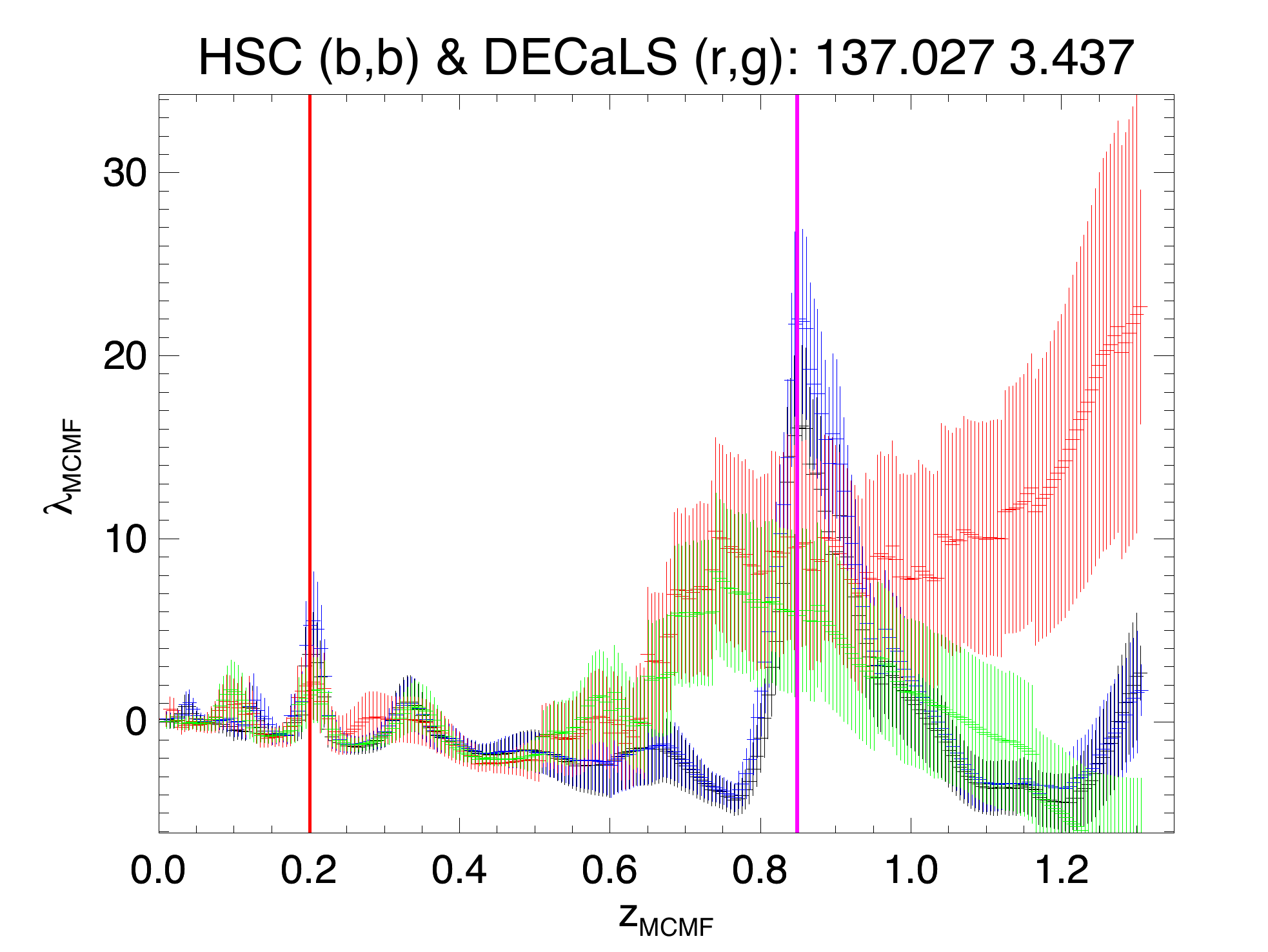}
\caption{eFEDS~J090806.5+032613: color coding is the same as in previous richness versus redshift plots.}
\label{fig:090806}
\end{center}
\end{figure}

From the available 2MASS based group catalogs, we use the one provided in \cite{Tully15} because it provides the largest number of matches within 120 arcsec with the eFEDS candidates. We find 7 positional matches, where 5 show offsets of less than 10 arcsec while the other two show offsets greater 100 arcsec. From the two outliers one is again associated with RXC~J0920.0+0102. The other match with a large offset is likely a chance association, because there seems to be no significant X-ray emission associated to the group center and the offset corresponds to more than 10 times the measured extent of that source.
All matches show \fcont>0.3 and are therefore not considered as MCMF confirmed, while all 5 matches with small positional offset can be considered as X-ray detected groups. Those missed systems show redshifts up to $z=0.05$ and reach X-ray luminosity based masses of up to $M_{500}\approx 6.0\times 10^{13}h^{-1}M_\odot$.

\subsubsection{Summary of the cross matching results}\label{sec:xmatchsum}
\begin{figure}
\begin{center}
\includegraphics[width=0.99\linewidth]{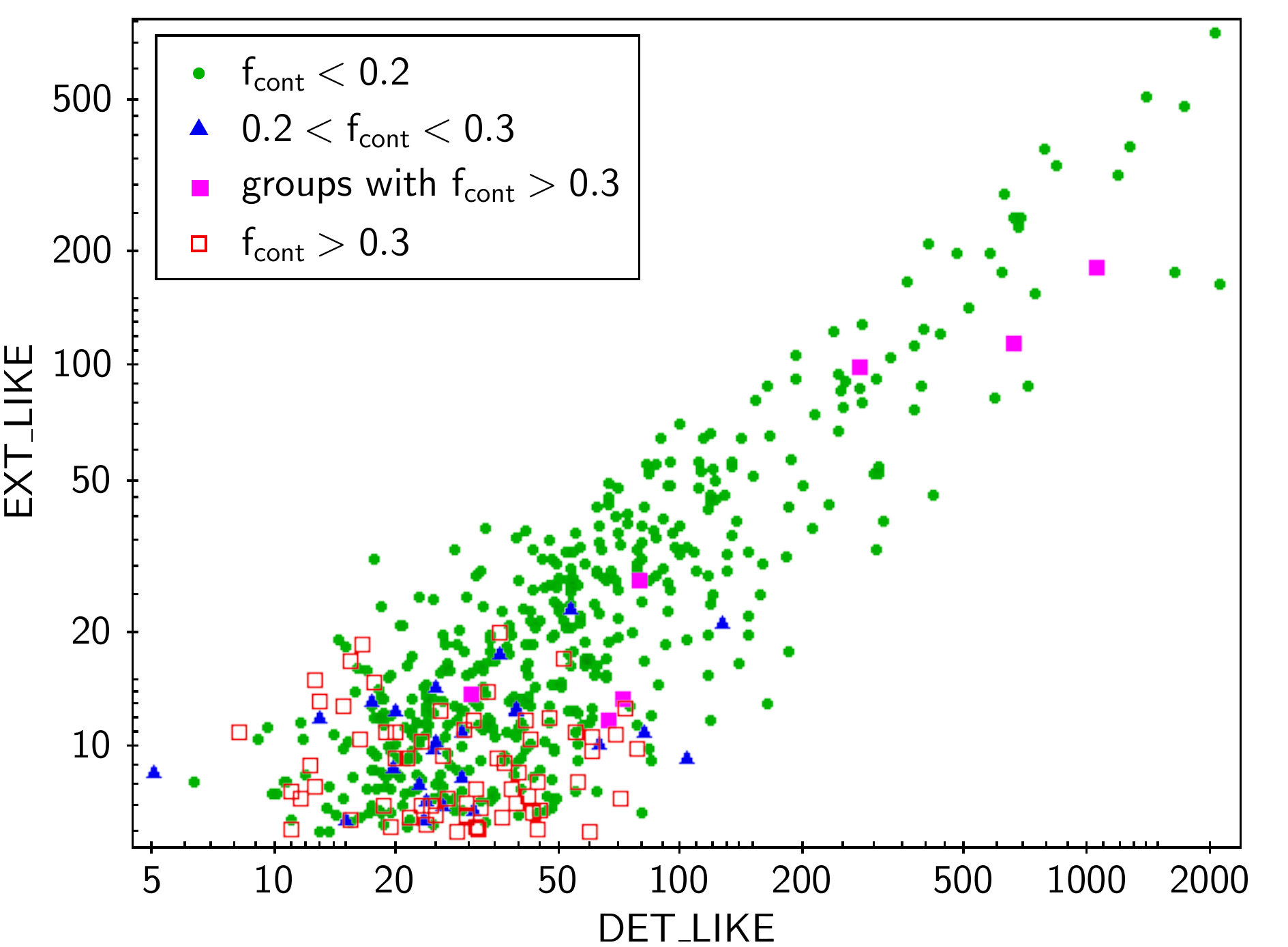}
\caption{Distribution of X-ray cluster candidates in X-ray extent likelihood versus detection likelihood, two of three key selection parameters. Clusters confirmed by MCMF are shown in green and blue. Unconfirmed clusters are shown in red. Systems not confirmed by MCMF but via cross-matching with group catalogs are shown as magenta squares. Unconfirmed systems lie at low likelihood values where contaminants are expected.}
\label{fig:extlikevsdetlike}
\end{center}
\end{figure}

The cross match with X-ray catalog resulted in the identification of one missing system and one system with two counterparts. All clusters matched to SZ selected systems are confirmed, but the redshift assignment of one cluster, eFEDS~J083120.5+030950, is changed to the initially second ranked cluster that show similar low \fcont. From the match with optical photometric cluster catalogs we identify one system, eFEDS~J084430.8+021737, that was missed by MCMF. The match to spectroscopic group an cluster catalogs yielded one redshift reassignment, eFEDS~J092821.1+041942. Further, it revealed five additionally missed groups that have a match to the 2MASS based group catalog by \citet{Tully15}.
Those in total seven missed systems are listed in Table~\ref{tab:missing} and are manually added into the cluster catalog. We assign dummy \fcont\ values of $-2$ for group matches where visual investigation of the optical and X-ray images yield X-rays from extended gas or from individual galaxies. A value of $-1$ is assigned for those group matches that show a clear indication of extended X-ray emission.
In Fig.~\ref{fig:extlikevsdetlike} we show the distribution of confirmed and unconfirmed eFEDS cluster candidates in extent likelihood (EXT\_LIKE) versus detection likelihood (DET\_LIKE), two of three key source selection variables. The third, source extent, modulates the width of the observed distribution. The systems found by the matching exercise and listed in Table~\ref{tab:missing} are shown as magenta squares. Besides the obvious scaling relation between EXT\_LIKE and DET\_LIKE it becomes obvious that unconfirmed systems cluster at low EXT\_LIKE and DET\_LIKE.
Three of the missed groups lie in the very high regime in EXT\_LIKE and DET\_LIKE where contamination by non-clusters is highly unlikely. In fact all MCMF unconfirmed systems that are not confirmed by cross matching with group catalogs lie within EXT\_LIKE$<20$ and DET\_LIKE$<80$.

\begin{table*}
\caption{List of MCMF unconfirmed systems (\fcont$>0.3$) that do have a counterpart in one of the matched catalogs. The \fcont\ entries are set to $-1$ (good) and $-2$ (less good) counterparts. The approximate X-ray based mass estimate is given in units of $10^{14} M_\odot/h$. The richness $\lambda$ is obtained at the given redshift.}\label{tab:missing}
\begin{center}
\begin{tabular}{l r r r r r r r r}
\hline
\hline
  \multicolumn{1}{c}{NAME} &
  \multicolumn{1}{c}{RA} &
  \multicolumn{1}{c}{DEC} &
    \multicolumn{1}{c}{DET\_LIKE} &  
    \multicolumn{1}{c}{EXT\_LIKE} &
  \multicolumn{1}{c}{z} &
  \multicolumn{1}{c}{$M_{500}$} &
  \multicolumn{1}{c}{\fcont} &
  \multicolumn{1}{c}{$\lambda$} \\
\hline
  eFEDS~J092002.2+010220 & 140.0090 & 1.0388 & 1049.64 & 179.60 & 0.017 & 0.169 & -1.0 & 1.30 \\
  eFEDS~J093744.2+024536 & 144.4340 & 2.7600 & 66.58 & 11.84 & 0.024 & 0.103 & -2.0 & 0.02 \\
  eFEDS~J090811.7-014811 & 137.0486 & -1.8032 & 72.52 & 13.37 & 0.04 & 0.146 & -2.0 & 0.00 \\
  eFEDS~J084034.6+023639 & 130.1440 & 2.6107 & 661.78 & 114.64 & 0.049 & 0.42 & -1.0 & 0.95 \\
  eFEDS~J084430.8+021736 & 131.1284 & 2.2935 & 79.31 & 27.49 & 0.0504 & 0.245 & -2.0 & 1.40 \\
  eFEDS~J084531.6+022831 & 131.3818 & 2.4753 & 275.71 & 98.86 & 0.0765 & 0.68 & -1.0 & 3.00 \\
  eFEDS~J093141.3-004717 & 142.9219 & -0.7883 & 30.46 & 13.71 & 0.093 & 0.556 & -1.0 & 3.00 \\
\hline\end{tabular}
\end{center}
\end{table*}

\subsection{Confirmation fraction and incompleteness}\label{sec:conffrac}

One way to evaluate the performance of MCMF cleaning is to compare the observed fraction of confirmed clusters to the expected confirmation fraction or purity
of the X-ray catalog based on simulations.
In the case of eFEDS, dedicated simulations were performed including details of the eFEDS footprint, background and exposure times. A detailed description of the simulations and their results appears in (Liu et al., submitted and Brunner et al., submitted).
One result is that the purity of the catalog is a strong function of extent likelihood (EXT\_LIKE), which allows one to not only compare overall confirmation fractions of simulation and observation but also check if its behaviour as a function of extent likelihood is as expected.

\begin{figure}
\begin{center}
\includegraphics[width=0.99\linewidth]{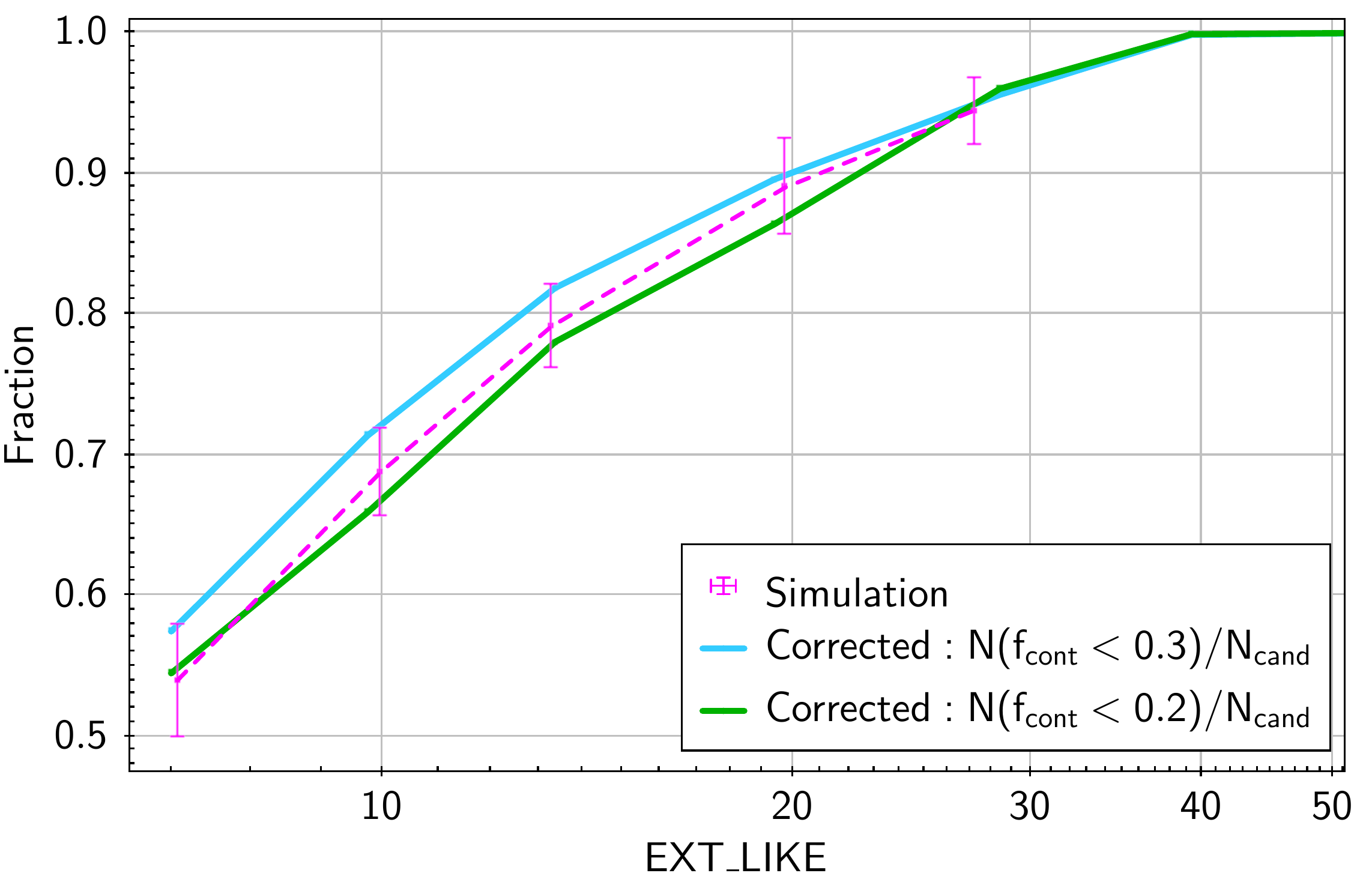}
\caption{Expected purity and recovered confirmation fraction as a function of extent likelihood. The expected fraction of detected clusters from dedicated eFEDS simulations is shown in magenta. The error bars indicate the standard deviation derived from 18 realisations of the simulations. Blue shows the fraction of \fcont$<0.3$ systems corrected for the expected number of residual contaminants. Green is similar but for \fcont$<0.2$. The difference between the green and red lines is an indication of the incompleteness caused by the stricter \fcont\ selection.}\label{fig:conffrac}
\end{center}
\end{figure}

In Fig.~\ref{fig:conffrac} we show the results of this comparison. The dashed magenta line shows the fraction of extent selected X-ray detections associated with clusters or cluster wings over number of all extent selected X-ray detections in the simulation. Sources classified as cluster wing are associated with a cluster but are not the primary detection. As we do not filter for such cases in the optical confirmation, we include them in the overall fraction of confirmed systems.
The blue line shows the fraction of \fcont$<0.3$ systems corrected for the residual contamination allowed,
\begin{equation}
    f_\mathrm{corrected,\mathrm{EXT\_LIKE}}=\frac{N_\mathrm{EXT\_LIKE}(f_\mathrm{cont}<c)}{N_\mathrm{EXT\_LIKE}}(1-c\,c_\mathrm{Sim}),
\end{equation}
where $N_\mathrm{EXT\_LIKE}$ is the number of candidates in a given bin in EXT\_LIKE, $N_\mathrm{EXT\_LIKE}(f_\mathrm{cont}<c)$ the subset with \fcont\ below threshold $c$ (here 0.3) and $c_\mathrm{Sim}$
the fraction of non-clusters in the simulation.
The green line shows the same but for \fcont$<0.2$ instead.

The difference between the lines derived using \fcont$<0.3$ and \fcont$<0.2$ indicate
excess incompleteness caused by the stricter cut. The difference between both cuts correspond to ~14 (3\%) of the real sources being missed with the stricter cut, which is within the scatter range seen within the simulations.

Using the curves in Fig.~\ref{fig:conffrac} and the number of sources given EXT\_LIKE we can estimate the contamination fraction of the original cluster candidate list, under the assumption of no incompleteness caused by the cleaning. Consequently using the curve including the identified groups missed by MCMF we expect in total 442 real sources; given the 542 candidates, this results in a contamination fraction in the original candidate list of 17\%.

\subsection{Estimate of the excess incompleteness induced by optical cleaning}\label{sec:incomp}

Any cluster survey that needs auxiliary information such as redshifts or cleaning of contaminants, likely suffers from additional incompleteness coming from the process of obtaining the needed information.
This is especially true if the initial cluster survey challenges the redshift or mass reach of the follow-up survey. In this particular case eFEDS challenges the optical confirmation with LS and HSC in the low mass and low redshift regime.
This is already reflected in the redshifts and masses of the systems that we find to be missed by MCMF (discussed in Sect.\ref{sec:xmatchsum}). 
The MCMF confirmation or cleaning of the X-ray candidate catalog is based on a redshift dependent richness cut that corresponds to a cut in \fcont. Because this is a systematic confirmation method, its impact can be modeled by using scaling relations connecting detection observable and richness.

The proper way to measure and account for the impact of the optical cleaning on the sample is to include the richness measure into the overall analysis and trace its scaling with mass and selection observable simultaneously. For an example using a MCMF based catalog, see \cite{MARDY3selec}.
Here, we perform a simplified estimation of the incompleteness induced by the MCMF cleaning by only using observed quantities.

As a first step, we measure the \LamMCMF - $M_{500}$ scaling relation and its scatter. We note here again that $M_{500}$ is the approximate mass estimate based on a simple model and therefore is close to the count rate and with that to the detection probability, just taking out the most pronounced dependencies such as on redshift and Galactic nH column density. 
We assume that the distribution of richnesses $\lambda$ at a given mass $M_{500}$ and redshift $z$ is given by a log-normal distribution
\begin{equation}
    P(\ln \lambda | M_{500}, z) = \mathcal{N}\big(\ln \lambda; \langle \ln \lambda \rangle(M_{500}, z), \sigma^2(M_{500}, z)\big), 
\end{equation}
with mean 
\begin{equation}
    \langle \ln \lambda \rangle(M_{500}, z) = \ln \lambda_0 + \alpha_0  + \alpha_M \ln\left(\frac{M}{M_0}\right) + \alpha_z \ln \left(\frac{1+z}{1+z_0}\right),
\end{equation}
and variance
\begin{equation}
    \sigma^2(M_{500}, z) = \exp\Big( \ln \zeta(z) - \langle \ln \lambda \rangle\Big) + \exp\Big(s + s_M \ln\Big(\frac{M}{M_0}\Big)\Big), 
\end{equation}
where $(\alpha_0,\, \alpha_M,\, \alpha_z,\, s,\, s_M)$ are the free parameters to be constrained by the likelihood analysis, $(\lambda_0=23,\,M_0=3\text{e}14\, \text{M}_\odot/h,\, z_0=0.35)$ are the pivots of the scaling relation and are chosen close to the medians in richness, mass, and redshift, respectively. $\zeta(z)$ is a redshift dependent factor which relates the number of actually measured galaxies to the reported richness (see equation 7 in \cite{MARDY3}). 
It modulates the first term of the variance, which capture the Poisson noise in measuring the richness. The second term of the variance models the intrinsic variance of the cluster population.
We find best fit parameters of $\alpha_0=0.192\pm0.025$, $\alpha_M=1.14\pm0.06$, $\alpha_z=-1.25\pm0.26$, $s=-1.69\pm0.10$ and $s_M=-0.12\pm0.17$. More details on the fitting including the impact of optical cleaning can be found in Appendix~\ref{app:scaling}.

Given the scaling relation, we can calculate for each confirmed cluster $i$ the probability of it not being confirmed as 
\begin{equation}\label{eq:pmiss}
    P_i = \int^{\ln\lambda_\text{cut}(z_i)}_{-\inf}\text{d}\ln \lambda\, P(\ln \lambda | M_{500,i}, z_i),
\end{equation}
where $\ln\lambda_\text{cut}(z_i)$ is the richness cut for a given \fcont\ threshold and cluster redshift $z_i$. And with that information we can derive a first estimate of the number of missed systems by summing over all clusters as,
\begin{equation}
    N_\text{missed} = \sum_i P_i.
\end{equation}
Using the posteriors from the scaling relation and taking the 6\% (4\%) residual contamination for the \fcont\ 0.3 (0.2) into account we end up with $10.4^{+2.6}_{-2.6}$ ($25.8^{+6.0}_{-4.7}$) missed systems.

Although those numbers fit quite well with the observed number of unconfirmed systems as well as the expected loss of systems when going from \fcont$<0.3$ to \fcont$<0.2$ (14 systems), they do not make use of the observed count rates and the number of unconfirmed candidates in the candidate list. 
It further relies on reasonably sampling the redshift range of true clusters, which might be violated given that we do not confirm any cluster below $z<0.05$, while the majority of missed systems are at that redshift. We therefore aim for a second, more sophisticated approach, making use of all candidates, the observed purity given extension likelihood and the redshift distribution.
We modify equation~\ref{eq:pmiss} by replacing the cluster redshift $z_i$ by the normalised smoothed redshift distribution $P_z(s)$ of confirmed clusters plus the missed systems that were added after cross comparison. The assumption here is that this distribution is close to the one of the true clusters in the X-ray catalog. To account for catalog contamination, we further multiply by that probability by $P_\text{true}$, the probability of a candidate being true given extent likelihood as shown in the previous section.
With that the probability of a system being missed is,
\begin{equation}\label{eq:pmiss2}
    P_j = P_{\text{true},j}\int_{0}^{\inf}dz\int^{\ln\lambda_\text{cut}(z)}_{-\inf}\text{d}\ln \lambda\, P(\ln \lambda | M_{500,j}, z)P_z(z).
\end{equation}
Summing now over all X-ray candidates yields $8.2_{-1.5}^{+2.2}$ and $22.5_{-3.0}^{+4.2}$ for \fcont$<0.3$ and 0.2 and a difference between both of $14.2^{+2.2}_{-1.3}$. This also fits well with the number of missed systems, the difference in number between when shifting the \fcont\ cuts and the simpler estimate before.

\begin{figure}
\begin{center}
\includegraphics[width=0.99\linewidth]{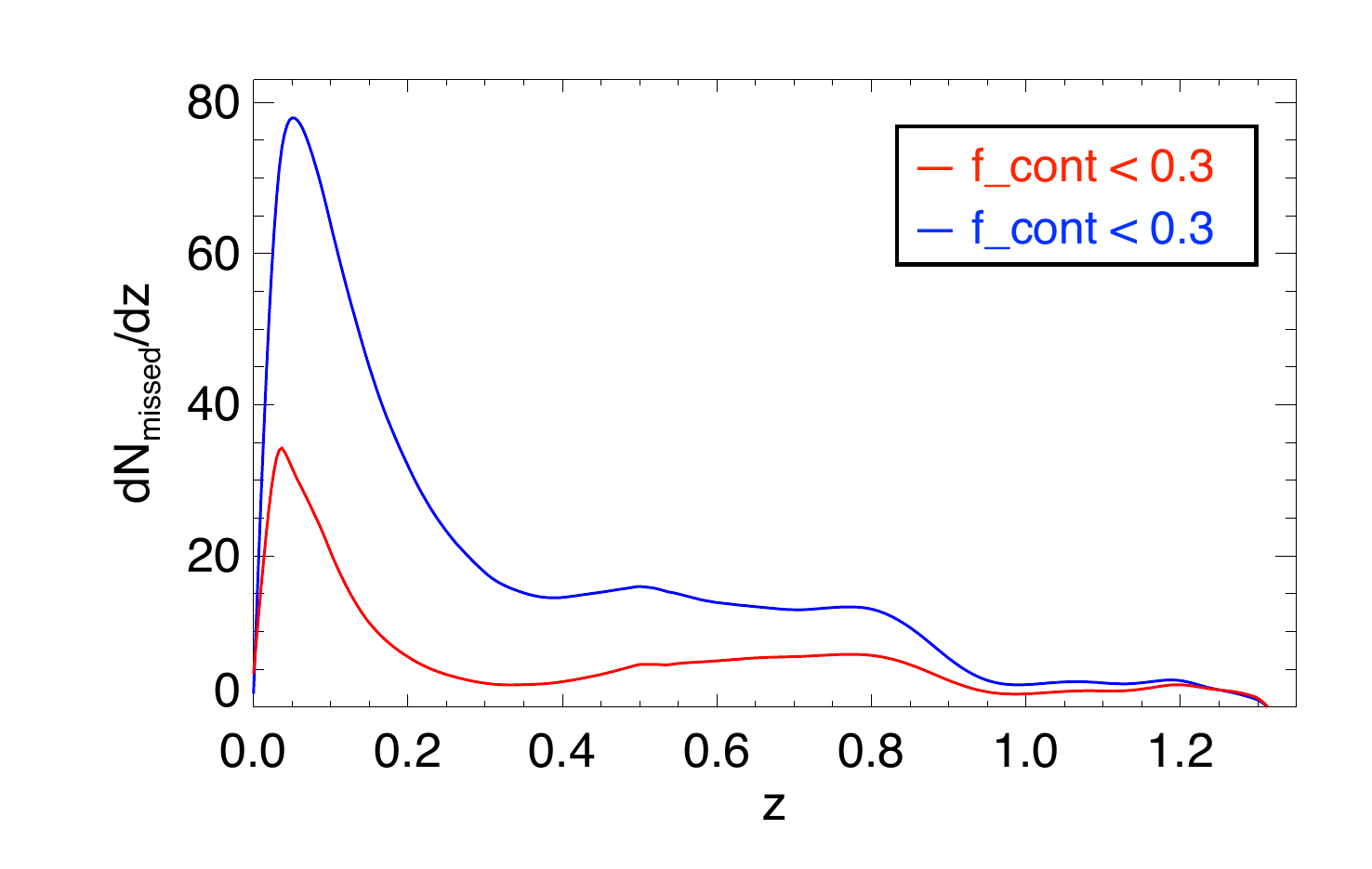}
\caption{Redshift distribution of missed systems according to the adopted selection: \fcont$<0.2$ (blue) and  \fcont$<0.3$ (red)}
\label{fig:redshiftmiss}
\end{center}
\end{figure}

\begin{figure}
\begin{center}
\includegraphics[width=0.99\linewidth]{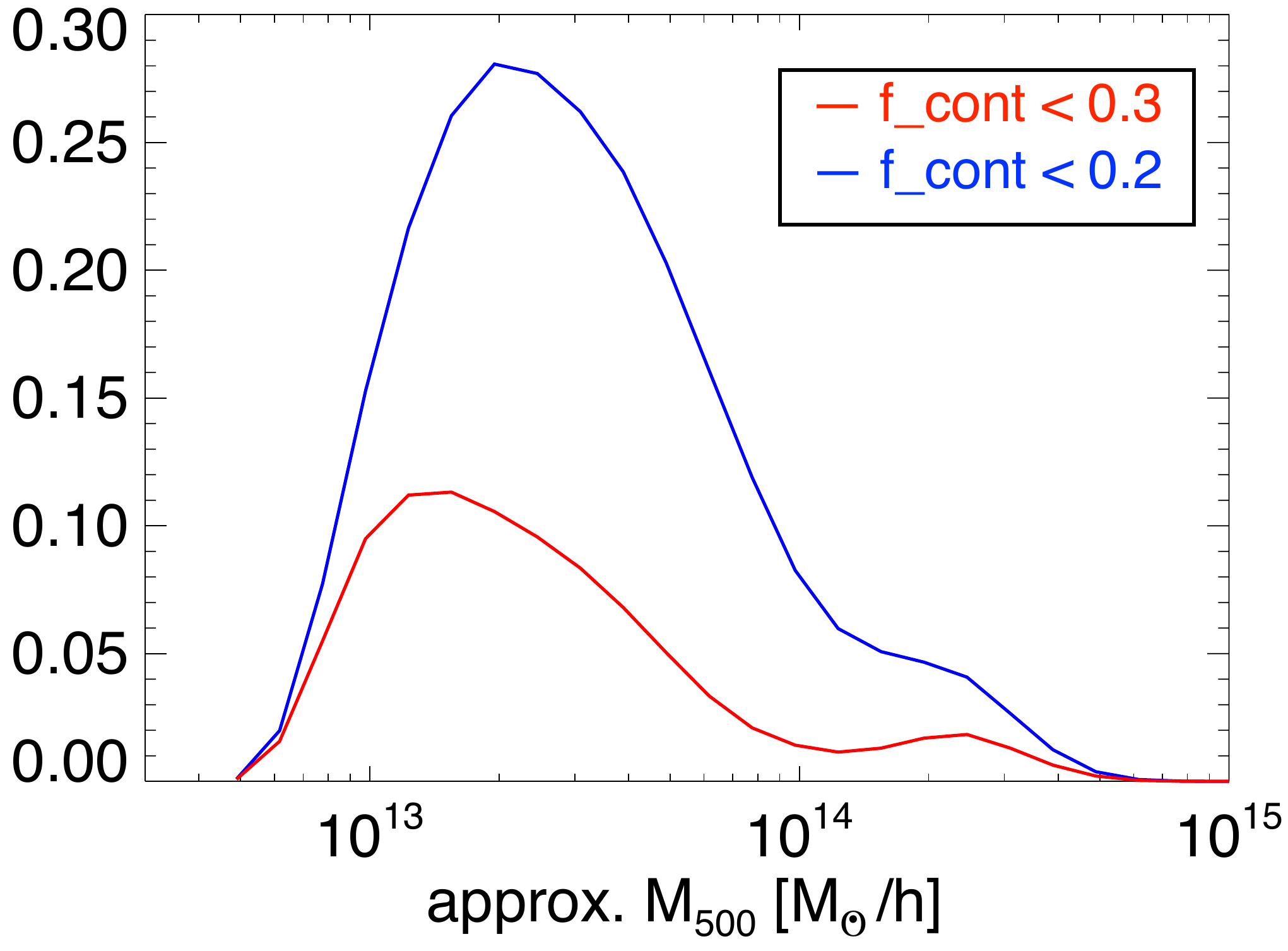}
\caption{Estimated $M_{500}$ distribution of missed systems according to the adopted selection: \fcont$<0.2$ (blue) and \fcont$<0.3$ (red).}
\label{fig:massmiss}
\end{center}
\end{figure}

As a side result we can further estimate the redshift and mass distribution of the missed systems by simply omitting the integral over redshift and summing over all candidates. The redshift distribution is shown in Fig.\ref{fig:redshiftmiss} and peaks at very low redshifts with a flat tail over the remaining redshift range. The missing systems later found by cross-matching are consistent with the peak of the redshift distribution, and the same is true for the mass distribution shown in Fig.\ref{fig:massmiss}.

In summary, this approach of estimating the incompleteness induced by the optical confirmation process provides consistent results between findings from individual catalog matches and in difference of confirmed systems for different \fcont\ cuts. Further, the predicted redshift distribution of missed systems is consistent with that of  the missed systems that are found by cross matching to other catalogs. We find that the optical cleaning induces an incompleteness of ~2\% (5\%) for \fcont$<0.3$ (0.2) cuts. Based on this, adding the identified MCMF unconfirmed but spectroscopically confirmed groups into the list of confirmed clusters would lead to a close to fully complete (>99\%) cluster catalog.

\subsection{Adding information from point source optimized source identification}

So far we call any non-cluster source a contaminant, making no distinction between it being a real non-cluster source or a spurious (noise) fluctuation. From simulations we do expect that almost all sources in the extent selected sample do have a true underlying astrophysical signal and are not noise fluctuations. The majority of other astrophysical sources fall under the category of point sources, mostly AGN and stars. The source identification for point like sources is treated separately in the dedicated companion paper (Salvato et al., submitted).
The size and variety of the training sample used to calibrate the point source identification method generally allows the identification of the counterparts independently on their nature. Thus that also cluster members near the X-ray position can be identified as correct counterpart.
On the other side not having a good point like counterpart does not mean that the X-ray emission stems from a cluster, either. A combination of multiple confused point sources causing the X-ray position to be off from a good counterpart is also possible.
Further, combinations of point and extended sources exist too, e.g. central AGN or miss-centered cluster due to a point source on top of a flat surface brightness source.
Without high resolution X-ray imaging many of these cases cannot be resolved by combining MCMF and point like identifications. 

The primary tool for finding the best point like counterparts is NWAY \citep{Salvato18}.
NWAY is based on a Bayesian method to find best counterparts given a set of photometric priors and the X-ray-to-optical relative astrometry.
In its latest iteration on eFEDS (Salvato et al., submitted), NWAY was complemented with priors in optical to MIR from LS and complemented with GAIA proper motion parameters. The tool provides probabilities of a certain source to have a LS counterpart (\texttt{p\_any}), which is the main output used for purity and completeness calibrations. The thresholds and purity derived for the point source catalog are likely not valid for the extent selected catalog as \texttt{p\_any} is directly related to X-ray morphological characteristics.


\begin{figure}
\begin{center}
\includegraphics[width=0.99\linewidth]{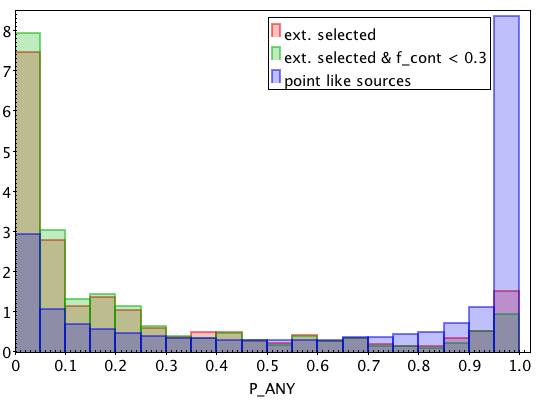}
\caption{Normalised distribution of \texttt{p\_any} for the extent selected and point like samples with and without a cut at \fcont$<0.3$. The distributions differ most for \texttt{p\_any}$<0.3$ and \texttt{p\_any}$>0.7$, reflecting the different mix of X-ray source types in the extent selected sample with respect to the point like sample.}
\label{fig:panyhist}
\end{center}
\end{figure}

\begin{figure}
\begin{center}
\includegraphics[width=0.99\linewidth]{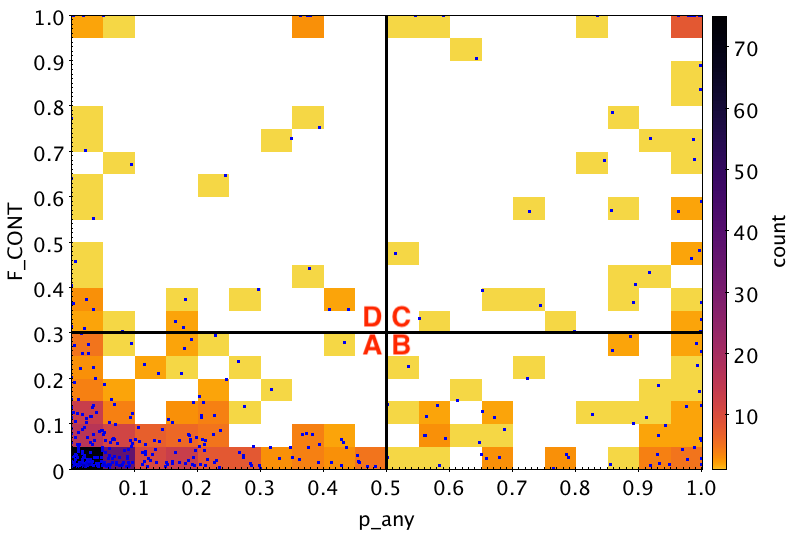}
\caption{Distribution of sample in two confirmation proxies \fcont\ and $p_\mathrm{any}$. Black lines splits the sources in four sectors (A, B, C, D). Most sources are at low $p_\mathrm{any}$ and low \fcont\ (sector A), suggesting that those are good clusters without a good point like counterpart. The second densest point is at high $p_\mathrm{any}$ and \fcont\ (sector C), suggesting good AGN. The bottom-right corner (low \fcont,high $p_\mathrm{any}$, sector B) is the region where both codes find counterparts and where further reconciliation is required, while the top-left corner (sector D) are sources without good counterpart in any of the codes.} 
\label{fig:fcontpany}
\end{center}
\end{figure}

This difference becomes visible in the different distribution of \texttt{p\_any} for the point like and the extent selected sample shown in Fig.~\ref{fig:panyhist}.

For this work we therefore choose \texttt{p\_any}=0.5 as boundary between good and less good point like counterpart
In Fig.\ref{fig:fcontpany} we show the distribution of sources in the \fcont-\texttt{p\_any} plane. There is an obvious over density at low \fcont\ and low \texttt{p\_any}, which is the region with the highest probability of being a cluster and with no good point like counterpart. On the other side  we find the most likely point sources at high \texttt{p\_any} and high \fcont.
Using \fcont=0.3 and \texttt{p\_any}=0.5 as division lines one can split the plane into four sectors, with rather clear cases for clusters in the lower left and good point sources in the upper right. The top left sector (sector D) is the region where no method finds a safe counterpart and the bottom right is the sector where both find good counterparts. 

Leaving the two sectors with clear counterparts aside we focus first on sector D. This is the sector where we expect to find most of the clusters lost by cleaning, as well as blended AGN or spurious sources.
We inspect all 29 sources by eye using the available optical images and the X-ray surface brightness maps. We find a few cases where the source are close to very bright X-ray sources, indicating that those might be spurious sources caused by the bright primary source.
We also find four potential cluster matches, although X-ray surface brightness maps do not show a convincing peak at the optical counterpart. From Sect.~\ref{sec:incomp} we expect about $8_{-2}^{+2}$ missing systems in total. In Sect.~\ref{sec:crossmatches} we find four good and three less good counterparts. 
Even when counting all those systems as truly missed clusters, the total number of 11 systems would still be within two sigma from our estimate of the incompleteness. Allowing some systems to be projections or AGN dominated will just improve the agreement between observation and prediction.

The second sector of interest is sector B, which contains 74 sources. As already outlined, a high \texttt{p\_any} as such does not exclude typical cluster galaxies as possible counterparts. We therefore aim to filter out clear cases where the best counterpart is actually a cluster member. To do so, we match the best counterparts found by NWAY with existing spec-$z$ catalogs. Further, we derive red sequence based photometric redshifts similar to redMaGiC \citep{redmagic} using the LS photometry. For sources where the red sequence model is a reasonable fit, we obtain $\sigma_{\Delta z/(1+z)}=0.014$ for $z_\text{phot}<0.9$ based on 578 spectroscopic redshifts in the joint point source and extended source catalog.

Using maximum offsets of $\Delta z/(1+z)<0.05$ between MCMF and spec-$z$ and $\Delta z/(1+z)<0.1$ between MCMF and photo-$z$, we find 43 systems with consistent redshifts between point like counterpart and cluster candidate. Thus leaving 31 sources with good counterparts in both but with discrepant redshifts.
We then visually inspect all systems using HSC, Legacy Survey images and smoothed X-ray surface brightness maps. From the 43 sources with consistent redshifts, we find 1 clear and one unclear contaminant (1.3-4.6\%). The vast majority are indeed obvious clusters where the X-ray PS counterpart is identical to the BCG. 

Repeating the same for the 31 cases with discrepant or non-valid redshift estimates the picture appears different and more complex compared to the other subset. Only eight systems appear as clear unaffected clusters, while nine systems appear point like. Another nine systems appear to be affected by both point like and extended emission. Those systems typically show X-ray emission at the optical cluster position, but the X-ray center seems to be shifted to a location of a good point like counterpart. Additionally we find 4 systems where a classification is unclear, typically associated with sources of lower \texttt{p\_any}. Summarising, we find 29-74\% contamination, just counting secure cluster or point sources. 

Based on these findings we add a additional flag to the catalog according to the sectors discussed. The flag corresponding to sector B is split into B1, where PS and cluster redshifts in agreement and B2 with redshift disagreement. This flag can be used to select even cleaner cluster sub-samples or to explicitly study AGN leaking into the clusters and group candidate catalog.

\subsection{Estimate on contamination in cluster catalog}

\begin{figure*}
\begin{center}
\includegraphics[width=0.33\linewidth]{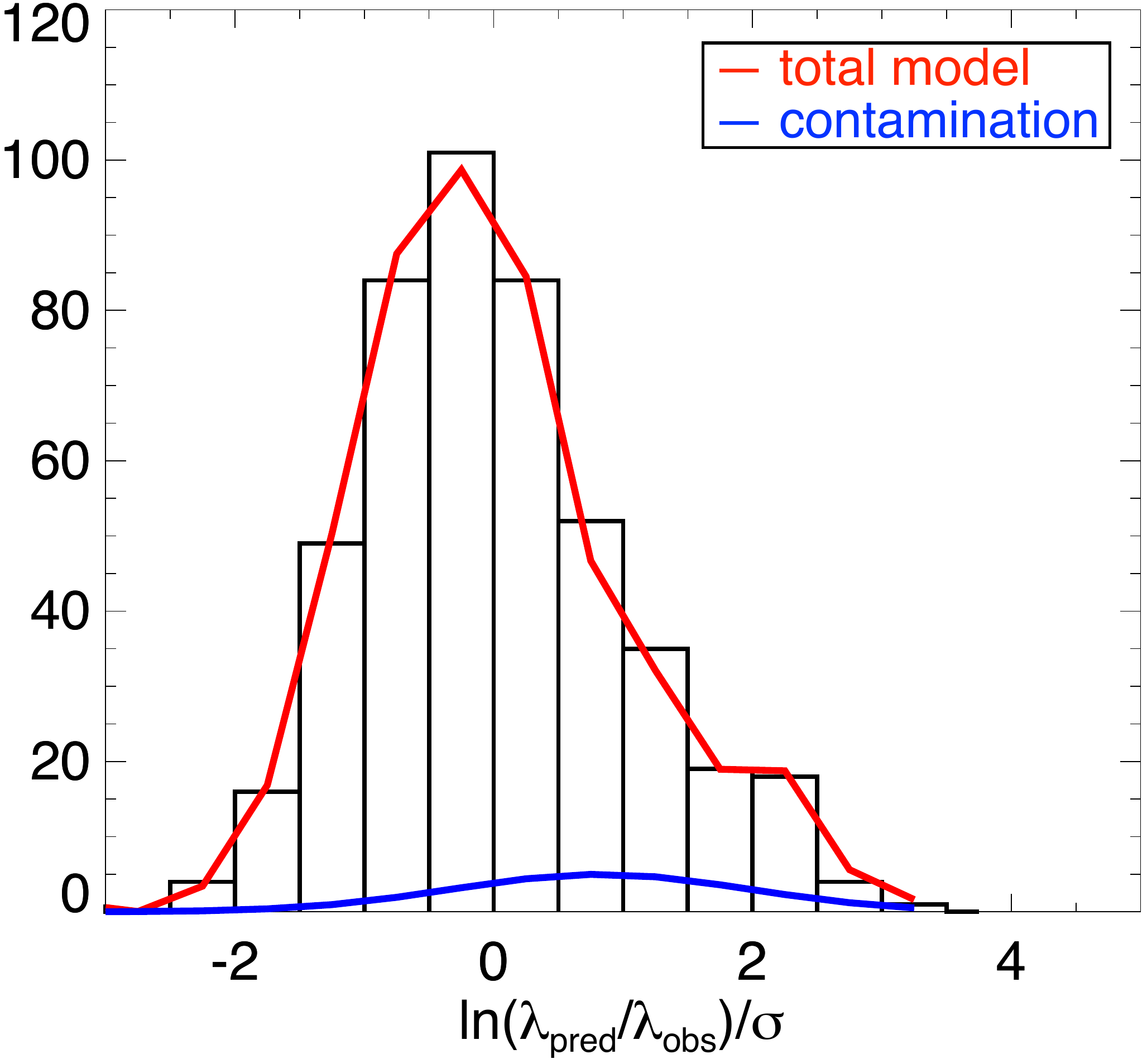}
\includegraphics[width=0.33\linewidth]{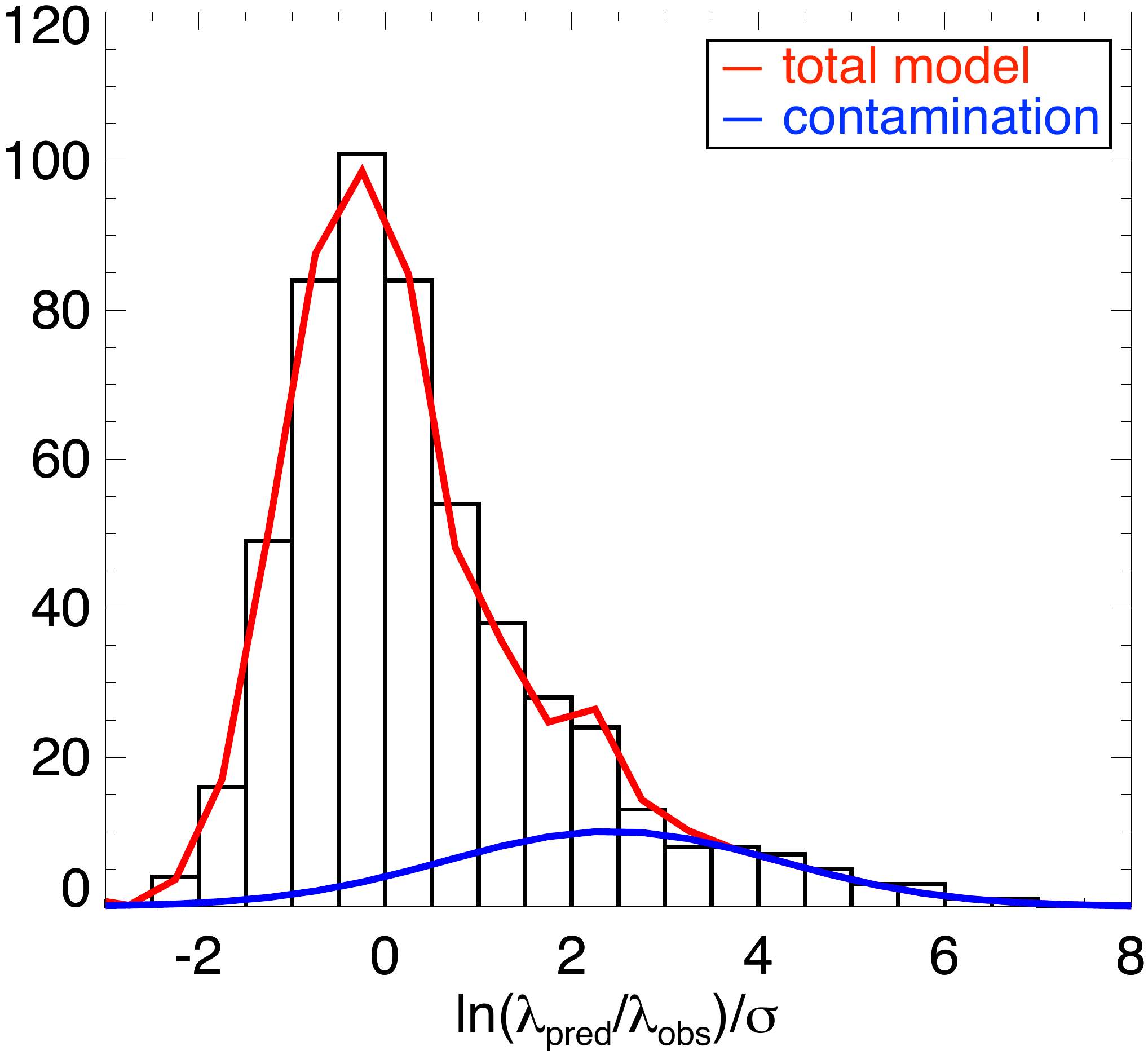}
\includegraphics[width=0.32\linewidth]{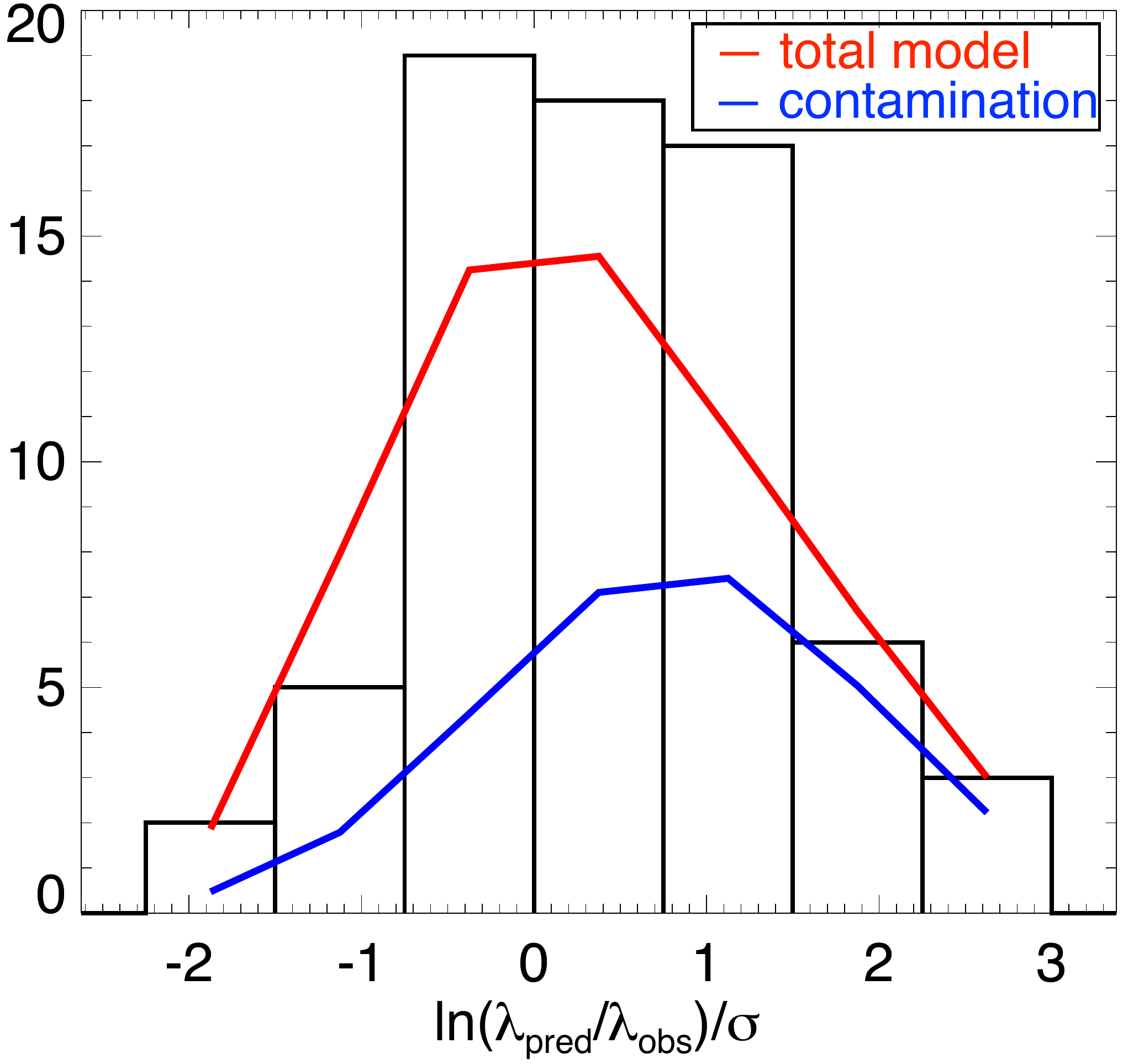}
\caption{Offset distribution around the best fit scaling relation in units of the standard deviation $\sigma$ for the \fcont$<0.3$ (left), the full (middle) and the \fcont$<0.3$ \& \texttt{p\_any}>0.5 sample (right). The red line show the best fit total model while the blue line shows the model of the contaminating population.} 
\label{fig:scalingoffsets}
\end{center}
\end{figure*}

In addition to the X-ray extent selected candidate catalog, MCMF was also run on the X-ray point like source list in eFEDS. More details on the run on point sources are provided in Appendix~\ref{app:mcmfonps}. The key difference between the runs is that the richness is estimated from optical information only and does not include the X-ray count rate based prior on the aperture within which the richness is extracted. 
Hence, the richness is constructed in that way that it is unbiased against the richness given in the run on extent selected clusters.
With that the MCMF run on the point like sample can act as a reference sample for the contaminants as it is dominated by real point like sources and shares a similar flux distribution as the expected point source contaminants in the extend selected sample. 

A reference sample for the \fcont$<0.3$ contaminants can be constructed by applying the same $\lambda$ versus redshift cut as done with the \fcont$<0.3$ on the extent selected sample. Second one has to apply a flux cut to mimic the selection of the extent selected sample. This is simply done my cutting at the minimum rate found in the extent selected sample. To further reduce the number of spurious sources and cluster signal coming from nearby detected clusters, we exclude 10~arcmin regions around extent selected clusters and select also \texttt{p\_any}>0.5.

From Fig.~\ref{fig:lambdavsmass}, we know that the main contamination is expected to be in the low-$\lambda$ high X-ray based mass corner of the scaling relation. Using the scaling relation derived in sec~\ref{sec:incomp} we can measure the offset distribution around the scaling relation measured in the sigma of the scatter. In Fig.\ref{fig:scalingoffsets} we show this distribution for the \fcont$<0.3$ as well as for the full sample. 
The offset distribution is then a combination of a the distribution of true clusters and of point like contaminants. We therefore aim for fitting a model for both populations to derive the fraction of contaminants in the sample.
The offset distribution for the reference contaminant sample is very well described by a Gaussian function. We therefore use the best fit Gaussian function as a model for the contaminant sample, leaving only the normalisation as free parameter.  For the clean clusters sample we use the \fcont$<0.3$ sample with additional cut on \texttt{p\_any}$<0.5$, to suppress the majority of contaminating point sources.
We then perform a MCMC fit to the observed offset distribution of the \fcont$<0.3$ sample, with the normalisations of the two sub populations as the only free parameters. We find a best fit contamination of $6_{-3}^{+3}\%$, which is in good agreement with the expected value of 6\%. The best fit model is shown as a red line in Fig.\ref{fig:scalingoffsets}.

Motivated by this, we repeat the same exercise for the full sample without an \fcont\ cut. The richness cut for the contaminant sample needs to be changed to a simple $\lambda>2$ cut, the typical lower limit in the default MCMF run. The best fit model is again shown in Fig.\ref{fig:scalingoffsets}. The best fit contamination is $17_{-3}^{+3}\%$, which fits well to the expectation of 19.4\% from the simulations and even better to our empirical estimate in Sect.\ref{sec:conffrac}.
Compared to Sect.\ref{sec:conffrac}, our results here do not rely on the simulations, the correct estimate of \fcont\ or on the completeness. 

The main assumption going into the estimate is that the \fcont$<0.3$ and \texttt{p\_any}$<0.5$ (class A) sub-sample is close to a clean sample and that the \texttt{p\_any} cut does not introduce an additional selection affecting the offset distribution besides increasing purity. To test this, we repeat the population modeling on the \fcont$<0.3$ and \texttt{p\_any}$>0.5$ sample.
If the assumption is true that the selected sub-sample is indeed clean, then the expected 6\% contaminants should be inside this sample. The result is shown in Fig.\ref{fig:scalingoffsets}. Despite the fact that this sub-sample is small, the composite model does not seem to match the data as nicely as in the previous cases. This is likely an indication that the typical scaling of the cluster population is different compared to the assumed model. One reason for that could be that this sample could host more relaxed cool core clusters than the clean reference sample. Cool core clusters show usually small X-ray to BCG offsets so that NWAY might identify the BCG as a good point source candidate. Another potential reason could be a significant contribution of a point source to the cluster emission. Both effects would shift the peak of the distribution to higher values, as suggested in the observed distribution. Leaving these issues aside, the fit suggests a significantly higher level of contamination than the other samples. The total number of contaminants is estimated to be $28.4\pm3.9$, which again is similar to the expected 28 systems from assuming 6\% contamination of the overall \fcont$<0.3$ sample. Although this number needs to be interpreted with care, it at least suggests that the majority of contaminants in the sample are indeed at \texttt{p\_any}>0.5.


\section{Conclusions}
The eROSITA eFEDS field is observed to a depth similar to that expected to be reached in that region at the end of the full, 4 year all-sky survey.
In this paper we describe the optical identification of X-ray selected galaxy cluster and group candidates from this field.
The optical identification yields a cluster catalog reaching out to $z=1.3$ and confirms groups out to $z\sim0.3$.
 We confirm and obtain redshifts for $>98\%$ of the true clusters and groups using only optical photometric data while simultaneously reducing the candidate catalog contamination by 70\%. 
 Using the richness to X-ray mass proxy scaling relation and the  X-ray source candidates we predict incompleteness induced by optical identification to be of 2\% (5\%) for the \fcont$<0.3$ (0.2) optical confirmation thresholds. The modeled incompleteness is consistent with the number of systems found by matching the eFEDS catalog to various cluster and group catalogs available in literature. The recovered fraction of confirmed systems as well as its dependency on source extent likelihood is in good agreement with predictions of dedicated simulations of the eFEDS field. The catalog contamination is estimated by modeling the impact of point like sources on the scatter distribution around the derived scaling relation. The final confirmed cluster catalog contamination of $6\pm3$\% is in good agreement with the expectation from the adopted cut in \fcont\ and the expected initial level of contamination. By adding the systems found to be missed by MCMF, we expect the catalog to include $>99$\% of the real clusters and groups in the X-ray cluster candidate list.
 Besides optical information related to cluster confirmation, we also provide optical estimators of the dynamical state of clusters and show its application for selecting galaxy cluster mergers.
 
These results provide a positive outlook for the future eROSITA all-sky surveys, as we could show that the vast majority of clusters and groups can be confirmed with good quality photometric data alone. By complementing the optical followup with cross-comparison to existing low redshift group catalogs, it should be possible to reach high completeness in optical confirmation of cluster candidates.

\begin{acknowledgements}

This work is based on data from eROSITA, the soft X-ray instrument aboard SRG, a joint Russian-German science mission supported by the Russian Space Agency (Roskosmos), in the interests of the Russian Academy of Sciences represented by its Space Research Institute (IKI), and the Deutsches Zentrum für Luft- und Raumfahrt (DLR). The SRG spacecraft was built by Lavochkin Association (NPOL) and its subcontractors, and is operated by NPOL with support from the Max Planck Institute for Extraterrestrial Physics (MPE).

The development and construction of the eROSITA X-ray instrument was led by MPE, with contributions from the Dr. Karl Remeis Observatory Bamberg \& ECAP (FAU Erlangen-Nuernberg), the University of Hamburg Observatory, the Leibniz Institute for Astrophysics Potsdam (AIP), and the Institute for Astronomy and Astrophysics of the University of Tuebingen, with the support of DLR and the Max Planck Society. The Argelander Institute for Astronomy of the University of Bonn and the Ludwig Maximilians Universitaet Munich also participated in the science preparation for eROSITA.
The eROSITA data shown here were processed using the eSASS/NRTA software system developed by the German eROSITA consortium.

The Hyper Suprime-Cam (HSC) collaboration includes the astronomical communities of Japan and Taiwan, and Princeton University.  The HSC instrumentation and software were developed by the National Astronomical Observatory of Japan (NAOJ), the Kavli Institute for the Physics and Mathematics of the Universe (Kavli IPMU), the University of Tokyo, the High Energy Accelerator Research Organization (KEK), the Academia Sinica Institute for Astronomy and Astrophysics in Taiwan (ASIAA), and Princeton University.  Funding was contributed by the FIRST program from the Japanese Cabinet Office, the Ministry of Education, Culture, Sports, Science and Technology (MEXT), the Japan Society for the Promotion of Science (JSPS), Japan Science and Technology Agency  (JST), the Toray Science  Foundation, NAOJ, Kavli IPMU, KEK, ASIAA, and Princeton University.
 
This paper makes use of software developed for the Large Synoptic Survey Telescope. We thank the LSST Project for making their code available as free software at  http://dm.lsst.org
 
This work was supported in part by the World Premier International
Research Center Initiative (WPI Initiative), MEXT, Japan, and JSPS
KAKENHI Grant Nos. JP18K03693 and JP19KK0076.  In addition, we acknowledge support from the Excellence Cluster ORIGINS, the Max Planck Society Faculty Fellowship program and the Ludwig-Maximilians-Universitaet. 
 
This paper is based on data collected at the Subaru Telescope and retrieved from the HSC data archive system, which is operated by Subaru Telescope and Astronomy Data Center (ADC) at NAOJ. Data analysis was in part carried out with the cooperation of Center for Computational Astrophysics (CfCA), NAOJ.
 
The Pan-STARRS1 Surveys (PS1) and the PS1 public science archive have been made possible through contributions by the Institute for Astronomy, the University of Hawaii, the Pan-STARRS Project Office, the Max Planck Society and its participating institutes, the Max Planck Institute for Astronomy, Heidelberg, and the Max Planck Institute for Extraterrestrial Physics, Garching, The Johns Hopkins University, Durham University, the University of Edinburgh, the Queen’s University Belfast, the Harvard-Smithsonian Center for Astrophysics, the Las Cumbres Observatory Global Telescope Network Incorporated, the National Central University of Taiwan, the Space Telescope Science Institute, the National Aeronautics and Space Administration under grant No. NNX08AR22G issued through the Planetary Science Division of the NASA Science Mission Directorate, the National Science Foundation grant No. AST-1238877, the University of Maryland, Eotvos Lorand University (ELTE), the Los Alamos National Laboratory, and the Gordon and Betty Moore Foundation.

The Legacy Surveys consist of three individual and complementary projects: the Dark Energy Camera Legacy Survey (DECaLS; Proposal ID \#2014B-0404; PIs: David Schlegel and Arjun Dey), the Beijing-Arizona Sky Survey (BASS; NOAO Prop. ID \#2015A-0801; PIs: Zhou Xu and Xiaohui Fan), and the Mayall $z$-band Legacy Survey (MzLS; Prop. ID \#2016A-0453; PI: Arjun Dey). DECaLS, BASS and MzLS together include data obtained, respectively, at the Blanco telescope, Cerro Tololo Inter-American Observatory, NSF’s NOIRLab; the Bok telescope, Steward Observatory, University of Arizona; and the Mayall telescope, Kitt Peak National Observatory, NOIRLab. The Legacy Surveys project is honored to be permitted to conduct astronomical research on Iolkam Du’ag (Kitt Peak), a mountain with particular significance to the Tohono O’odham Nation.

NOIRLab is operated by the Association of Universities for Research in Astronomy (AURA) under a cooperative agreement with the National Science Foundation.

This project used data obtained with the Dark Energy Camera (DECam), which was constructed by the Dark Energy Survey (DES) collaboration. Funding for the DES Projects has been provided by the U.S. Department of Energy, the U.S. National Science Foundation, the Ministry of Science and Education of Spain, the Science and Technology Facilities Council of the United Kingdom, the Higher Education Funding Council for England, the National Center for Supercomputing Applications at the University of Illinois at Urbana-Champaign, the Kavli Institute of Cosmological Physics at the University of Chicago, Center for Cosmology and Astro-Particle Physics at the Ohio State University, the Mitchell Institute for Fundamental Physics and Astronomy at Texas A\&M University, Financiadora de Estudos e Projetos, Fundacao Carlos Chagas Filho de Amparo, Financiadora de Estudos e Projetos, Fundacao Carlos Chagas Filho de Amparo a Pesquisa do Estado do Rio de Janeiro, Conselho Nacional de Desenvolvimento Cientifico e Tecnologico and the Ministerio da Ciencia, Tecnologia e Inovacao, the Deutsche Forschungsgemeinschaft and the Collaborating Institutions in the Dark Energy Survey. The Collaborating Institutions are Argonne National Laboratory, the University of California at Santa Cruz, the University of Cambridge, Centro de Investigaciones Energeticas, Medioambientales y Tecnologicas-Madrid, the University of Chicago, University College London, the DES-Brazil Consortium, the University of Edinburgh, the Eidgenossische Technische Hochschule (ETH) Zurich, Fermi National Accelerator Laboratory, the University of Illinois at Urbana-Champaign, the Institut de Ciencies de l’Espai (IEEC/CSIC), the Institut de Fisica d’Altes Energies, Lawrence Berkeley National Laboratory, the Ludwig Maximilians Universitat Munchen and the associated Excellence Cluster Universe, the University of Michigan, NSF’s NOIRLab, the University of Nottingham, the Ohio State University, the University of Pennsylvania, the University of Portsmouth, SLAC National Accelerator Laboratory, Stanford University, the University of Sussex, and Texas A\&M University.

BASS is a key project of the Telescope Access Program (TAP), which has been funded by the National Astronomical Observatories of China, the Chinese Academy of Sciences (the Strategic Priority Research Program “The Emergence of Cosmological Structures” Grant \# XDB09000000), and the Special Fund for Astronomy from the Ministry of Finance. The BASS is also supported by the External Cooperation Program of Chinese Academy of Sciences (Grant \# 114A11KYSB20160057), and Chinese National Natural Science Foundation (Grant \# 11433005).

The Legacy Survey team makes use of data products from the Near-Earth Object Wide-field Infrared Survey Explorer (NEOWISE), which is a project of the Jet Propulsion Laboratory/California Institute of Technology. NEOWISE is funded by the National Aeronautics and Space Administration.

The Legacy Surveys imaging of the DESI footprint is supported by the Director, Office of Science, Office of High Energy Physics of the U.S. Department of Energy under Contract No. DE-AC02-05CH1123, by the National Energy Research Scientific Computing Center, a DOE Office of Science User Facility under the same contract; and by the U.S. National Science Foundation, Division of Astronomical Sciences under Contract No. AST-0950945 to NOAO.

JW acknowledges support by the Deutsche Forschungsgemeinschaft (DFG, Ger-man Research Foundation) under Germany’s Excellence Strategy - EXC-2094 -390783311.

 \end{acknowledgements}

%
%
\bibliography{optid_refs}



\appendix

\section{Richness--mass scaling relation}\label{app:scaling}

\begin{figure}
\begin{center}
\includegraphics[width=0.99\linewidth]{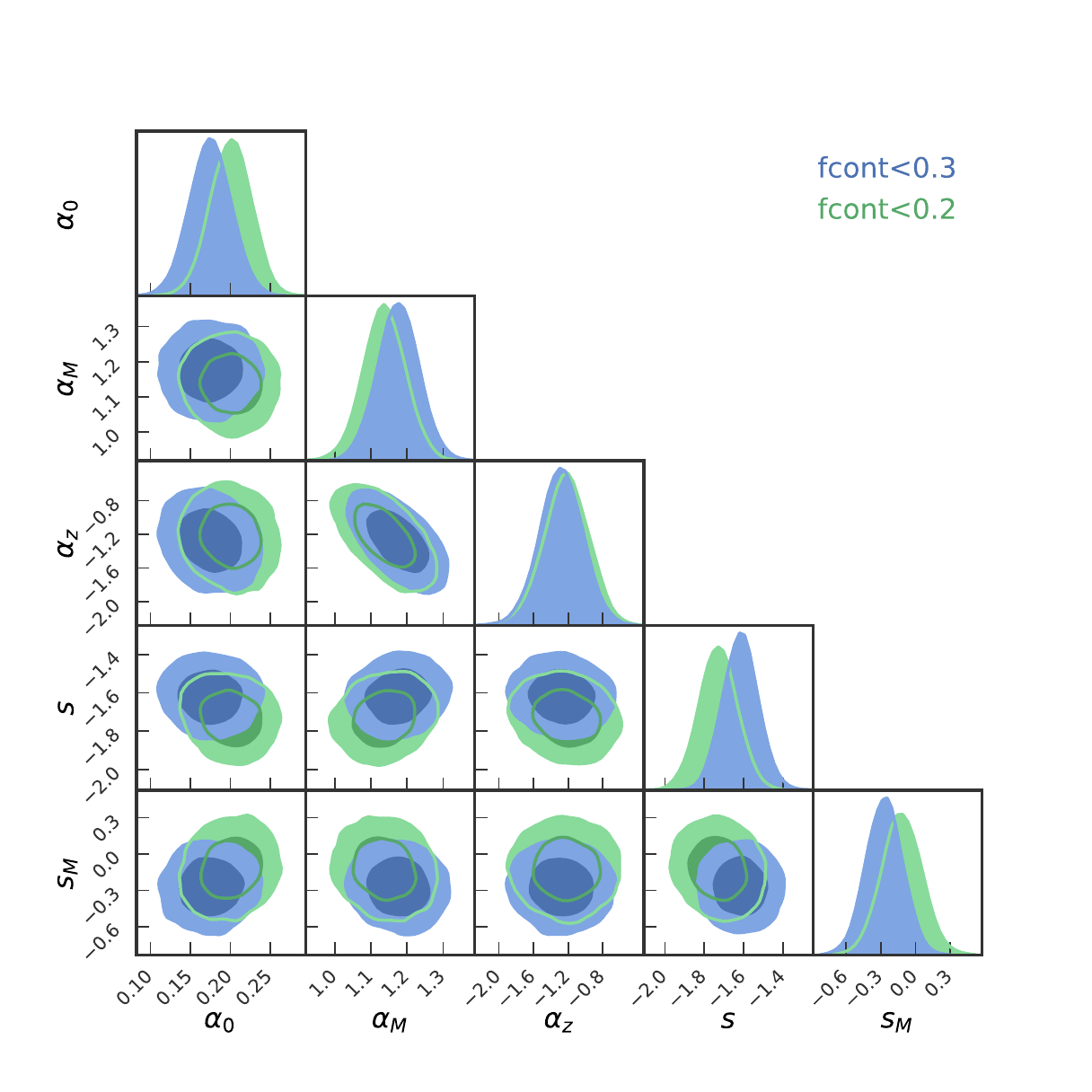}
\caption{Plots of the posteriors for parameters of the richness--mass scaling relation fit using the sample with \fcont$<0.2\,(0.3)$ in green (blue). The two sets of posteriors are consistent. We expect that the minor differences are due to the larger number of contaminating random line-of-sight superpositions in the \fcont$<0.3$ case. }\label{fig:lam-mass_contour}
\end{center}
\end{figure}

\begin{figure}
\begin{center}
\includegraphics[width=0.99\linewidth]{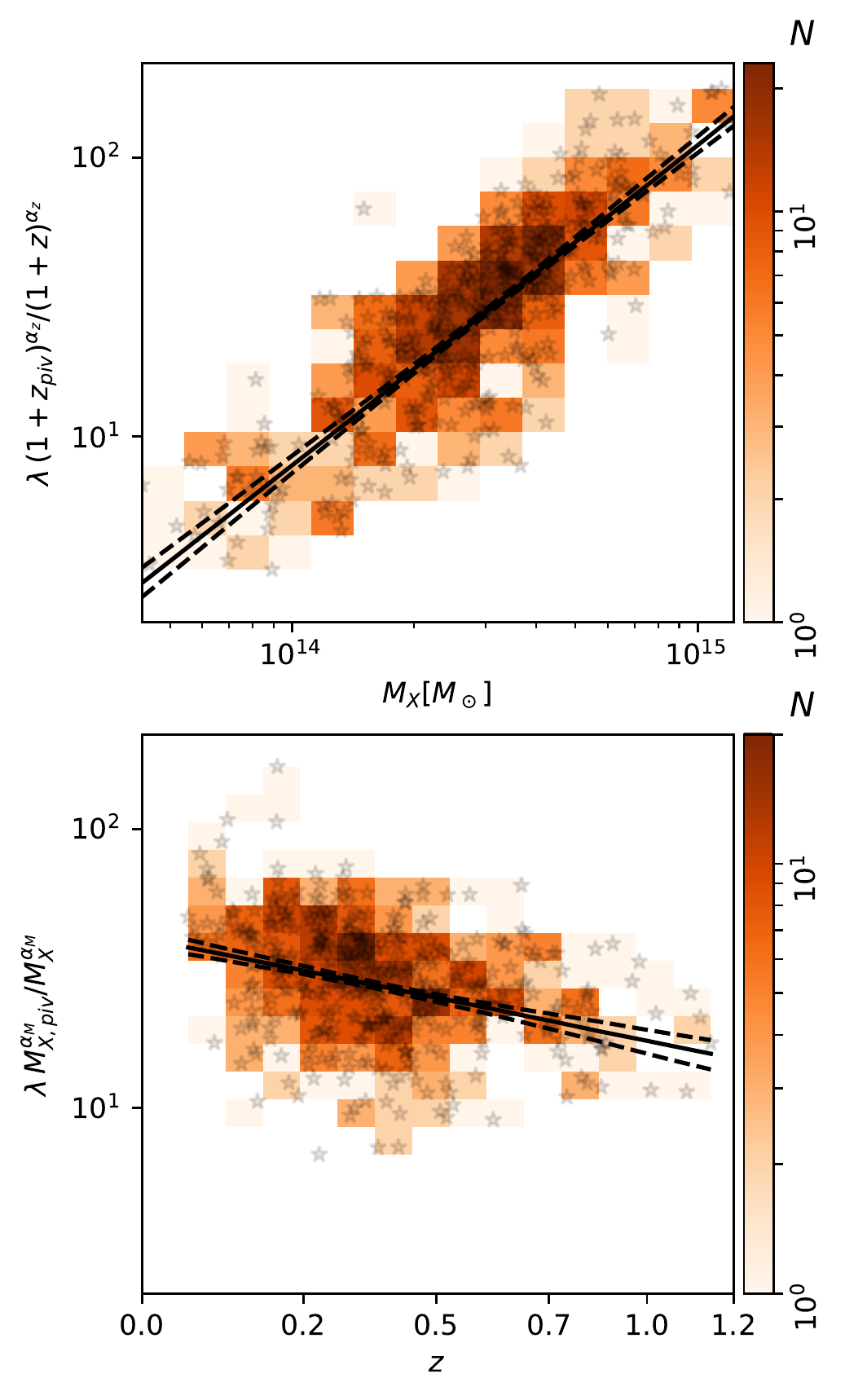}
\caption{\textit{Upper panel:} Redshift trend corrected richness versus mass for the \fcont$<0.2$. The best fit mass trend (solid black) and 1-$\sigma$ uncertainty (dashed black) are shown. \textit{Lower panel:} Mass trend corrected richness versus redshift for the \fcont$<0.2$ sample with best fit redshift trend and uncertainty as above. }\label{fig:lam-mass_trends}
\end{center}
\end{figure}

In this section we provide some further details and results on the scaling relation used in Sect.\ref{sec:incomp}.
To fit the free parameters of the scaling relation, we set up a likelihood for each cluster $i$ to conform with the assumed relation,
\begin{equation}\label{eq:SR_single_cl_like}
    \mathcal{L}_i = C_i^{-1} P(\ln \lambda_i | M_{500,i}, z_i),
\end{equation}
where $(\lambda_i,\, M_{500,i},\, z_i)$ are the richness, mass, and redshift of the cluster, respectively. The normalisation $C_i$ is required to ensure that the likelihood is normalised over the range of possible richnesses.  We remind the reader that the likelihood is the probability of data (in this case $\lambda_i$) given the model (here $(M_{500,i},\, z_i, \,\alpha_0,\, \alpha_M,\, \alpha_z,\, s,\, s_M)$). As such it needs to be normalised over all possible data. In our case, that means that the richness has to be larger than the richness cut associated with the optical cleaning, i.e.  $\lambda>\lambda_\text{cut}(z_i)$. Thus,
\begin{equation}
    C_i = \int_{\ln\lambda_\text{cut}(z_i)}^{\inf}\text{d}\ln \lambda\, P(\ln \lambda | M_{500,i}, z_i).
\end{equation}
This term ensures proper accounting for the incompleteness induced by the optical cleaning. 

Note also, that the X-ray selection function does not impact this calculation. $M_{500}$ is an analytical function of the measured count rate. Any X-ray selection can thus be expressed as some function $\mathcal{S}_X(M_{500}, z)$. If we explicitly account for the X--ray selection function, it appears in both the numerator and the demoninator (normalisation condition) of Eq.~\ref{eq:SR_single_cl_like}, and therefore the selection cancels out.

Assuming a wide, flat prior for all parameters, we sample a posterior of the scaling relation parameters for the sample selected by \fcont$<0.2\,(0.3)$, with the additional redshift cut of $z<1.2$. The marginal contours of the posterior plot are shown in Fig.~\ref{fig:lam-mass_contour}. The \fcont$<0.2$ sample is by construction a subset of the \fcont$<0.3$ sample. As such, the two posteriors are far from being independent. The difference between the posteriors are much smaller than 1-$\sigma$. We interpret the small difference to be due to the residual contamination that is included by the more lenient \fcont$<0.3$ cut. For this reason, in the following we use the posterior derived from the \fcont$<0.2$ sample.

To investigate the adequacy of our best fit mass and redshift trends, we plot in Fig.~\ref{fig:lam-mass_trends} the mass/redshift-trend corrected richnesses to highlight the redshift/mass-trend. We also overplot the best fit mass/redshift trends. They match the data nicely, showing the adequacy of our best fit. Note that the mass proxy used here is nothing other than a modified count rate. From our eFEDS simulations we know that measured count rates are biased with respect to true count rates where the bias depends on the input count rate, size and signal to noise. This could bias the redshift trend. In fact, preliminary results using weak gravitational lensing (Chiu et al., in prep.) suggest no redshift evolution of the lambda-mass relation.

\section{MCMF X-ray point like sources}\label{app:mcmfonps}

\begin{figure}
\begin{center}
\includegraphics[width=0.99\linewidth]{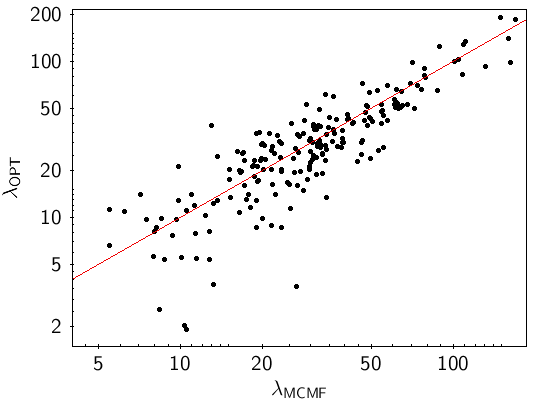}
\includegraphics[width=0.99\linewidth]{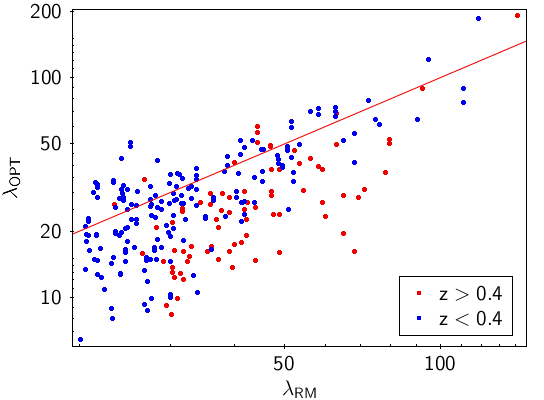}
\caption{Top: Comparison between the optical only estimated richness $\lambda_\mathrm{OPT}$ and the default MCMF richness (which uses X-ray prior). Bottom: Similar but comparing to richnesses from matches to the SDSS redMaPPer sample. The red line indicates the one-to-one relation.}
\label{fig:richcomb}
\end{center}
\end{figure}

The MCMF run on the point like X-ray candidate catalog was performed in two steps. First, MCMF was run in normal mode using the X-ray source counts as a mass proxy. To avoid unrealistically large apertures, the maximum count rate was limited to that of a typical ext\_like=50 source. True clusters brighter than that would very likely be detected as extended sources in eFEDS.
The main aim of that MCMF run is to provide redshifts for possible counterparts. In the second step these redshifts and the X-ray positions are used to 
calculate richnesses. Similar to the approach in redMaPPer \citep{redmapper}, we define our richness via a scaling relation between aperture $r_\lambda$ and the optical only based richness $\lambda_\mathrm{OPT}$ as
\begin{equation}
    \lg r_\lambda = \left(\lg \lambda_\mathrm{OPT}\right)m +b,
\end{equation}
with a slope $m$ and a normalisation $b$. Our goal here is to calibrate our optical only richness to those obtained using X-ray priors that use a proxy of $r_{500}$. We therefore simply fit for this relation using the MCMF results on the extent selected sample and find $m=0.29$ and $b=-0.496$ as suitable parameters to define our relation.
Due to the high point source density and source splitting of extended sources there is a significant overlap between the extent selected sample and the point source selected sample. We find 206 matches within 2 arcminutes and within a redshift offset of $\Delta z<0.04$. The richnesses of the optical only versus the default MCMF richnesses are shown in the left panel of Fig.\ref{fig:richcomb}.
We also match the point source sample with the SDSS redMaPPer \citep{redmapper} sample using a 90~arcsecond search radius and the same maximum redshift offset as before. The comparison to this sample is shown in the right panel of Fig.\ref{fig:richcomb}. Incompleteness in SDSS starts to impact the richness estimate in redMaPPer for $z>0.4$, causing increased scatter and a shift in the richness compared to the MCMF optical only richness $\lambda_\mathrm{OPT}$. For redshifts below $z=0.4$ the richnesses between both methods seem to agree.

\section{Additional results on dynamical state estimators}\label{app:dyns}

\begin{figure*}
\begin{center}
\includegraphics[width=0.99\linewidth]{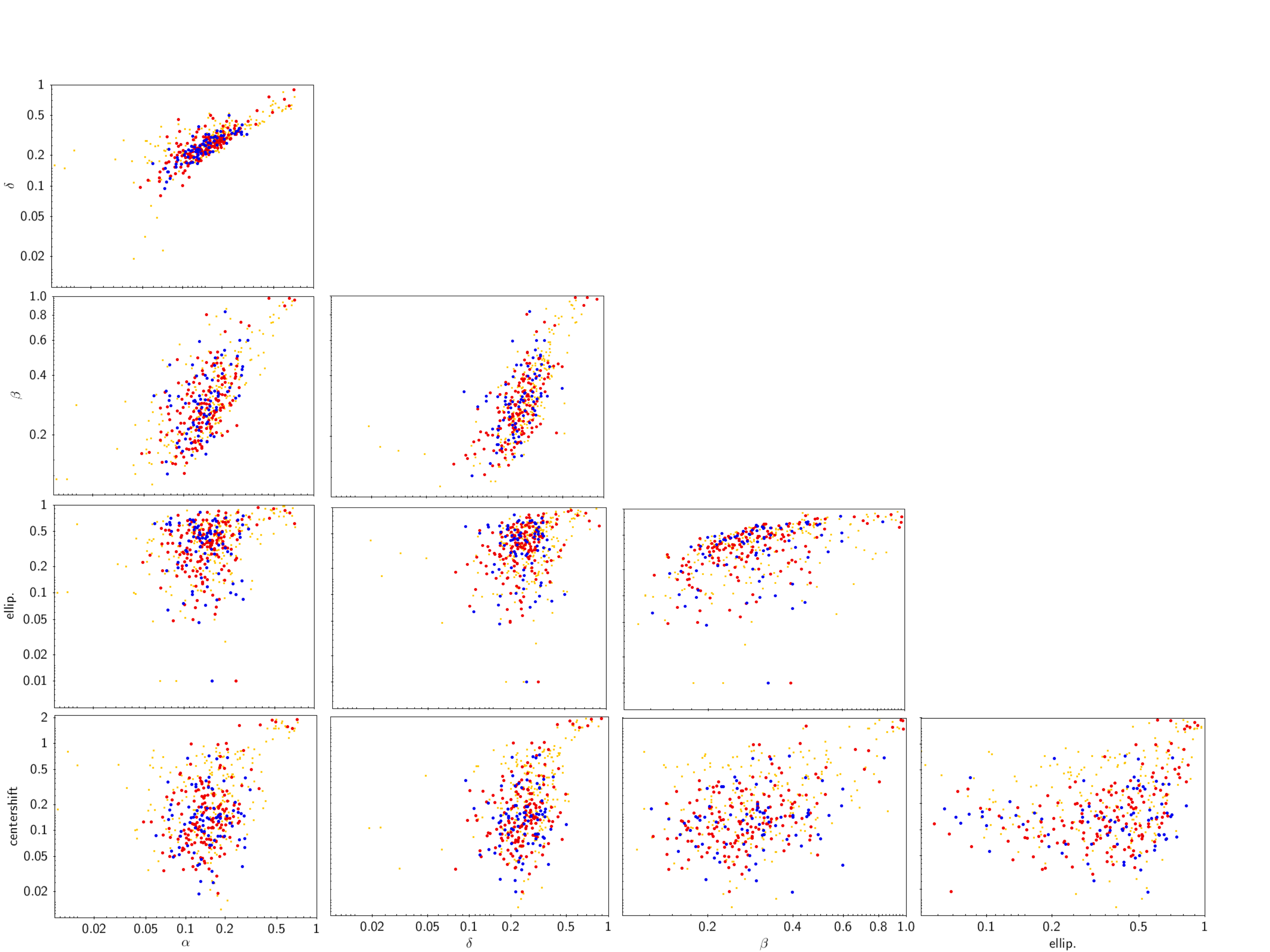}
\caption{Comparison of the five different optical estimators of cluster dynamical state that are based on model fitting to the red sequence galaxy density map. Color coded are same richness thresholds as in Fig.~\ref{fig:dynestimatorcorr}: yellow for $\lambda<25$, red for $25\lambda<50$ and blue for $\lambda>50$. 
}
\label{fig:dynestimatorcorr}
\end{center}
\end{figure*}

In Fig.~\ref{fig:dynestimatorcorr} we show the correlation between different estimators that probe the cluster morphology. All estimators are based on the galaxy density maps and are therefore prone to similar systematics such as projections, masks or other artifacts. The estimators themselves probe different properties that are associated with unrelaxed systems. The three \citet{Wen13} based estimators (Greek letters), focus on deviation from symmetry or from the adopted 2D-model (see Sect.~\ref{sec:optdynest}). The ellipticity and the center shift with respect to the X-ray position are mostly independent of the first set of estimators as they also trigger on perfectly symmetric, well modeled clusters but with unusual shape or offset. Estimates become more noisy and therefore less reliable with lower richness and higher redshifts. We therefore suggest that these estimators are most useful when one adopts a rather high richness threshold, e.g., $\lambda>50$, and restricts to redshifts $z<0.8$.

In Table~\ref{tab:Merglgt50} we show the 25 most disturbed eFEDS clusters with $\lambda>50$ according to the sum of the estimators $\alpha$, $\delta$, $\beta$, ellipticity and center shift, called $S_\mathrm{DYN}$. We further provide a flag if another eFEDS cluster is nearby. The flag is 1 if another eFEDS cluster is nearby, 2 if the nearby cluster has a higher disturbance estimate and 3 if there are two other nearby clusters with higher disturbance estimates. With that, Table~\ref{tab:Merglgt50} shows 21 individual merging systems, where two systems appear as three eFEDS sources in the table. Furthermore, the table shows that 10 systems do have at least one other eFEDS cluster nearby.

Table~\ref{tab:Mergpairs} shows cluster pairs and potential mergers using the requirement of having another optical and X-ray detection within 2.5*$r_{500}$, as suggested in Sect.\ref{sec:mergers}. The list is sorted by $S_\mathrm{DYN}$. We find 29 eFEDS clusters fulfilling the selection, the list contains 17 individual systems.

\begin{table*}
\caption{Top 25 most unrelaxed eFEDS clusters with $\lambda>50$, according to $S_\mathrm{DYN}$. }\label{tab:Merglgt50}
\begin{center}
\begin{tabular}{l  c  c  c  c  c  c  c  c  c  c c }
\hline
\hline
  \multicolumn{1}{ l }{NAME} &
  \multicolumn{1}{c }{RA} &
  \multicolumn{1}{c }{DEC} &
  \multicolumn{1}{c }{z} &
  \multicolumn{1}{c }{$\lambda$} &
  \multicolumn{1}{c }{$\alpha$} &
  \multicolumn{1}{c }{$\delta$} &
  \multicolumn{1}{c }{$\beta$} &
  \multicolumn{1}{c }{Ellip.} &
  \multicolumn{1}{c }{cent. shift} &
  \multicolumn{1}{c }{$S_\mathrm{DYN}$} &
  \multicolumn{1}{c }{Flag} \\
\hline
eFEDS~J093431.3-002310	&	143.630	&	-0.386	&	0.335	&	54.6	&	0.28	&	0.34	&	1.04	&	0.75	&	0.26	&	2.67	&	1	\\
eFEDS~J092220.4+034806	&	140.585	&	3.802	&	0.269	&	51.9	&	0.21	&	0.29	&	0.84	&	0.71	&	0.41	&	2.46	&	1	\\
eFEDS~J085620.8+014650	&	134.086	&	1.780	&	0.732	&	70.9	&	0.11	&	0.23	&	1.04	&	0.69	&	0.32	&	2.38	&	1	\\
eFEDS~J091302.2+035001	&	138.259	&	3.834	&	0.455	&	59.3	&	0.31	&	0.32	&	0.60	&	0.53	&	0.31	&	2.07	&	1	\\
eFEDS~J084910.6+024117	&	132.294	&	2.688	&	0.830	&	68.0	&	0.13	&	0.21	&	0.59	&	0.82	&	0.29	&	2.05	&		\\
eFEDS~J092209.4+034629	&	140.539	&	3.775	&	0.268	&	95.6	&	0.26	&	0.35	&	0.50	&	0.67	&	0.26	&	2.04	&	2	\\
eFEDS~J084823.3+041205	&	132.097	&	4.201	&	0.872	&	90.8	&	0.24	&	0.33	&	1.04	&	0.13	&	0.29	&	2.03	&		\\
eFEDS~J092046.2+002849	&	140.193	&	0.480	&	0.400	&	55.5	&	0.27	&	0.37	&	0.60	&	0.43	&	0.26	&	1.92	&		\\
eFEDS~J084459.3-011903	&	131.247	&	-1.317	&	0.447	&	53.1	&	0.28	&	0.32	&	0.43	&	0.59	&	0.26	&	1.89	&	2	\\
eFEDS~J091305.9+035022	&	138.275	&	3.839	&	0.454	&	93.9	&	0.18	&	0.30	&	0.49	&	0.65	&	0.27	&	1.89	&		\\
eFEDS~J090328.7-013622	&	135.870	&	-1.606	&	0.443	&	72.3	&	0.17	&	0.28	&	0.45	&	0.70	&	0.28	&	1.87	&		\\
eFEDS~J092202.3+034520	&	140.510	&	3.756	&	0.268	&	59.6	&	0.20	&	0.34	&	0.51	&	0.50	&	0.33	&	1.87	&	3	\\
eFEDS~J092339.1+052655	&	140.913	&	5.449	&	0.373	&	56.1	&	0.17	&	0.30	&	0.38	&	0.67	&	0.26	&	1.78	&		\\
eFEDS~J083933.8-014044	&	129.891	&	-1.679	&	0.279	&	78.6	&	0.12	&	0.25	&	0.45	&	0.67	&	0.28	&	1.77	&	1	\\
eFEDS~J084637.1-002257	&	131.655	&	-0.382	&	0.293	&	52.7	&	0.08	&	0.23	&	0.45	&	0.71	&	0.28	&	1.75	&		\\
eFEDS~J091315.0+034850	&	138.313	&	3.814	&	0.444	&	88.8	&	0.15	&	0.32	&	0.39	&	0.43	&	0.42	&	1.71	&	3	\\
eFEDS~J084223.1+003341	&	130.596	&	0.561	&	1.077	&	52.8	&	0.17	&	0.30	&	0.33	&	0.50	&	0.41	&	1.71	&		\\
eFEDS~J083125.9+015533	&	127.858	&	1.926	&	0.684	&	61.6	&	0.12	&	0.31	&	0.30	&	0.65	&	0.32	&	1.69	&	1	\\
eFEDS~J082820.5-000722	&	127.086	&	-0.123	&	0.845	&	61.5	&	0.27	&	0.35	&	0.32	&	0.47	&	0.28	&	1.68	&	1	\\
eFEDS~J092049.5+024514	&	140.206	&	2.754	&	0.285	&	79.0	&	0.21	&	0.32	&	0.54	&	0.32	&	0.28	&	1.66	&	1	\\
eFEDS~J090915.3-010104	&	137.314	&	-1.018	&	0.822	&	54.6	&	0.23	&	0.35	&	0.45	&	0.28	&	0.28	&	1.60	&		\\
eFEDS~J092212.1-002731	&	140.550	&	-0.459	&	0.318	&	108.1	&	0.18	&	0.31	&	0.27	&	0.55	&	0.27	&	1.59	&	1	\\
eFEDS~J084246.9-000917	&	130.696	&	-0.155	&	0.415	&	61.3	&	0.29	&	0.40	&	0.44	&	0.08	&	0.33	&	1.55	&		\\
eFEDS~J093207.6-021317	&	143.032	&	-2.221	&	0.666	&	57.5	&	0.11	&	0.19	&	0.36	&	0.61	&	0.28	&	1.55	&		\\
eFEDS~J090129.2-013854	&	135.372	&	-1.648	&	0.318	&	104.7	&	0.20	&	0.31	&	0.41	&	0.36	&	0.27	&	1.55	&	1	\\
\hline\end{tabular}
\end{center}
\tablefoot{The table is sorted by the value $S_\mathrm{DYN}$, the linear combination of the dynamical state estimators $\alpha$, $\delta$, $\beta$, ellipticity and center shift. It lists the 25 eFEDS clusters with \fcont$<0.3$ and $\lambda>50$. Meaning of the flag values: 1) at lease one other eFEDS source is nearby, 2) another eFEDS source with $\lambda>50$ is nearby and has higher $S_\mathrm{DYN}$. 3) same as 2) but having two eFEDS clusters with higher $S_\mathrm{DYN}$. The postions are given in J2000 system.}
\end{table*}

\begin{table*}
\caption{Close cluster pairs or mergers, selected by having  an optical and X-ray neighbour within 2.5$r_{500}$.}\label{tab:Mergpairs}
\begin{center}
\begin{tabular}{l c c c c c c c c}
\hline
  \multicolumn{1}{c}{NAME} &
  \multicolumn{1}{c}{RA} &
  \multicolumn{1}{c}{DEC} &
  \multicolumn{1}{c}{z} &
  \multicolumn{1}{c}{$\lambda$} &
  \multicolumn{1}{c}{$D_\mathrm{OPT}$} &
  \multicolumn{1}{c}{$D_\mathrm{X}$} &
  \multicolumn{1}{c}{$S_\mathrm{DYN}$} & 
  \multicolumn{1}{c}{Flag} \\
    \multicolumn{1}{c}{} &
  \multicolumn{1}{c}{} &
  \multicolumn{1}{c}{} &
  \multicolumn{1}{c}{} &
  \multicolumn{1}{c}{} &
  \multicolumn{1}{c}{($R_{500}$)} &
  \multicolumn{1}{c}{($R_{500}$)} &
  \multicolumn{1}{c}{} & 
  \multicolumn{1}{c}{} \\
\hline
\hline
eFEDS~J085541.3+002740	&	133.922	&	0.461	&	0.156	&	8.7	&	0.94	&	1.63	&	3.59	&	1	\\
eFEDS~J091415.0+022710	&	138.562	&	2.453	&	0.330	&	19.8	&	0.66	&	0.89	&	3.09	&	1	\\
eFEDS~J090146.2-013756	&	135.443	&	-1.632	&	0.304	&	38.5	&	1.39	&	1.23	&	2.77	&	1	\\
eFEDS~J091354.8+025323	&	138.478	&	2.890	&	0.423	&	24.5	&	0.87	&	1.55	&	2.69	&	1	\\
eFEDS~J092220.4+034806	&	140.585	&	3.802	&	0.269	&	51.9	&	2.12	&	0.90	&	2.46	&	1	\\
eFEDS~J085620.8+014650	&	134.086	&	1.780	&	0.732	&	70.9	&	1.93	&	1.89	&	2.38	&	1	\\
eFEDS~J083921.1-014149	&	129.838	&	-1.697	&	0.278	&	39.5	&	0.93	&	0.84	&	2.19	&	1	\\
eFEDS~J091302.2+035001	&	138.259	&	3.834	&	0.455	&	59.3	&	1.06	&	0.37	&	2.07	&	1	\\
eFEDS~J092209.4+034629	&	140.539	&	3.775	&	0.268	&	95.6	&	2.05	&	0.43	&	2.04	&	2	\\
eFEDS~J090137.7+030254	&	135.407	&	3.048	&	0.188	&	16.4	&	1.23	&	0.68	&	2.01	&	1	\\
eFEDS~J091305.9+035022	&	138.275	&	3.839	&	0.454	&	93.9	&	0.99	&	0.34	&	1.89	&	2	\\
eFEDS~J092202.3+034520	&	140.510	&	3.756	&	0.268	&	59.6	&	0.91	&	0.55	&	1.87	&	3	\\
eFEDS~J091213.4-021621	&	138.056	&	-2.273	&	0.160	&	25.5	&	1.28	&	0.29	&	1.80	&	1	\\
eFEDS~J083933.8-014044	&	129.891	&	-1.679	&	0.279	&	78.6	&	0.86	&	0.64	&	1.77	&	2	\\
eFEDS~J091851.7+021432	&	139.716	&	2.242	&	0.280	&	28.4	&	1.15	&	0.82	&	1.76	&	1	\\
eFEDS~J091315.0+034850	&	138.313	&	3.814	&	0.444	&	88.8	&	0.78	&	0.83	&	1.71	&	3	\\
eFEDS~J083930.3-014349	&	129.876	&	-1.730	&	0.271	&	11.3	&	0.59	&	1.03	&	1.70	&	3	\\
eFEDS~J091358.2+025707	&	138.492	&	2.952	&	0.435	&	19.0	&	1.95	&	1.38	&	1.63	&	2	\\
eFEDS~J090131.2+030057	&	135.380	&	3.016	&	0.194	&	62.3	&	0.47	&	0.43	&	1.53	&	2	\\
eFEDS~J090750.2+025006	&	136.959	&	2.835	&	0.648	&	16.0	&	1.56	&	1.51	&	1.53	&	1	\\
eFEDS~J085751.7+031039	&	134.465	&	3.178	&	0.198	&	97.9	&	1.70	&	2.13	&	1.52	&	1	\\
eFEDS~J083806.9-003601	&	129.529	&	-0.600	&	0.434	&	18.0	&	1.15	&	0.71	&	1.43	&	1	\\
eFEDS~J085436.6+003835	&	133.653	&	0.643	&	0.110	&	31.8	&	0.74	&	0.24	&	1.41	&	1	\\
eFEDS~J093009.0+040144	&	142.538	&	4.029	&	0.342	&	21.6	&	1.44	&	1.54	&	1.39	&	1	\\
eFEDS~J093003.3+035630	&	142.514	&	3.942	&	0.330	&	34.9	&	1.43	&	1.45	&	1.38	&	2	\\
eFEDS~J093513.1+004758	&	143.805	&	0.799	&	0.364	&	169.0	&	1.45	&	1.55	&	1.35	&	1	\\
eFEDS~J085433.0+004009	&	133.638	&	0.669	&	0.113	&	20.4	&	1.05	&	0.35	&	1.28	&	2	\\
eFEDS~J085627.2+014218	&	134.113	&	1.705	&	0.732	&	100.5	&	1.08	&	1.67	&	1.26	&	2	\\
eFEDS~J093500.8+005417	&	143.753	&	0.905	&	0.381	&	54.8	&	1.93	&	2.04	&	1.26	&	2	\\
\hline\end{tabular}
\end{center}
\tablefoot{The table is sorted by the value $S_\mathrm{DYN}$, the linear combination of the dynamical state estimators $\alpha$, $\delta$, $\beta$, ellipticity and center shift. Beside \fcont$<0.3$, the list is selected by the distance to the next optical and X-ray to be $D_\mathrm{OPT}<2.5$ and $D_\mathrm{X}<2.5$. Flags are the same as in Table~\ref{tab:Merglgt50}. The postions are given in J2000 system.}
\end{table*}

\section{Column description of the results tables}
Because the tables containing the main results are too long and include too many columns, we provide them only in electronic form on CDS and the official eROSITA web site.
To give the reader at least a brief overview of available information we provide here a short description of the key entries of these tables.

In Table~\ref{tab:columns} we list the key entries of the results table. This includes the most essential entries from
the X-ray catalog that were used in this work. Further, we list entries related to the best optical counterpart, based on the combination of the MCMF runs on HSC and LS. We note here that similar entries also exist for the second best optical counterpart, which contains "2BEST" in their column names.


\begin{table*}
\caption{Column names and description of the entries of the main results table.}\label{tab:columns}
\begin{center}
\begin{tabular}{l l}
\hline
\hline
  \multicolumn{1}{l}{Column name} &
  \multicolumn{1}{l}{Description} \\
 \hline
Name	&	eROSITA source name	\\
ID\_SRC	&	Source ID of detection pipeline	\\
RA\_CORR & RA of the X-ray center \\
DEC\_CORR & DEC of the X-ray center \\
EXT	&	Source extent in X-ray	\\
EXT\_ERR	&	Error in source extent	\\
EXT\_LIKE	&	Extent likelihood	\\
DET\_LIKE\_0	&	Detection likelihood	\\
ML\_RATE\_0	&	X-ray count rate	\\
ML\_RATE\_ERR\_0	&	error on count rate	\\
F\_CONT\_BEST\_COMB	&	\fcont\ of best optical counter part: -1, -2 for manually added systems	\\
Z\_BEST\_COMB	&	Photometric redshift of best counter part	\\
SIGMA\_Z\_BEST\_COMB	&	Uncertainty of the photometric redshift	\\
LAMBDA\_BEST\_COMB	&	Richness of best counter part	\\
ELAMBDA\_BEST\_COMB	&	Uncertainty richness	\\
SURV\_BEST\_COMB	&	Survey used: -1 group catalog, 1 LS, 2 HSC	\\
RA\_OPTCEN\_BEST\_COMB	&	Ra of optical center of best counter part	\\
DEC\_OPTCEN\_BEST\_COMB	&	Same but DEC	\\
SPEC\_Z\_BEST\_COMB	&	Spec-z	\\
N\_SPEC	&	Number of redshifts used for spec-z	\\
MASSPROX\_BEST\_COMB	&	Approx. M500 used for richness estimate [1/h]	\\
FLAG\_OPTICAL\_X\_POS	&	Value of the optical footprint map at X-ray position	\\
MASKFRAC\_3\_FOOT	&	Fraction of the area not int the optical footprint within a 3 arcmin radiuss	\\
MASKFRAC\_3\_FLAGGED	&	Similar but including flagged area due to bright star masks	\\
MASKFRAC\_R500\_FOOT\_BEST\_COMB	&	Same as MASKFRAC\_3\_FOOT but for a r\_500 sized region	\\
MASKFRAC\_R500\_FLAGGED\_BEST\_COMB	&	Same as MASKFRAC\_3\_FLAGGED but for a r\_500 sized region	\\
ALPHA	&	Dynamical state estimator alpha	\\
BETA	&	Dynamical state estimator beta	\\
DELTA	&	Dynamical state estimator delta	\\
ELLIP	&	Dynamical state estimator: model ellipticity	\\
CENTERSHIFT	&	Shift of the centre of the 2D density model with respect to X-ray position	\\
DIST\_NEXT\_OPT	&	Distance to next optical structure in the galaxy density map	\\
DIST\_NEXT\_XCLUST	&	Distance to next extent selectect eFEDS cluster	\\

\hline\end{tabular}
\end{center}
\end{table*}

%

\end{document}